\RequirePackage{fix-cm}
\documentclass[twocolumn]{svjour3}          
\smartqed  
\usepackage{graphicx}
\usepackage{amssymb}
\usepackage[draft]{hyperref}

\usepackage[dvipsnames]{xcolor} 

\newcommand{\lam}{$\lambda$}
\newcommand{\1}{\footnotesize I\normalsize}
\newcommand{\2}{\footnotesize II\normalsize}
\newcommand{\3}{\footnotesize III\normalsize}
\newcommand{\4}{\footnotesize IV\normalsize}

\newcommand{\6}{\footnotesize VI\normalsize}
\newcommand{\7}{\footnotesize VII\normalsize}

\newcommand{\kms}{km\,s$^{-1}$}

\newcommand{\feii}{\mbox{Fe\,{\sc ii}}}
\newcommand{\znii}{\mbox{Zn\,{\sc ii}}}
\newcommand{\niii}{\mbox{Ni\,{\sc ii}}}
\newcommand{\mgii}{\mbox{Mg\,{\sc ii}}}
\newcommand{\oii}{\mbox{O\,{\sc ii}}}

\begin{document}

\title{The CUBES Science Case
}


\author{Chris Evans \and
Stefano Cristiani \and
Cyrielle Opitom \and
Gabriele Cescutti \and
Valentina D'Odorico \and
Juan Manuel Alcal\'{a} \and
Silvia H. P. Alencar \and  
Sergei Balashev \and
Beatriz Barbuy \and
Nate Bastian \and
Umberto Battino \and
Pamela Cambianica \and
Roberta Carini \and
Brad Carter \and
Santi Cassisi \and
Bruno Vaz Castilho \and
Norbert Christlieb \and
Ryan Cooke \and
Stefano Covino \and
Gabriele Cremonese \and
Katia Cunha \and
Andr\'e R.~da Silva \and
Valerio D'Elia \and 
Annalisa De Cia \and
Gayandhi De Silva \and
Marcos Diaz \and
Paolo Di Marcantonio \and
Heitor Ernandes \and
Alan Fitzsimmons \and
Mariagrazia Franchini \and
Boris T.~G\"{a}nsicke \and 
Matteo Genoni \and
Riano E.~Giribaldi \and
Andrea Grazian \and
Camilla Juul Hansen \and
Fiorangela La Forgia \and
Monica Lazzarin \and
Wagner Marcolino \and
Marcella Marconi \and
Alessandra Migliorini \and
Pasquier Noterdaeme \and
Claudio Pereira \and
Bogumil Pilecki \and
Andreas Quirrenbach \and
Sofia Randich \and 
Silvia Rossi \and
Rodolfo Smiljanic \and
Colin Snodgrass \and
Julian St\"{u}rmer \and
Andrea Trost \and
Eros Vanzella \and
Paolo Ventura \and
Duncan Wright \and
Tayyaba Zafar}

\authorrunning{Evans et al.} 

\institute{C. Evans\at
European Space Agency (ESA), ESA Office, Space Telescope Science Institute, 3700 San Martin Drive, Baltimore, MD 21218, USA \\
UK Astronomy Technology Centre, Royal Observatory, Blackford Hill, Edinburgh, EH9 3HJ, UK\\
\email{chevans@stsci.edu}           \\
S. Cristiani, V. D'Odorico, P. Di Marcantonio, M. Franchini and A. Trost \at
 INAF - Osservatorio Astronomico di Trieste, Via G. B. Tiepolo 11, 34143, Trieste, Italy\\
C. Opitom and C. Snodgrass\at
Institute for Astronomy, University of Edinburgh, Royal Observatory, Blackford Hill, Edinburgh, EH9 3HJ, UK\\
G. Cescutti\at 
Dipartimento di Fisica, Sezione di Astronomia, Università di Trieste, Via G. B. Tiepolo 11, 34143 Trieste, Italy\\
J. M. Alcal\'a and M. Marconi \at
INAF-Osservatorio Astronomico di Capodimonte, via Moiariello 16, 80131 Napoli, Italy\\
S. H. P. Alencar\at
Departamento de Fisica - ICEx - UFMG, Av. Ant\^onio Carlos, 6627, 30270-901, Belo Horizonte, MG, Brazil\\
S.~A. Balashev\at
 Ioffe Institute, Polyteknicheskaya 26, 194021 Saint-Petersburg, Russia\\ 
 HSE University, Saint-Petersburg, Russia \\
B. Barbuy, M. Diaz, H. Ernandes and S. Rossi \at
Universidade de S\~ao Paulo, IAG, Rua do Mat\~ao 1226, Cidade Universit\'aria, S\~ao Paulo 05508-900, Brazil\\
N. Bastian \at
Donostia International Physics Center (DIPC), Paseo Manuel de Lardizabal, 4, 20018, Donostia-San Sebastián, Guipuzkoa, Spain \\ IKERBASQUE, Basque Foundation for Science, E-48013 Bilbao, Spain\\
U. Battino \at
E. A. Milne Centre for Astrophysics, Department of Physics and Mathematics, University of Hull, Hull, HU6~7RX, UK\\
P. Cambianica, G. Cremonese and A. Grazian \at
INAF-Osservatorio Astronomico di Padova, 
Vicolo dell'Osservatorio 5, 35122, Padova, Italy\\
R. Carini and P. Ventura \at
INAF, Osservatorio Astronomico di Roma, Via Frascati 33, 00078, Monte Porzio Catone (RM), Italy \\
B. Carter \at 
Centre for Astrophysics, University of Southern Queensland, University Drive, Springfield Central, QLD 4300, Australia\\
S. Cassisi \at
INAF, Osservatorio Astronomico di Abruzzo, Via M. Maggini, I-64100 Teramo, Italy \\
INFN - Sezione di Pisa, Largo Pontecorvo 3, I-56127 Pisa, Italy \\
B. V. Castilho \at
Laboratório Nacional de Astrofísica MCTI, Rua Estados Unidos 154, Itajubá, MG 37504364, Brazil\\
N. Christlieb, A. Quirrenbach and J. St\"urmer \at
Landessternwarte, Zentrum f\"ur Astronomie der Universit\"at Heidelberg, 
K\"onigstuhl 12, 69117 Heidelberg, Germany\\
R. Cooke \at
Centre for Extragalactic Astronomy, Durham University, Durham DH1 3LE, UK \\
S. Covino and M. Genoni \at
INAF - Osservatorio Astronomico di Brera, via Bianchi 36, 23807, Merate (LC), Italy \\
K. Cunha \at
Observat\'orio Nacional, Rua Gen. Jos\'e Cristino 77, S\~ao Crist\'ov\~ao, 20921-400, Rio de Janeiro, Brazil\\
Steward Observatory, University of Arizona, 950 N. Cherry Ave., Tucson, AZ, 85719 \\
A.~R.~da Silva, R.~E.~Giribaldi, B.~Pilecki and R.~Smiljanic \at
Nicolaus Copernicus Astronomical Center, Polish Academy of Sciences, ul.~Bartycka 18, 00-716, Warsaw, Poland\\
V. D'Elia \at
Italian Space Agency - Space Science Data Centre, via del Politecnico snc 00133 Rome, Italy\\
INAF - Osservatorio Astronomico di Roma, via di Frascati 33, 00040, Monteporzio Catone, Rome, Italy\\
A. De Cia \at
Department of Astronomy, University of Geneva, Chemin Pegasi 51, 1290 Versoix, Switzerland\\
G. De Silva and T. Zafar \at
Australian Astronomical Optics, Macquarie University, 105 Delhi Road, North Ryde, NSW 2113, Australia \\
A. Fitzsimmons \at 
Astrophysics Research Centre, School of Physics and Astronomy, Queen's University Belfast, Belfast BT7 1NN, UK\\
B. G\"ansicke \at University of Warwick, Department of Physics, Gibbet Hill Road, Coventry, CV7 4AL, UK\\
C. J. Hansen \at Goethe University Frankfurt, Institute for Applied Physics, Max-von-Laue-Str. 12, 60438 Frankfurt am Main, Germany\\
F. La Forgia and M. Lazzarin\at
Dipartimento di Fisica e Astronomia-Padova University, Vicolo dell'Osservatorio 3, 35122 Padova, Italy\\
W. Marcolino \at 
Observat\'orio do Valongo, Universidade Federal do Rio de Janeiro, Ladeira Pedro Ant\^onio, 43, Rio de Janeiro, Brazil\\
A. Migliorini \at
INAF - Institute of Space Astrosphysics and Planetology, via Fosso del Cavaliere 100, 00133, Roma, Italy\\
P. Noterdaeme \at
Franco-Chilean Laboratory for Astronomy, Camino El Observatorio 1515, Las Condes, Santiago, Chile \\
3 Institut d’Astrophysique de Paris, CNRS-SU, UMR 7095, 98bis bd Arago, 75014 Paris, France \\
C. B. Pereira \at Observat\'orio Nacional, Rua Gen. Jos\'e Cristino 77, S\~ao Crist\'ov\~ao, 20921-400, Rio de Janeiro, Brazil\\
S. Randich\at 
INAF - Osservatorio Astrofisico di Arcetri, Largo E. Fermi 5, 50125, Firenze, Italy\\
E. Vanzella\at
INAF - OAS, Osservatorio di Astrofisica e Scienza dello Spazio di Bologna, Via Gobetti 93/3, 40129, Bologna, Italy \\
D. Wright \at Centre for Astrophysics, University of Southern Queensland, West Street, Toowoomba, QLD 4350, Australia\\
}

\date{Received: 20 April 2022 / Accepted: 09 July 2022}

\maketitle

\begin{abstract} 
  We introduce the scientific motivations for the development of the
  Cassegrain U-Band Efficient Spectrograph (CUBES) that is now in
  construction for the Very Large Telescope. The assembled cases span
  a broad range of contemporary topics across Solar System, Galactic
  and extragalactic astronomy, where observations are limited by the
  performance of current ground-based spectrographs shortwards of
  400\,nm. A brief background to each case is presented and specific
  technical requirements on the instrument design that flow-down from
  each case are identified. These were used as inputs to the CUBES
  design, that will provide a factor of ten gain in efficiency for
  astronomical spectroscopy over 300-405\,nm, at resolving powers of
  $R$\,$\sim$\,24,000 and $\sim$7,000. We include performance
  estimates that demonstrate the ability of CUBES to observe sources
  that are up to three magnitudes fainter than currently possible at
  ground-ultraviolet wavelengths, and we place its predicted
  performance in the context of existing facillities.
\end{abstract}

\section{Introduction}
\label{intro}
The four 8.2m telescopes of the Very Large Telescope (VLT) at the European Southern Observatory (ESO) are the world’s most scientifically productive ground-based observatory in the visible and infrared. Looking to the future of the VLT there is a long-standing aspiration for an optimised ultraviolet (UV) spectrograph (e.g. \cite{bar14}). Here we introduce the scientific motivations for the Cassegrain U-Band Efficient Spectrograph (CUBES), that has been designed to provide high-efficiency observations in the near UV (300-405\,nm) at spectral resolving powers of $R$\,$\ge$\,20,000 and $R$\,$\sim$\,7,000. 

CUBES will provide the VLT with a unique capability and will open-up new discovery space for a broad range of astrophysics. The near UV contains a tremendous diversity of iron-peak and heavy elements in stellar spectra, as well as some lighter elements (notably beryllium) and light-element molecules (CO, CN, OH). The near-UV is also important for extragalactic science, such as studies of the interstellar and circumgalactic medium (CGM) of distant galaxies, in measuring the contribution of different types of galaxies to the cosmic UV background, and follow-up of explosive transients. Closer to home, significantly improved sensitivity at the shortest (ground-based) wavelengths will also provide exciting new opportunities in solar system science.

The CUBES project completed a Phase~A conceptual design in June 2021 and has now entered the detailed design and construction phase; first science operations are planned for 2028. Alongside an overview of the CUBES design in this Special Issue \cite{ExA_Zanutta}, this article introduces the science cases developed during Phase~A that motivated the instrument design and informed its top-level requirements (TLRs). The cases assembled by the Science Team were grouped into four fields: solar system, Galactic, extragalactic and transients. Table~\ref{sci_summary} summarises the topics within each field, and highlights those that are presented in more depth elsewhere in this Special Issue. While the design was informed by the cases outlined here, we envisage CUBES more generally as providing the VLT with an exciting new capability that will enable an even broader range of applications and serve a diverse user community once in operation.

In Sections~2, 3, 4 and 5 we briefly introduce the cases considered during Phase~A and some of their key instrument requirements.
In Section~6 we identify the science cases used to specify the TLRs, and in Section~7 we compare the predicted performance of CUBES with other facilities worldwide.

\begin{table*}[h]
\begin{center}
\caption{Summary of high-level science cases developed during the CUBES Phase~A study. Entries in the third column refer to the expanded articles presented elsewhere in this Special Issue (SI).}
\label{sci_summary}
\begin{tabular}{lll}
\hline\noalign{\smallskip}
Field & Science Case & SI Contrib.\\
\noalign{\smallskip}\hline\noalign{\smallskip}
Solar System & S1: Cometary Science & \cite{ExA_Opitom} \\
& S2: Icy Satellites & \dots \\
\noalign{\smallskip}
Galactic & G1: Accretion, winds \& outflows in YSOs & \cite{ExA_Alcala} \\
& G2: Exo-planet composition & \dots \\
& G3: Stellar astrophysics \& exoplanets & \ldots \\
& G4: Beryllium in metal-poor stars and stellar clusters & \cite{ExA_Smiljanic}, \cite{ExA_Giribaldi} \\
& G5: Lithium production in novae & \ldots \\
& G6: Metal-poor stars \& light elements & \cite{ExA_Bonifacio}, \cite{ExA_Hansen} \\
& G7: Neutron-capture elements & \cite{ExA_Ernandes} \\
& G8: Precise metallicities of metal-poor pulsators & \ldots \\
& G9: Horizontal branch stars in Galactic GCs & \ldots \\
& G10: Early-type companions in binary Cepheids & \ldots \\
& G11: Extragalactic massive stars & \cite{ExA_Evans} \\
\noalign{\smallskip}
Extragalactic & E1: Primordial deuterium abundance & \ldots \\
& E2: Missing baryonic mass in the high-$z$ CGM & \cite{ExA_DOdorico} \\
& E3: Cold gas at high redshift & \cite{ExA_Balashev} \\
& E4: Reionisation & \ldots \\
\noalign{\smallskip}
Transients & T1: GRBs & \ldots \\
& T2: Kilonovae & \ldots \\
& T3: Superluminous supernovae & \ldots \\
\noalign{\smallskip}\hline
\end{tabular}
\end{center}
\end{table*}

\section{Solar System Cases}

\subsection{S1: Cometary science}
\begin{figure}
\begin{center}
  \includegraphics[width=8cm]{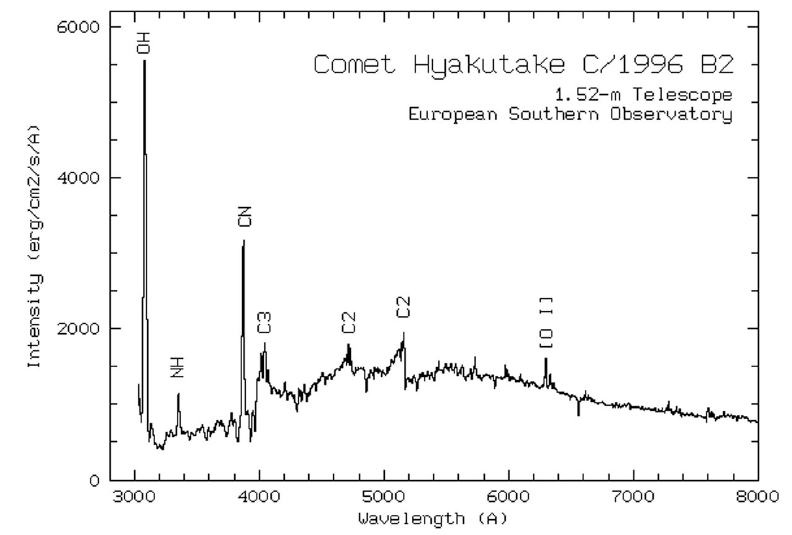}
  \caption{OH emission at 308\,nm is a potentially powerful probe of water outgassing from main-belt comets. (Image credit: ESO.)}
\label{comet_ESO}
\medskip

  \includegraphics[width=8cm]{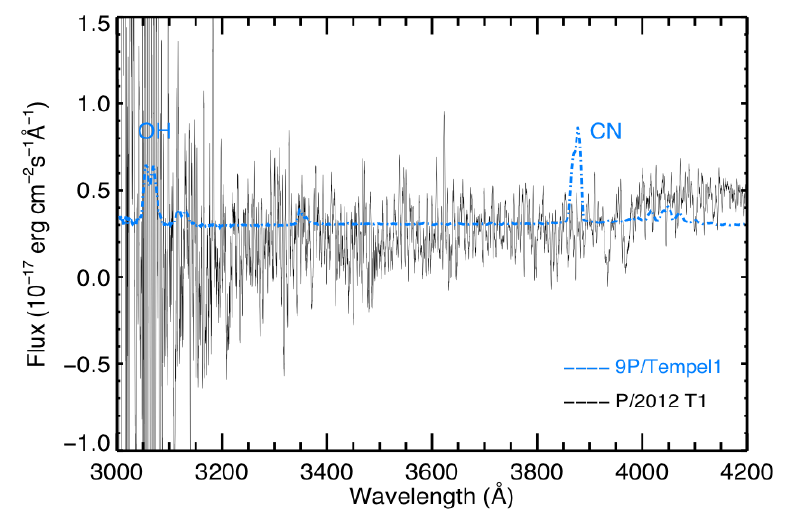}
  \caption{VLT/X-Shooter spectrum of main-belt comet P/2012 T1 (black) compared with an active Jupiter family comet (9P/Tempel~1, in blue). Significantly better sensitivity is required for the OH line for such objects \cite{snodgrass17b}.}
\label{comet_XSh}
\end{center}
\end{figure}

The search for water in our solar system is far from complete and is tremendously difficult with ground-based facilities given the water content of Earth's atmosphere. Infrared spectroscopy can detect ice on the surface of distant bodies, but only for the largest/brightest objects. For smaller bodies the ice is sub-surface and we must look for outgassing water escaping into space, e.g. cometary comae and tails. The most powerful probe of this outgassing water from the ground is OH emission at 308 nm (Fig.~\ref{comet_ESO}). While observations in the far-UV or IR from space could be more sensitive, they would also be substantially more expensive, and ground-based OH observations are the most compelling next step (see \cite{snodgrass17a}); CUBES will provide this capability.

Observation of the OH line is only possible with current facilities for a few active comets while they are near the Sun and Earth. This severely limits studies of water production in comets around their orbits and we miss seasonal effects that the Rosetta mission revealed to be important (e.g. \cite{kramer17}). It also prevents study of the large majority of comets which are simply too faint. Even more tantalising are studies of main-belt comets – bodies in asteroidal orbits that undergo activity (usually detected by a dust tail or trail) that is thought to arise from sublimation (e.g. recurrent activity near perihelion). As shown by the X-Shooter spectrum in Fig.~\ref{comet_XSh}, constraining the OH emission of such objects is well beyond our current capabilities. Main-belt comets have typical sizes that are very common in the asteroid belt, so detection of outgassing water would point to a potentially large population of icy bodies, hence a large reservoir of water, of considerable interest in the context of models of the formation and evolution of the inner solar system (e.g. \cite{obrien18}).

Enhanced sensitivity in the near UV will also allow us to measure the deuterium to hydrogen ratio, using the same OH band, in a much wider range of comets than currently possible, providing invaluable constraints to investigate the fate of water in the solar system and the origin of terrestrial water. Finally, CUBES will enable us to detect and measure the abundance of critical species in the comae of comets, and constrain their composition; with its greater sensitivity than current instrumentation it will also give us access to more distant comets. This capability will be particularly timely in the late 2020s and early 2030s, as it will enable characterisation of potential targets for the ESA {\em Comet Interceptor} mission, which is due to launch in 2029 \cite{snodgrass19}.

Quantitative simulations from the end-to-end simulator have demonstrated that CUBES observations will be able to detect outgassing water down to a level ten times lower than the current upper limit possible with X-Shooter. For details on these calculations
and studies of the chemistry of cometary comae (D/H ratio, N$_2$/CO ratio, and detection of metals), we refer the reader to \cite{ExA_Opitom} in this Special Issue.

\subsubsection{S1: Key requirements}
\begin{itemize}
\item{{\it Spectral resolution:} Essential: $R$\,$>$\,20,000 (with a goal of $R$\,$\sim$\,40,000). For OH detections in the faintest targets, $R$\,$>$\,5,000 will suffice (and provide even greater sensitivity), see \cite{ExA_Opitom}.}\smallskip
\item{{\it Spectral coverage:} Essential: 305-400\,nm; Goal: 300-420\,nm. Ensures well sampled continuum on both sides of the OH(0,0) and CO$^+$(3,0) bands.}\smallskip
\item{{\it Simultaneous coverage at longer \lam:} Desirable with UVES to place constraints on the general composition of the coma. 
Note that observations close in time (rather than simultaneous) would be almost as valuable.}\smallskip
\item{{\it Flux calibration:} Required to $\pm$10\% accuracy (goal of $\pm$5\%).}\smallskip
\item{{\it Exposure times:} Long exposures (of up to 2\,hr) will be needed to detect faint signatures.}\smallskip
\item{{\it Position angle:} It is important to be able to configure the slit position angle to sample different parts of the coma, e.g. jets or tails for active comets.}
\end{itemize}

\subsection{S2: Icy satellites}
\subsubsection{Surface composition}

The near-UV offers a unique opportunity to investigate the surface properties of the satellites of our solar system, including the Galilean and Saturnian satellites, Triton and several Kuiper Belt Objects (KBOs). 

Sulphur-bearing materials are quite common on the surface of Europa, the third Galilean satellite. Europa is one of the targets of the ESA {\em JUICE} mission (together with Ganymede and Callisto), and the target of the NASA {\em Europa Clipper} mission. Sulphur is expected to be found on Europa’s surface from Io’s plasma torus \cite{lane81} and its presence was inferred in the past from ground-based observations in the near-UV and visible \cite{jm70,wamsteker72,mcfadden80,nelson87,noll95,spencer95}.
A spatially-varying surface distribution of SO$_2$ on Europa was also reported, with a clear absorption band at 280\,nm on the trailing face, and less intense (or maybe absent) absorption on the leading face \cite{lane81}. This absorption band has not been formally identified as condensed SO$_2$ but the detection of polymeric sulphur around 360\,nm with CUBES, due to its improved sensitivity in the near-UV, would provide confirmation.

Other peculiar features that have been identified in the UV spectral range could be ascribed to organic compounds, as reported for the South Pole of Callisto with the Galileo/UVS spectrograph \cite{hendrix08}. The red slope observed in the UV-visible in the spectra of the icy satellites of Saturn is not typical of a water-dominated surface, and hence it is an open question which species or process is responsible. If subject to radiolytic/photolytic or thermal processing, hydrocarbons in the outer Solar System develop a very distinctive absorption feature at about 220\,nm, as a consequence of H loss \cite{hendrix16}. They also present a distinctive spectral slope longwards of 300\,nm that can be used with CUBES to better constrain the surface composition of these objects. 

The Galilean satellites have exospheres \cite{mcgrath04}, or thin atmospheres, which are thought to be the result of ion-induced sputtering and sublimation of the surface materials. Any further information we can derive from observations of these exospheres will provide
insights into the processes acting on the surface, as well as important constraints on the surface compositions. 
The search for exospheric compounds is a key scientific goal in the near future for satellites of the outer planets, with particular attention on the Galilean moons. Greater sensitivity to features from Al (396.1\,nm) and Fe (371.9, 374.5 and 395.9\,nm), similar to those identified in Mercury's exosphere \cite{bk17}, will provide the opportunity to detect these elements for the first time around a Galilean moon.

UV observations of the Galilean and Saturnian satellites have revealed a lot about their surface compositions and exospheric properties. Unfortunately, the {\em JUICE} mission will not cover the 300-370\,nm range with its selected payload. A similar UV spectrometer to that on {\em JUICE} will be part of the {\em Europa Clipper} mission, with the aim of investigating the atmospheric and surface properties of Europa. Future observations at 300-400\,nm with a facility such as CUBES will therefore be essential to complement measurements by missions to the Jovian system. 

\subsubsection{S2: Key requirements}
\begin{itemize}
\item{{\it Spectral resolution:} Essential: $R$\,$>$\,10,000; Goal: $R$ $>$\,20,000.}\smallskip
\item{{\it Spectral coverage:} Essential: 300-400\,nm; Goal: 300-420\,nm.}\smallskip
\item{{\it Simultaneous coverage at longer \lam:} Desirable with UVES to enable identification of other species.}
\end{itemize}

\section{Galactic Cases}
\subsection{G1: Accretion, winds \& outflows in YSOs}

Studies of protoplanetary disks play a pivotal role in determining the initial conditions for planet formation, and many of their characteristics are only now being unveiled by the new observational facilities at high angular and spectral resolution, over a wide range of wavelengths, from X-ray to radio. The way in which circumstellar disks evolve and form protoplanets is strongly influenced by the processes of mass accretion onto the star, ejection of outflows and photo-evaporation in winds of the disk material. A proper understanding of the impact of these phenomena requires comprehensive, multi-wavelength studies of different physical processes throughout the first 10\,Myr of the star--disk evolution. Classical T Tauri stars (CTTS) are key objects to investigate these processes observationally. They are young (a few Myr), low- to solar-mass stars that are actively accreting mass from planet-forming disks.

\begin{figure}
\begin{center}
  \includegraphics[width=8.5cm]{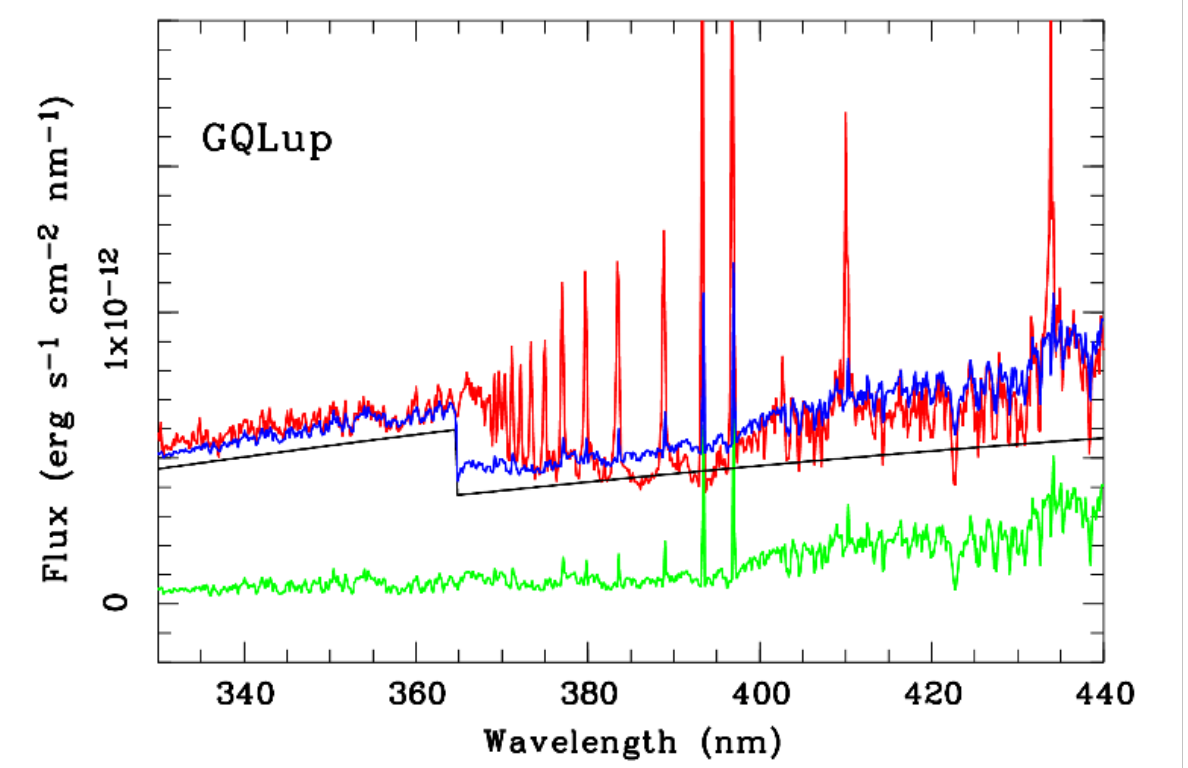}
  \caption{Example X-Shooter spectrum (red) of the classical T Tauri star GQ Lup in the region of the Balmer jump. Note the blending of the high Balmer series due to the spectral resolving power of only $R$\,$\sim$\,5000, which shifts the Balmer limit from 364.6\,nm to an apparent jump at $\sim$370\,nm. The template spectrum of a non-accreting YSO of the same spectral type as GQ Lup is overplotted (green). The continuum emission from a hydrogen slab is shown by the black continuous line. The best fit to the data with the emission predicted from the slab model is given by the blue line. (Adapted from \cite{alcala17}.)}
\label{GQLup}
\end{center}
\end{figure}

Spectroscopic surveys of CTTS in nearby (d\,$<$\,500\,pc) star-forming regions, have been used to study the mutual relationships between accretion, jets and disk structure (e.g. \cite{giannini15,alcala17,manara17,nisini18}) but further aspects remained uncertain, mainly because of the low sensitivity and spectral resolution so far available in the UV. A more expansive discussion of how CUBES will enable important contributions to the investigation of accretion/winds-outflows in young stellar objects (YSOs) is given by \cite{ExA_Alcala} in this Special Issue. Here we briefly highlight some of the key points:
\begin{itemize}
    \item{High throughput observations at R\,$\sim$\,20,000 will enable more detailed study of the accretion process than currently possible, via precise modelling of the Balmer jump, and by investigating the high-order Balmer lines (e.g. Fig.~\ref{GQLup}). The latter and Ca~\2 H\&K lines provide diagnostics of the accretion funnel flows and heated chromosphere in the post-accretion shock region (e.g. \cite{alencar12}).}\smallskip
    \item{Studies of YSOs with low accretion rates will help to disentangle the contribution from chromospheric emission, which is difficult to remove solely using e.g. X-Shooter spectra, particularly for YSOs with spectral types earlier than K3. Accretion has been confirmed in a few such objects using {\em HST} data (e.g. \cite{alcala19}), and we anticipate strong synergies with the {\em HST} ULLYSES Director's Discretionary Program, that has observed $\sim$60 CTTS (see \cite{ExA_Alcala} for details).}\smallskip
    \item{Studies of CTTS in star-forming regions at larger distances and with sub-solar metallicity and/or in different environments to current Galactic targets will only be possible with much greater sensitivity in the near UV. This will enable studies of accretion at low metallicities, as well as the effects of local UV fields on accretion and winds – weak fields should not influence the disks, whereas strong fields may modify the ionisation rate in the disk with an important impact on mass-loss rates.}\smallskip
    \item{Emission lines from [O~\2] 372\,nm and [S~\2] 407\,nm are useful diagnostics of the winds of protoplanetary disks. Also, by combining observations of [O~\2] 372\,nm with [O~\1] 630\,nm (from e.g. UVES), we will be able to measure the degree of ionisation of the wind.}  
    The [O~\1] 630\,nm line has provided the historical foundation for identifying and measuring jets and disk winds, but the accuracy of mass-loss rates improves significantly when paired with high-resolution spectra of the [S~\2] 406.8\,nm line \cite{fang18}. Much of the remaining uncertainty relates to how jets and winds are excited, and the little-studied [Ne~\3] 386.9\,nm line has the potential to discriminate between shocks and X-ray heating \cite{liu14}.\smallskip
    \item{There are bright [Fe~\2] and [Fe~\3] lines in the near UV with higher Einstein coefficients than the [Fe~\2] 1.25 and 1.64\,$\mu$m lines that are commonly observed in YSO jets. The [Fe~\3] lines are particularly important as they are produced from high energy levels so are good tracers of the temperature of the ionised gas. Observations of [Fe~\3] combined with [Fe~\2] allow simultaneous determination of temperature, density and degree of ionisation, as well as a precise estimate of the Fe abundance in the gas.}\smallskip
   \item{Beryllium abundances for older stars will help to constrain evolutionary models of pre-main sequence stars, complementing the Li (670.7\,nm) line.}
   \end{itemize}

\subsubsection{G1: Key requirements}
\begin{itemize}
\item{{\it Spectral resolution:} $R$\,$>$\,20,000 for detailed quantitative analysis; $R$\,$\gtrsim$\,5,000 for observations of the faintest targets (see \cite{ExA_Alcala} for further discussion).}\smallskip
\item{{\it Spectral coverage:} Essential: 310-400\,nm; Goal: 310-430\,nm (includes [S~\2] 407\,nm and Ca~\2 422\,nm).}\smallskip
\item{{\it Position angle:} Non-parallactic observations are essential so the slit can be aligned along the outflow/jet direction and/or perpendicular to it.}\smallskip
\item{{\it Simultaneous observations at longer \lam:} Given the variable nature of YSOs, simultaneous observations with UVES are essential to correctly characterise the CUBES observations.}\smallskip
\item{{\it Flux calibration:} Required to better than $\pm$10\% (goal), $\pm$15\% (essential).}
\end{itemize}

\subsection{G2: Composition of exo-planets}
In the past 25 years we have gone from not knowing if the Solar System is unique to understanding that it is totally normal for stars to have planets. Planetary systems have been identified in increasingly complex architectures, including rich multi-planet systems around stars both more and less massive than the Sun, circumbinary planets, and wide binaries with planets orbiting one or both stellar components. Characterising exo-planets and understanding their formation and evolution are now major research areas, but the fundamental question of what these other worlds are made of cannot be answered from studies of planets orbiting main-sequence hosts. We can only estimate the bulk densities of transiting planets (with significant uncertainties), and considerations of internal structure and composition are model dependent (e.g. \cite{rs10}). 

The key to answering the above question draws on a method well-established in planetary science: measuring the abundances of the building blocks of planets, and their fragments. We know more about the composition of the terrestrial planets in the Solar System from abundance studies of meteorites than from surface samples collected by astronauts and robots. Similarly, it has been demonstrated (\cite{z07}) that we can accurately measure the bulk composition of exo-planetary systems (analogous to studies of Solar System meteorites) from spectroscopic analysis of white dwarfs which are accreting debris from tidally-disrupted planetesimals \cite{jura03}.

White dwarfs are moderately hot stars so most of the atomic transitions that can be used for abundance studies are located in the ground-based UV, in particular for refractory and transitional elements (Sc, Ti, V, Cr, Mn, Fe, Ni). White dwarfs, being Earth-sized, are intrinsically faint, so high-throughput high-resolution spectroscopy ($R\ge20,000$) is essential. So far only about twenty systems have precise abundances for more than five chemical elements, the bulk of which were observed with Keck-HIRES (e.g. \cite{klein10}).

\begin{figure}[h]
\begin{center}
  \includegraphics[width=8.33cm]{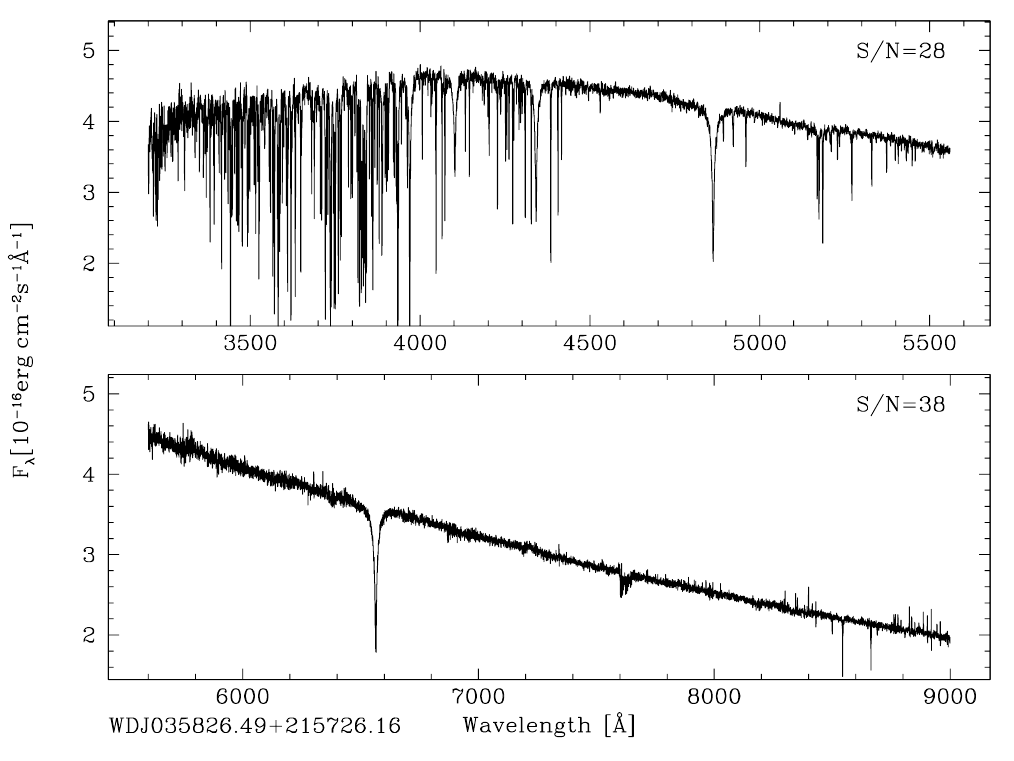}
  \caption{Example 1\,hr X-Shooter spectrum for a white dwarf with $V$\,$=$\,16.5\,mag. }
\label{WD}
\end{center}
\end{figure}

The current workhorses for studies of the bulk compositions of exo-planets are VLT-UVES (south) and Keck-HIRES (north), but they are limited to targets with $V$\,$<$\,16 to 16.5\,mag for such observations. High S/N data are possible with X-Shooter (see Fig.~\ref{WD}) but we then lack the spectral resolution required to study the detailed chemistry. CUBES will dramatically advance this new area of exo-planet research, pushing $\simeq$3 mag. fainter than Keck-HIRES or VLT-UVES and opening up a vastly larger sample of targets. For instance, $\simeq260,000$ white dwarfs were identified from {\em Gaia} DR2, almost half of which were new discoveries in the southern hemisphere \cite{gf19}. Low-resolution spectroscopic confirmation will be done with SDSS-V and later with 4MOST. Statistically, $\simeq$1\% of white dwarfs are suited to detailed abundance studies, so CUBES will be able to increase the sample of known exo-planetesimal compositions to a level comparable to the number of meteorite samples with known abundances, thus providing exquisitely detailed abundance information for the next generation of planet formation models.

\subsubsection{G2: Key requirements}
\begin{itemize}
\item{{\it Spectral resolution:} $R$\,$>$\,20,000.}\smallskip
\item{{\it Spectral coverage:} Essential: 305-400\,nm; Goal: 305-420\,nm.}\smallskip
\item{{\it Position angle:} Parallactic observations are sufficient (unless avoiding a nearby bright star).}\smallskip
\item{{\it Simultaneous observations at longer \lam:} Desirable.}
\end{itemize}

\subsection{G3: Stellar astrophysics \& exoplanets}

Stellar astrophysics faces significant questions as to how stellar magnetic fields are generated by an internal dynamo, emerge as flux tubes, and energise the outer atmospheres and activity, including impulsive phenomena such as flares \cite{tsd15}.  Even for the Sun, our theory of atmospheric energy transport and release is incomplete, yet Solar physics necessarily remains the key to our explanations of the physics of stellar atmospheres and activity. Conversely, stars provide proxies for studying otherwise unobservable Solar evolution, and testing Solar models under more extreme conditions. 

Two questions arise: 1) Can stellar activity be better understood by new connections with Solar physics? 2) How can this new understanding of stellar activity advance exoplanet studies? \medskip

Exoplanet surveys of young or otherwise active stars are important to investigate the shared evolution of host stars and their planetary systems. However, the interpretation of high-resolution visible spectroscopy (with e.g. ESPRESSO) can be limited by our understanding of stellar activity. To address this we need a better understanding of stellar chromospheres: the region where  stellar activity is prominent due to the dominance of non-radiatively heated plasma, with variability from flares and other transient magnetic activity. To characterise chromospheric activity we require near-UV observations in support
of e.g. ESPRESSO programmes, and with better sensitivity than is available from current instruments (particularly given the relatively rapid timescales of variability).

The ideal programme would use contemporaneous CUBES and ESPRESSO observations to:
\begin{itemize}
    \item{Monitor the Ca~\2 H\&K lines that are diagnostics of quiescent chromospheres and transient flares \cite{klocova17}.}\smallskip
    \item{Include the Balmer jump as an added diagnostic of stellar chromosphere physics, for both the quiescent chromosphere and any major flares that are expected to change the UV continuum \cite{kowalski15}.}\smallskip
    \item{Accumulate chromospheric spectra to expand our knowledge of activity trends for Solar-type stars. This would aid forecasting of future Solar activity level extremes and their impact on space weather in the Solar System, while also giving insights into space weather in exoplanetary systems to investigate its impact on habitable zones.}\smallskip
    \item{Monitor radial velocities (ESPRESSO) and chromospheric activity (CUBES) simultaneously to help separate genuine stellar reflex motion due to exoplanets from jitter arising from stellar activity.} 
\end{itemize}

In short, CUBES and ESPRESSO together would provide a powerful combination for stellar and exoplanet science, including an enhanced search for planetary systems around the youngest, most active stars. Future observations would focus on young, active and rapidly-rotating stars, spanning cool G to M dwarfs over a wide range of $V$\,$=$\,8-16\,mag. Alongside radial-velocity studies with ESPRESSO, CUBES would be used to monitor the Ca~\2 H\&K lines (and other near-UV chromospheric variability), as well as monitoring flares to constrain models of flare physics (e.g. \cite{k13}).

Such a programme would be similar to ESPRESSO observations of inactive stars studied in current exoplanet surveys, but would need to include enough exposures to provide phase coverage of the stellar rotation rate. As a guide, a minimum set of ESPRESSO data would comprise two to four exposures per night of a given target star (reaching S/N\,$=$\,10 per pixel), for two to four near-contiguous nights. This would enable tomographic active-region modelling to be applied, with CUBES observations obtained roughly contemporaneously to disentangle activity effects. If flare monitoring was also included, the CUBES exposure times are best kept as short as possible (1-10 min).
For example, a single young active high-priority target could be continuously monitored with ESPRESSO across a rotation period of several nights to improve the detection of planets around the youngest, most active stars (e.g. \cite{yu17}). Adding contemporaneous CUBES monitoring of chromospheric variability would provide firmer constraints on stellar activity, hence a clearer understanding of the radial-velocity changes due to a planet.

Tomographic mapping of chromospheres requires observations at $R$\,$\ge$\,20,000 of the Ca~\2 H\&K and Balmer lines (e.g. \cite{donati97}). Both the CUBES and ESPRESSO data could also be used for tomographic imaging of different layers of the active regions of the host star, which further helps to separate stellar activity from planetary signals in the spectra (as used already to e.g. detect hot Jupiters around stars as young as T-Tauri stars \cite{donati16}), and is of intrinsic scientific interest to those studying stellar magnetic activity and dynamos.

\subsubsection{G3: Key requirements}
\begin{itemize}
\item{{\it Spectral resolution:} Essential: $R$\,$>$\,20,000; Desirable: $R$ of up to 100,000.}\smallskip
\item{{\it Spectral coverage:} Essential: 305-400\,nm; Goal: 300-420\,nm.}\smallskip
\item{{\it Simultaneous coverage at longer \lam:} Desirable to have contemporaneous observations with ESPRESSO (using another UT).}
\end{itemize}

\subsection{G4: Beryllium in metal-poor stars and stellar clusters}

Although one of the lightest, simplest elements, there remain profound questions regarding the production of Be in the early Universe. For instance, the upper limit on its abundance of log(Be/H)\,$<$\,$-$14.0 in an extremely metal-poor star with [Fe/H]\,$=$\,$–$3.84 is consistent with no primordial production \cite{spite19}, but larger samples of very metal-poor stars are required to constrain its formation channels. Beryllium is also a powerful tracer of early nucleosynthesis in a range of different contexts, including stellar evolution, the formation of globular clusters, and the star-formation rate and chemical evolution of the Galaxy \cite{smiljanic14}.

Furthermore, $^9$Be is thought to be mostly produced by spallation of Galactic cosmic rays \cite{reeves70}. During the early stages ($<$0.5-1 Gyr) of Galactic evolution, cosmic rays were generated and transported on a Galactic scale, diffusing over the whole Galaxy. This implies that the production site was widespread and, to a first approximation, the Be abundance should have one value at a given time over the whole Galaxy  \cite{beers00,sy01}. Beryllium could then be used as an ideal cosmic clock \cite{pasquini05,pasquini07,smiljanic09} and is therefore a key element to trace star-formation histories. In particular, its abundance can potentially be used to identify accretion of external systems into the early halo, complementing astrometry from {\em Gaia} \cite{molaro20,smiljanic21}.

Beryllium is destroyed in stars at the relatively low temperature of 3.5\,$\times$\,10$^6$\,K (compared to 2.5\,$\times$\,10$^6$\,K for lithium). Measurements of both Li and Be in the same star therefore provide valuable probes of stellar interiors and, in particular, of the non-standard mechanisms at work. Lithium estimates are now available from recent large spectroscopic surveys (e.g. in several open clusters) but Be has only been studied for the brighter stars in a few clusters (down to V\,$\sim$\,14) \cite{smiljanic10,smiljanic11,boesgaard20,boesgaard22}. CUBES will provide Be abundances for many more stars, in a range of clusters, complementing the Li measurements, and enabling study of internal mixing as a function of stellar age, mass, and metallicity. \medskip

The only Be lines available from the ground for abundance estimates are the Be~\2 313.04 and 313.10\,nm resonance lines, which require good S/N ($>$50) and sufficient resolution ($R$\,$\gtrsim$\,20,000) to clearly discern them from nearby, relatively strong V~\@ (313.03\,nm) and Ti~\2 (313.08\,nm) absorption. 

There are only $\sim$200 stars with estimated Be abundances (from Keck-HIRES and VLT-UVES), which span near-solar metallicities down to [Fe/H]\,$<$\,$–$3
\cite{smiljanic09,boesgaard99,primas00a,primas00b,boesgaard11}. The limiting magnitude of current observations is V $\sim$ 12\,mag., for observations of a few hours. An ambitious UVES programme with a total integration of 20\,hrs has observed a metal-poor star ([Fe/H]\,$=$\,$-$3.0) with $V$\,$=$\,13.2\,mag., resulting in a S/N around the Be doublet of $\sim$200 \cite{smiljanic21}, but such long exposures are impractical to assemble large samples of objects.

In general, CUBES will be able to provide estimates of Be in stars belonging to different populations for the first time, complementing results for Li, metallicities, and $\alpha$-elements from surveys such as {\em Gaia}-ESO and 4MOST, and {\em Gaia} astrometry. Some example target magnitudes in this context include:
\begin{itemize}
    \item {{\it Essential:} $V$\,$=$\,16\,mag.: Turn-off and main-sequence stars (FGK-type) in open clusters where Li abundances are available from other surveys.}\smallskip
    \item{{\it Essential:} $G$\,$=$\,16\,mag.: Turn-off stars in the Gaia-Enceladus population and/or other kinematic structures discovered by {\em Gaia} up to 2.5\,kpc from the Sun.}\smallskip
    \item{{\it Goal:} $V$\,$=$\,16.7\,mag.: Solar-type stars in one of the oldest known globular clusters.}\smallskip
    \item{{\it Goal:} $G$\,$=$\,17.5\,mag.: Turn-off stars up to 5\,kpc from the Sun.}
\end{itemize}

We now briefly expand on two key areas for studies of Be abundances.

\subsubsection{Beryllium abundances in extremely metal-poor stars}

Abundances of beryllium in stars with [Fe/H]\,$<$\,$–$3.0 seem to indicate an increased scatter or even a flattening of the relation between [Fe/H] and log(Be/H) (e.g. \cite{primas00b}). Possible explanations include: Be production by inhomogeneous primordial nucleosynthesis, significant production of Be by pre-Galactic cosmic-rays in the intergalactic medium, or the inhomogeneous nature of the early halo having formed from mergers of stellar systems with different [Fe/H] vs. log(Be/H) relations \cite{smiljanic21}.

Ongoing multi-band imaging surveys are finding excellent candidates for very metal-poor stars (e.g. Pristine \cite{pristine}; SkyMapper \cite{skymapper}), as well as wide-field spectroscopic surveys (e.g. LAMOST and the upcoming WEAVE and 4MOST surveys). However, stars with such low metallicity are faint and hard to observe in the near-UV with current spectrographs (see \cite{ExA_Bonifacio} and \cite{ExA_Hansen} in this Special Issue). Only 10-15 stars have been investigated in this region and with only upper limits to the Be abundances in several cases. Improved near-UV sensitivity is required to access a larger sample over the metallicity range $-$4.0\,$<$\,[Fe/H]\,$<$\,$-$3.0 to investigate if there is a plateau or a larger scatter in the [Fe/H] vs. log(Be/H) relation.

Even with enhanced near-UV sensitivity, it will be particularly challenging to detect the Be lines in such extremely metal-poor stars, as the lines become so weak. Detailed simulations, presented elsewhere in this Special Issue \cite{ExA_Smiljanic}, show that this will be possible with CUBES for stars that have $\log$(Be/H) $\gtrsim$ $-$13.6, as long as a S/N of order 400 can be achieved. Given the anticipated high efficiency of CUBES, stars down to V $\sim$ 14\,mag. will be able to be observed within a couple of nights of observations \cite{ExA_Smiljanic}, and a dedicated Large Programme could more than double the precious sample of extremely metal-poor unevolved stars to be investigated.

\subsubsection{Multiple populations in globular clusters}\label{gc_pops}

Globular clusters (GCs) are not the simple stellar populations that we once thought there were. They exhibit star-to-star variations in specific elements (e.g., He, C, N, O, Na, Mg, Al) that bear the hallmark of high-temperature H-burning \cite{gratton12}; variations are also seen in fluorine abundances (e.g. \cite{ymc08,gcs19}). These abundance variations can be observed spectroscopically and also photometrically, with the appropriate choice of filters, due to the changing of spectral features within the band pass. This phenomenon is observed in nearly all of the ancient GCs and has recently been found in many younger clusters as well.

Many scenarios have been suggested to explain this phenomenon, with most invoking multiple epochs of star formation within the cluster, in which a second generation of stars are formed from material contaminated by proton-capture nucleosynthesis products. Pro\-ton-cap\-ture reactions can decrease the abundances of Li, C, O and Mg, and enhance the abundances of N, Na and Al. The observed correlations and anti-correlations of these light-element abundances are thought to arise from mixing pristine and processed material, with potential polluters thought to be first-generation massive asymptotic-giant-branch stars (AGBs) or first-generation rapidly-rotating massive stars. However, the proposed scenarios all fail to reproduce key observations.

An important constraint on the problem has come from lithium, that is easily destroyed at high temperatures (like those required to burn H through the CNO cycle). However, Li is a complicated element with potential nucleosynthetic production happening in evolved stars through the Cameron-Fowler effect. Beryllium, on the other hand, could provide new insights as it is not produced in significant quantities through nuclear burning and only destroyed inside stars. Thus, the polluting material from the first generation of evolved stars, regardless of the exact nature of the polluting star, should be completely devoid of Be. A star formed with only pristine material should therefore have the original Be abundance of that material. Stars with different amounts of polluted material should have diluted the surface Be abundance to different levels. Stars with bigger fractions of polluted material should have stronger correlations and anti-correlations between the light elements and should be strongly depleted in Be.

The objective is to look for differences in the Be abundances between turn-off stars of the same cluster. Different stars should have been polluted to different levels, so if different Be abundances are detected we can use them to quantify the fraction of polluting material. An expanded study of this case is presented elsewhere in this Special Issue \cite{ExA_Giribaldi}, where simulated observations of 4\,hr per star suggest that depletions of Be of $\sim$0.6\,dex will be possible in nearby GCs with turn-off magnitudes as faint as $V$\,$=$\,18\,mag.

Furthermore, it is not clear whether the multiple populations phenomenon is limited to light-elements, as some studies have suggested that K, as well as neutron-capture elements (e.g. Eu, La) may also vary. The diverse set of elements made available through CUBES will offer the exciting potential of discovering new variations and correlations that may finally allow us to solve this long-standing puzzle.

\subsubsection{Key requirements}
\begin{itemize}
    \item{{\it Spectral resolution:} $\ge$\,20,000 to sufficiently resolve the Be~\2 doublet (see e.g. \cite{smiljanic14}).}\smallskip
    \item{{\it Position angle:} Observations at non-parallactic angles are desirable.}
\end{itemize}

\subsection{G5: Lithium production in novae}

The detection of the beryllium UV doublet in nova outburst ejecta \cite{tajitsu15} confirmed the long-standing theoretical prediction of substantial Li production in nova thermonuclear runaways in accreting white dwarfs. Such runaways soon develop density instabilities and convection plumes, allowing the decay of processed $^7$Be. Any lithium brought by accretion from the secondary star sinks to deep layers, being destroyed at the onset of the process. Therefore, Li enhancement recently observed in novae is associated with production during the thermonuclear runaway itself. Once the $^7$Be line-formation conditions and radiative transfer are known, measuring the UV $^7$Be lines during outburst provides a gauge of the Li enhancement in the ejecta (see, e.g. \cite{ExA_Izzo} in this Special Issue).

While the $^{}$Li nucleosynthesis process was successfully verified, two fundamental questions remain unanswered. The first concerns the determination of the total ejected mass and the corresponding Li mass contribution to the ISM from a single nova outburst. Equally important, the long-term contribution from the diverse nova population (classical, recurrent and symbiotic) remains to be effectively constrained by observations.

The inhomogeneous nature of novae ejecta implies geometry-dependent line optical depths, calling for synoptic and/or large sample observations. A statistically robust sample of Li abundances in recent novae as well as a precise modelling of the atmosphere and ejecta are needed to evaluate the current and past contribution of novae to the Galactic Li abundance.

\subsubsection{G5: Key requirements}
\begin{itemize}
\item{{\it Spectral resolution:} Essential: $R$\,$>$\,20,000 (to study the Be doublet).}\smallskip
\item{{\it Spectral coverage:} Essential: 300-400\,nm; Goal: 300-420\,nm.}\smallskip
\item{{\it Simultaneous coverage at longer \lam:} Useful but not essential.}\smallskip
\item{{\it Position angle:} A configurable slit direction is essential to study resolved shells and to enable differential techniques.} 
\end{itemize}

\subsection{G6: Metal-poor stars \& light elements}

Characterising the first stars that formed from primordial material produced by Big Bang Nucleosynthesis is a vibrant field. The chemical imprints of the first stars are detectable in the second generation of stars, including long lived low-mass stars. While these are exceptionally rare, the tell-tale sign is the extremely low metal content. Only a handful of these so-called ultra-metal-poor stars are known; these are the most chemically ancient objects known. They exhibit an enormous range in their relative chemical abundance ratios, which demand a wide variety in the properties of the first supernovae, e.g., mass, explosion energy, rotation, mass cut, explosion mechanism. More data at the lowest metallicities are needed to better constrain the diversity of properties, and relative frequency, amongst the first supernovae and characterise the properties of the parent first stars.

There are many surveys planned for the future that promise large number statistics on ultra-metal-poor stars. These are identified using photometric and lower resolution surveys, and are typically selected using broad indications of metallicity rather than precise measurements. Accurate and detailed chemical abundance patterns of such candidates are essential to gain the full insight into the range of nucleosynthesis processes that were available in the early Universe, as well as to probe the physical properties and mass distribution of the very first stars.

Many potential targets in this context will be drawn from the so-called Carbon-enhanced metal-poor (CEMP) stars, with [C/Fe]\,$>$\,$+$1.0 (see \cite{bc05}). Although rare, they have a diverse range of abundances of neutron-capture elements, commonly grouped as: `CEMP-no' (no over-abundance of r-process elements), `CEMP-r' and `CEMP-s' (stars with over-abundances of r- and s-processed elements, respectively) and `CEMP-r/s' (with contributions from both processes enriching their photospheres). 

\begin{figure}[h]
\begin{center}
  \includegraphics[width=8cm]{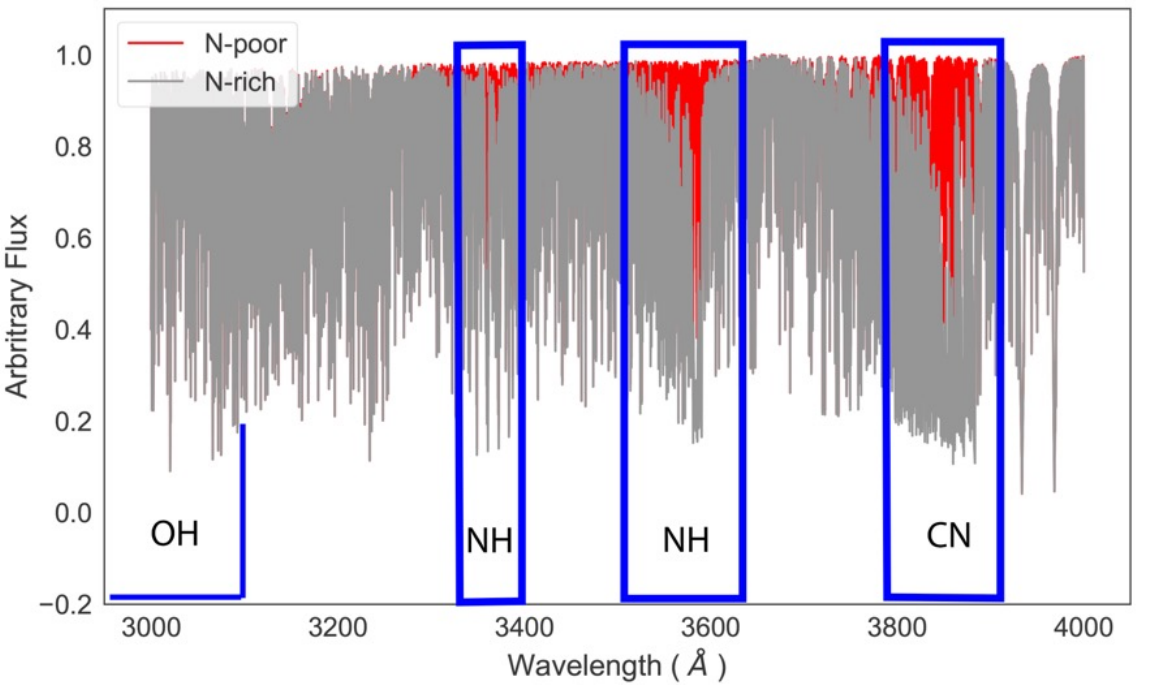}
  \caption{{\it Red:} Synthetic spectrum of a metal-poor star generated using the {\sc turbospec} radiative transfer code \cite{plez12}, adopting physical parameters as for CS~31082–001 \cite{cayrel01,hill02} including [Fe/H]\,$=$\,$–$2.9. {\it Grey:} Synthetic spectrum of the same star, but now N-rich ($\Delta$[N/H]\,$=$\,$+$\,2.0\,dex). }
\label{CUBES_CNO}
\end{center}
\end{figure}

Abundances of CNO bring a wealth of information on stellar evolution and the chemical evolution of the Galaxy. In contrast to the atomic transitions of the elements discussed in the other Galactic cases, CNO features in the near UV are dominated by a series of molecular bands (see Fig.~\ref{CUBES_CNO}) which include the A-X OH transitions at the shortest wavelengths that can be used to estimate oxygen abundances.

A range of scenarios have been explored to investigate these patterns, including rotational mixing in rapidly rotating, low-metallicity stars (e.g. \cite{chiappini13,choplin16}) and supernova models which include both mixing and fallback of material to yield the observed abundance ratios (e.g. \cite{un02,un05,t14}). In short, CEMP stars are perfect probes to investigate nucleosynthesis from the first stars (including production of neutron-capture elements) as well as mass transfer in binary systems (e.g. \cite{abate15}). However, comprehensive near-UV spectroscopy of CEMP stars from the ground has been limited to date by the sensitivity of current facilities to just a few targets (e.g. \cite{placco15,hansen15,hansen19}).

A more expansive discussion of these objects in the context of CUBES is given elsewhere in this Special Issue (\cite{ExA_Bonifacio,ExA_Hansen}).

\subsubsection{G6: Key requirements}
\begin{itemize}
    \item{{\it Spectral resolution:} $R$\,$\ge$\,20,000; Goal: $R$\,$\ge$\,30,000.}\smallskip
    \item{{\it Spectral coverage:} Essential: 305-400\,nm; Goal: 300-400\,nm.}\smallskip
    \item{{\it Simultaneous observations at longer \lam:} Desirable in terms of observational efficiency, but not essential.}
\end{itemize}

\subsection{G7: Neutron-capture elements}
\begin{figure}[h]
\begin{center}
  \includegraphics[width=9cm]{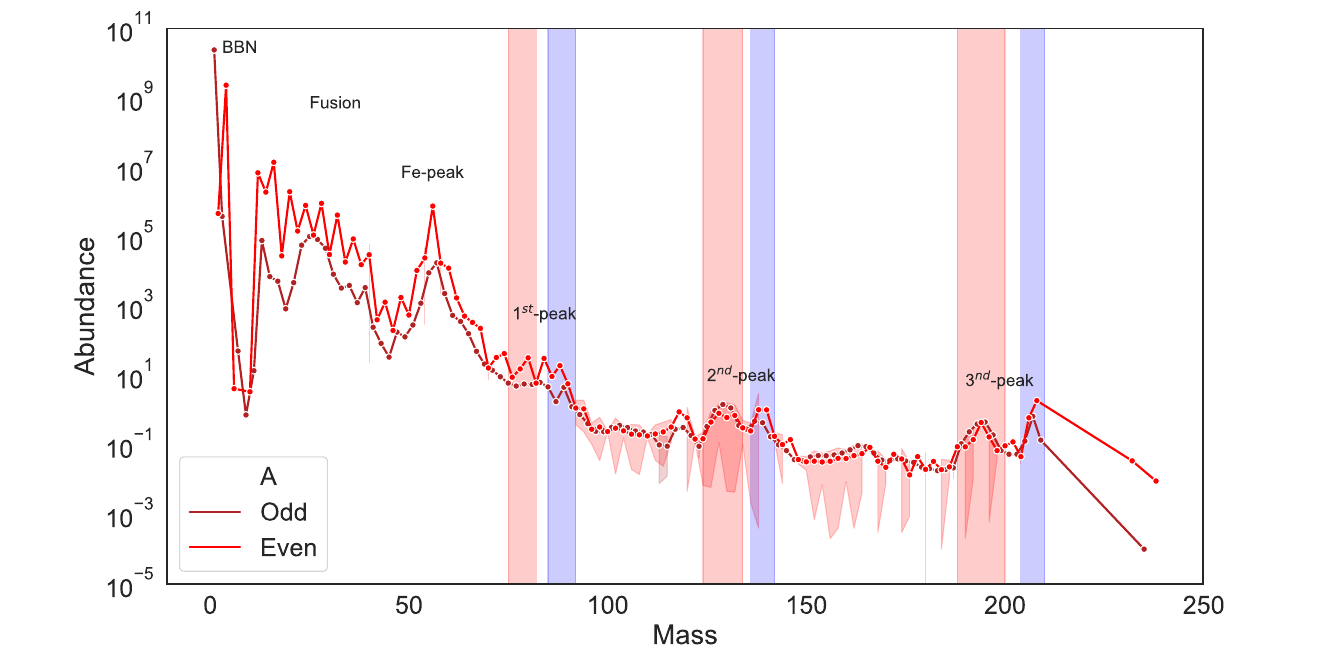}
  \caption{Abundances of isotopes in the Solar System vs. mass number (A, where A = Z + N, in which N is the neutron number) normalised by the abundance of $^{28}$Si to 10$^6$ (from \cite{lodders03}). The iron peak and the regions of the three peaks of r-process and s-process are indicated.}
\label{mass_peaks}
\end{center}
\end{figure}

Abundances of the heavy elements are often described in terms of three mass peaks (see Fig.~\ref{mass_peaks}), within which the respective r- and s-process elements are thought to have common formation channels. These peaks are connected to the nuclear magic numbers 50, 82 and 126. Nuclei with these ‘closed’ nuclear shells are especially stable and are produced in greater abundance during n-capture reactions. The s-process path closely follows the valley of $\beta$-stability, whereas the r-process encounters the closed shells in nuclei of lower proton number, Z, than the s-process. The abundance peaks of the r-process are therefore shifted to lower-mass atoms relative to the s-process. The s-process elements are highlighted in blue in Fig.~\ref{mass_peaks}. The main s-process, responsible for Zr to Pb, is associated with stars on the Asymptotic Giant Branch (AGB, e.g. \cite{busso99}), while the production of lighter s-process elements (Fe to Zr) is thought to be dominated by the weak s-process component during core He-burning and C-shell burning phases of massive stars ($M$\,$>$\,10\,$M_{\odot}$) (e.g. \cite{p10}).

The formation channels for the r-process are particularly topical following the detection of the GW170817 kilonova from a binary neutron-star merger \cite{pian17,smartt17,watson19}. The r-process is thought to occur both during the merging and in the milliseconds afterwards (e.g. \cite{bovard17}) and is thought to play a major role in the chemical evolution of the Galaxy \cite{matteucci14,cescutti15}. Other predicted sites of r-process nucleosynthesis included low-mass (8-10\,M$_{\odot}$) core-collapse supernovae \cite{mc90,iw99}. However, according to nucleosynthesis calculations, normal core-collapse SNe do not appear to host the proper conditions for the production of r-process \cite{arcones07}. In contrast, the r-process is thought to take place in a special family of core-collapse SNe which have  magneto-hydrodynamically-driven jets arising from rapidly-rotating massive stars with a strong magnetic field \cite{winteler12,ntt15}.

Determining the abundances of neutron-capture elements in metal-poor stars is fundamental to understand the physics of the different processes outlined above (slow, rapid and possibly intermediate), as well as the chemical evolution of the Galaxy. Neutron-capture elements have been measured in over 1000 stars in the Galactic halo (with at least one neutron-capture element measured, typically Ba). Looking ahead, 4MOST and WEAVE will provide Ba estimates (at least) for several 10$^5$ stars. Most of these will be in the Galactic disk, but their samples will also include halo stars. However, a clear understanding of the different processes can only be reached with detailed measurements of all the key elements, many of which are in the UV domain (e.g. Hf, Sn, Ag, Bi) and have only been measured for a tiny sample of stars. CUBES will be the instrument to fill this gap. 

The full pattern of neutron-capture elements (coupled with stellar kinematic data from {\em Gaia}) could also be fundamental to understand the origin of the Galactic halo (e.g. \cite{r18}). In the wider context of Galactic Archaeology and chemical evolution of the Milky Way, it is important to have a sample that evenly covers the range of metallicities of halo giants (not only the most metal poor) with as much information on the neutron-capture elements as possible. {\em Gaia} and 4MOST will help with target selection, and CUBES will provide the UV spectra to estimate the abundances.
\medskip

To illustrate the expected performance of CUBES in the determination of heavy elements, detailed simulations of the Ge~\1 303.9\,nm and Hf~\2 340.0 and 371.9\,nm lines are presented elsewhere in this Special Issue \cite{ExA_Ernandes}. These demonstrate the gain of up to three magnitudes that CUBES will provide compared to current observations.

\begin{figure*}
    \centering
    \includegraphics[width=6.0in]{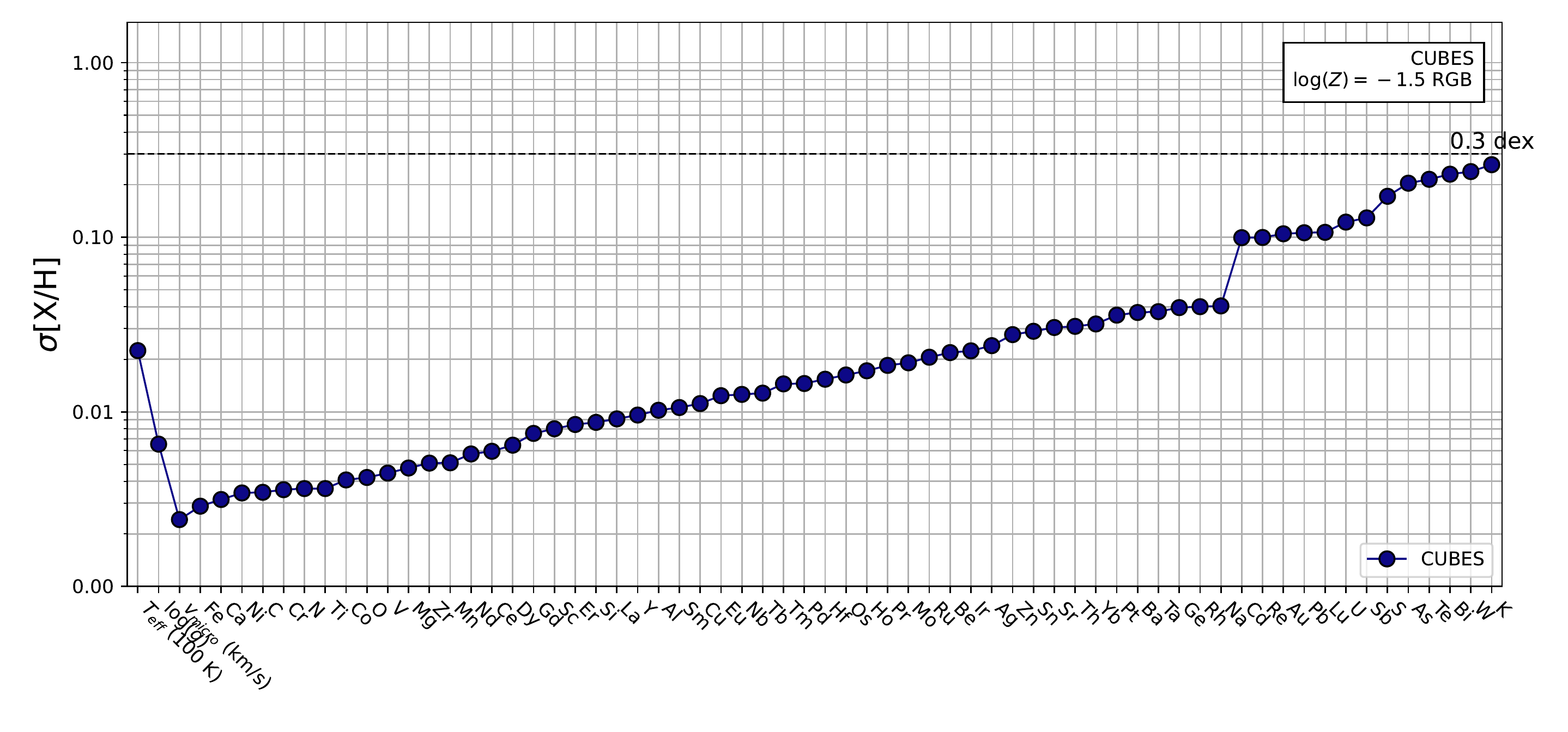}
    \includegraphics[width=6.0in]{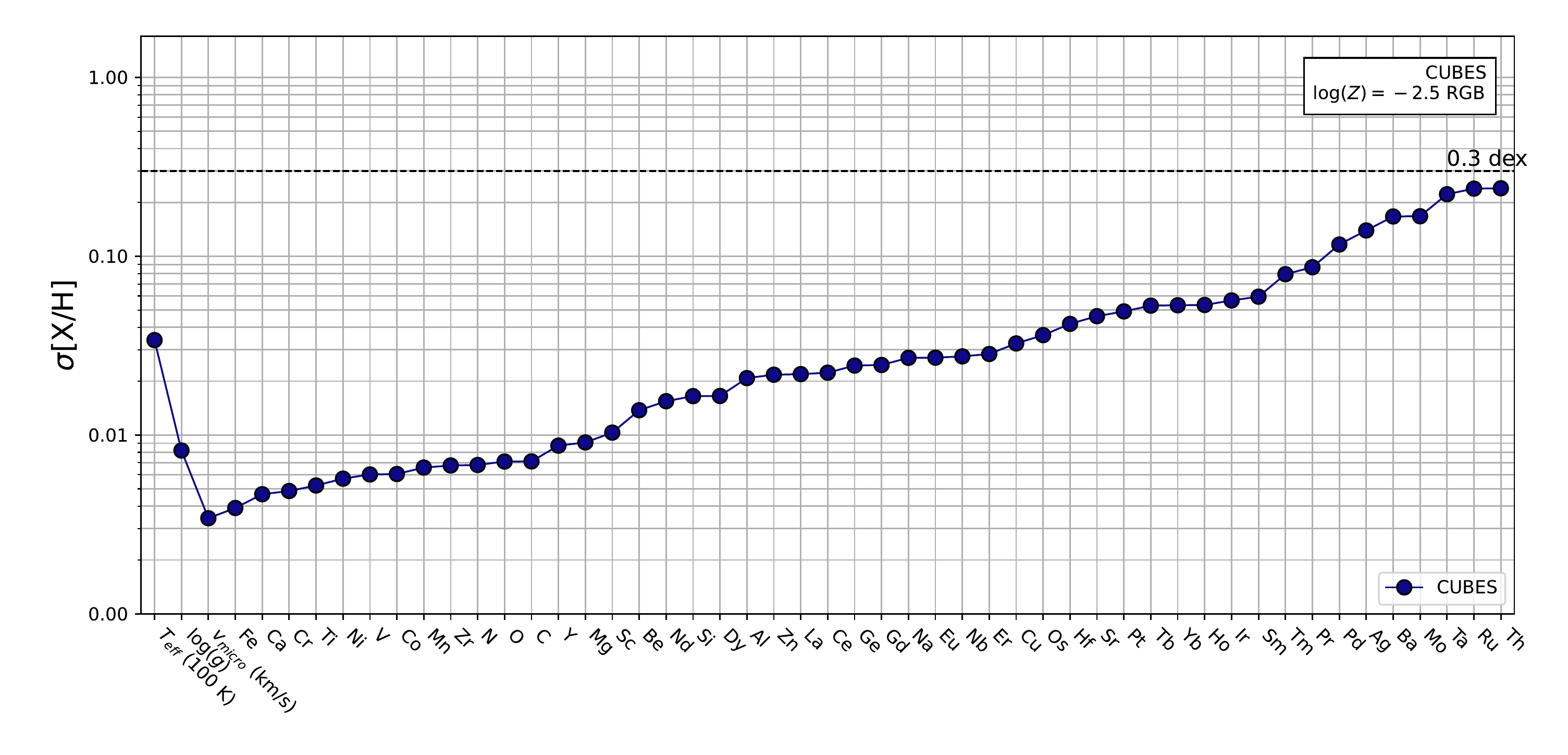}
    \caption{Simulations using the {\tt Chem-I-Calc} package \cite{swt20} to investigate the precision to which stellar effective temperature (T$_{\rm eff}$), gravity (log($g$)), microturbulence ($v_{\rm micro}$) and chemical abundances ($\sigma$[X/H]) can be recovered for observations of stars on the red giant branch (RGB) with CUBES (at $R$\,$=$\,20,000, S/N\,$=$\,100), for models with [Fe/H]\,$=$\,$-$1.5 and $-$2.5 in the upper and lower panels, respectively.}
    \label{chemcalc}
\end{figure*}

Building on analysis undertaken before the Phase~A study \cite{ernandes20}, we also explored the precision limits for stellar parameters and abundances at $R$\,$=$\,20,000 more quantitatively using the {\tt Chem-I-Calc} package \cite{swt20}. This estimates the precision using a Fisher information matrix and the Cram\'{e}r-Rao inequality to find the Cram\'{e}r-Rao Lower Bound, which is a method to find the lower bound to the root-mean-square of an unknown, non-biased variable. We used the full-spectrum fitting capability of {\tt Chem-I-Calc}, adopting the 1D LTE models provided by the package. Simulations for observations with S/N\,$=$\,100 for two template models for RGB stars with [Fe/H]\,$=$\,$-$1.5 and $-$2.5 are shown in the upper and lower panels of Fig.~\ref{chemcalc}, respectively.

Fig.~\ref{chemcalc} shows that for the stellar parameters (T$_{\rm eff}$, log($g$), $v_{\rm micro}$) and the majority of the elements we can expect to recover a precision to better than 0.1\,dex at $R$\,$=$\,20,000 for observations with S/N\,$=$\,100. For many of the heavier elements a precision of  $\sim$0.2\,dex is predicted. We note that for elements with relatively few and weak lines, it is preferable to use fits to single lines rather than the full spectral fitting.

\subsubsection{G7: Key requirements}
\begin{itemize}
    \item{{\it Spectral resolution:} $R$\,$\ge$\,20,000.}\smallskip
    \item{{\it Spectral coverage:} Essential: 300-400\,nm; Goal: 300-420\,nm.}\smallskip
    \item{{\it Simultaneous coverage at longer \lam:} Desirable for observational efficiency, but not essential.}
\end{itemize}

\subsection{G8: Precise metallicities of metal-poor pulsators}

Pulsating stars such as Classical Cepheids and RR Lyrae stars are traditionally used to derive distances within the Local Group and to calibrate the extragalactic distance scale, and secondary distance indicators, at cosmologically interesting distances. Interest in their properties has been renewed by the revolution in astrometry from {\em Gaia} and recent tensions in the Hubble constant between local estimates based on distance indicators (e.g. \cite{riess19}) and early Universe measurements and predictions (see \cite{verde19}).

Classical Cepheids have Period-Luminosity (PL) and Period-Luminosity-Colour (PLC) relations. Due to the finite width of the instability strip, PL relations are better defined towards near-IR filters, whereas using optical filters entails a colour term via PLC relations or (reddening-free) Period-Wesenheit (PW) relations (e.g. \cite{marconi05,marconi10,ripepi12,fiorentino13,ripepi17,ripepi19}. In contrast, lower-mass RR Lyrae stars do not obey PL relations in the optical, but they are accurate Population~II distance indicators via near-IR PL relations and optical/near-IR PW relations. 

In this context, the universality of PL, PLC and PW relations for both Cepheids and RR Lyrae is uncertain. Given the many theoretical and observational effects, there is no consensus on the metallicity dependence of the Cepheid PL and PW relations (e.g. \cite{marconi05,ripepi19,romaniello08,bono10}).  Metallicity differences between the calibrating sample and the target galaxy could induce systematic effects of $>$10\% in calibration of the extragalactic distance scale.

RR Lyrae stars are a promising (Population II) alternative for calibration of the cosmic distance ladder, with several authors concluding that the $K$-band PL relation includes a metallicity term (e.g. \cite{bono01,bono03,sollima06,marconi15}) but without agreement on its magnitude. Moreover, the relative abundance of CNO elements in metal-poor RR Lyrae stars can simulate a different global metallicity, giving variations in the pulsation properties and in the relation between period, magnitudes and colours. In particular, CNO abundances can affect the intrinsic luminosity of Horizontal Branch pulsating stars and also the coefficients of the traditionally adopted absolute visual magnitude versus [Fe/H] relation.

The CNO abundance had long been considered as constant among stars in the same cluster, but within the discoveries of multiple stellar populations in GCs, the summed CNO abundances have also been seen to vary in some clusters. To support the dramatic improvement in astrometry from {\em Gaia} and the chance to study variables out to cosmologically-interesting distances, we need accurate abundance determinations of individual pulsating stars in external galaxies to quantify the impact of such abundance variations. In particular, CUBES will allow us to investigate iron-peak elements in metal-poor stars, as well as estimating the CNO abundances from the molecular bands in the UV range (see Case G6).

Another interesting application at UV wavelengths is to investigate pulsating low-mass horizontal branch (HB) stars. During the pulsational cycle their effective temperatures undergo large variations, where they can become hotter (cooler) than the blue (red) boundary of the instability strip. The effective temperatures of RR Lyrae stars are usually obtained from colour relations that have uncertainties of at least $\pm$100K. Spectroscopic determinations of temperature would avoid using such relations. Moreover, accurate sampling of the temperature variations during their pulsational cycles would also allow us to investigate (via comparisons with suitable pulsation models) the impact of hydrodynamical processes such as convective shocks on their stellar envelopes.

\subsubsection{G8: Key requirements}
\begin{itemize}
    \item{{\it Spectral resolution:} $R$\,$\ge$\,20,000.}\smallskip
    \item{{\it Spectral coverage:} Essential: 305-400\,nm.}
\end{itemize}

\subsection{G9: Horizontal branch stars in Galactic GCs}
Observations of HB stars provide another route to investigate the origins of multiple populations in GCs. In addition to the spectroscopic results that highlight star-to-star differences in GCs \cite{gratton12}, high-quality space photometry has further reinforced the idea of multiple populations. This is manifested by the spread in the main sequence in GCs, which cannot be explained by photometric uncertainties alone. For instance, the {\em HST} study of NGC~2808 was particularly important as it revealed three discrete components, that were interpreted as due to different helium abundances of the stars in each sequence, with the bluest stars the most helium-rich \cite{piotto07}.

These results argue that the stars with anomalous chemistry (on the blue side of the main sequence) formed from gas processed by proton-capture nucleosynthesis. This is supported by the finding of a  significant spread in the Al content in NGC~2808, with stars on the bluer sequence being Al-enriched \cite{bragaglia10}. As noted in Section \ref{gc_pops}, such results have stimulated theories regarding possible self-enrichment mechanisms. Possible sources of this polluted material from the first generation of stars include massive AGB stars \cite{dercole08}, rapidly-rotating massive stars \cite{decressin07}, super-massive stars \cite{dh14} and interacting massive binaries \cite{sdm09}. Each of these scenarios have their own difficulties when compared to the observational results (see e.g. \cite{renzini15}).

\subsubsection{HB stars in the context of multiple populations}

The HB feature in the colour-magnitude diagram (CMD) of GCs is populated by stars brighter than the main-sequence turn-off and bluer than their counterparts evolving through the red giant branch (RGB). HB stars are currently in the core He-burning phase, which starts in the He-flash modality once the stars reach the tip of the RGB, and continues in a quiescent mode after the electron degeneracy in the core is removed. The CMDs of GCs obtained so far offer a variety of situations regarding the HB extension and morphology. Some GCs are characterised by narrow and red HBs (e.g. 47~Tuc; \cite{sarajedini07}), others exhibit very extended HBs, harbouring groups of stars significantly fainter and bluer than the main HB component (e.g. NGC~2808, \cite{bedin00}; NGC~2419, \cite{dc15}). 

The location of HB stars in the CMD is primarily related to the mass of the envelope on the He-burning core. This is influenced by factors such as the age and metallicity of the cluster, and the mass-loss history during the RGB phase. Unfortunately, some of these factors counterbalance each other, which introduces degeneracies in the interpretation of the observed HB morphologies. In addition to the mass distribution, other physical properties such as chemical composition have also been suggested to explain the peculiar HBs of some GCs. 

\subsubsection{He abundances of HB stars}
To illustrate the potential contribution of CUBES, we consider the GC NGC~6441. Its CMD has a well developed blue HB, with a strong slope upwards from the red clump to the blue of the RR~Lyrae region \cite{piotto02}. The peculiar features of the HB of NGC~6441 are not explained by conventional evolutionary models, which is a strong indication of the co-existence of at least two generations of stars. One suggestion is that
a spread in the helium content (of $\Delta Y$\,$\sim$\,0.1) between the stars on the red and blue side of the HB might account for the morphology \cite{dantona02,cda07}. Direct measurement of the He content of the HB stars is required to test this -- if confirmed then the difference in He abundance would indicate that the most contaminated stars formed from gas with $Y$\,$\sim$\,0.35, which is consistent with contamination from super-AGB stars.

\begin{figure}[h]
\begin{center}
  \includegraphics[width=8.5cm]{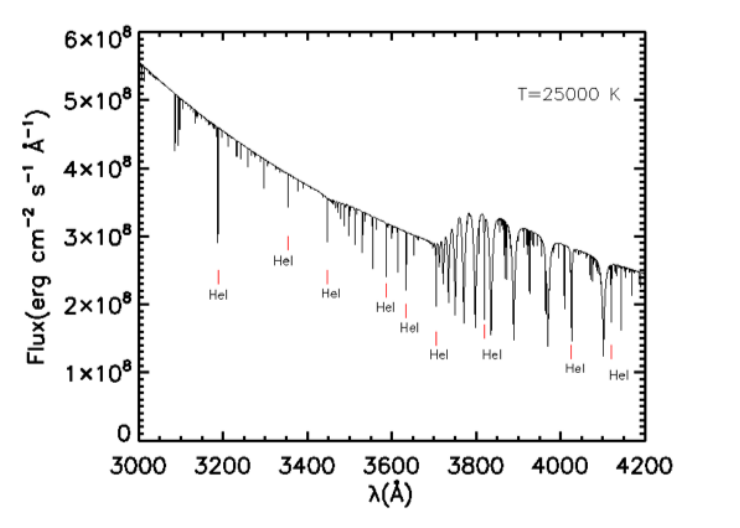}
  \includegraphics[width=8.5cm]{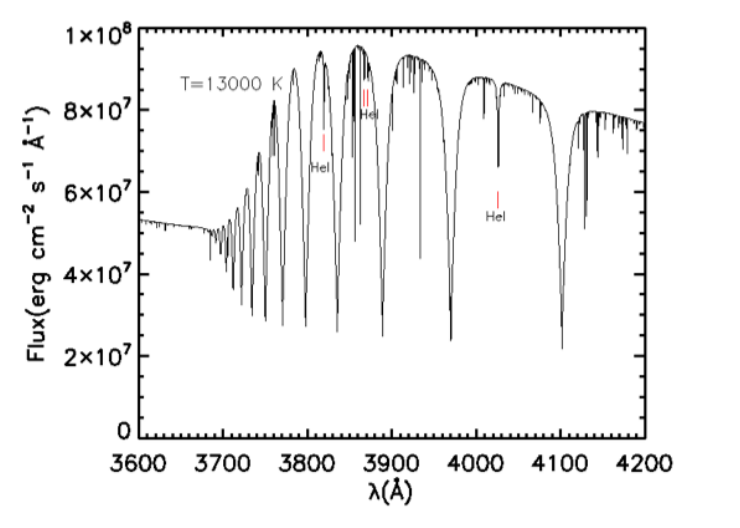}
  \caption{Synthetic spectra of stars with effective temperature 25\,kK (upper panel) and 13\,kK (lower panel) in the 300-420\,nm range. The wavelength scale in the lower panel is limited to \lam\,$>$\,360\,nm (3600\,\AA), as there are no significant helium lines at shorter wavelengths at this temperature.}
\label{HB_models}
\end{center}
\end{figure}

Synthetic model spectra ($R$\,$=$\,20,000) over the 300-420\,nm range of stars with log($g$)\,$=$\,4.0 and T$_{\rm eff}$\,$=$\,25 and 13\,kK are shown in Fig.~\ref{HB_models}. These examples correspond to stars on the hot and cool side of the blue HB in NGC~6441. There are multiple He~\1 lines in this wavelength region that can be used to estimate the He-mass fraction. Specifically, in the hotter star these include: He~\1 318.8, 335.5, 344.8, 358.7, 363.4, 370.5, 382.0, 386.9 and 387.2\,nm. In the cooler star only the latter three lines are useful, and the use of He~\1 402.6\,nm becomes more important. Thus, extension of the CUBES range to 400-420 nm is desirable for this case.

As for NGC~6441, the majority of GCs investigated so far have extended HBs (e.g. the case of NGC~2808, see \cite{piotto07,bedin00}). By obtaining estimates of surface helium abundances CUBES will provide important information towards understanding the multiple populations in GCs, with no other instrument capable of achieving the same result for our targets. 

To achieve the desired precision, we require S/N\,$\gtrsim$\,50 to estimate the He abundance. Using the model spectra shown in Fig.~\ref{HB_models} as input templates to the CUBES ETC for a target with $V$\,$=$\,16\,mag., the predicted exposure times to reach S/N\,$=$\,50 are 1 and 4\,hrs for the 13 and 25\,kK models, respectively. In contrast, UVES observations would require integrations of 8 and 24\,hrs for the same stars. 

The He~\1 lines at longer wavelengths can also be used to partly address this case, but UV observations with CUBES will take advantage of the rising spectral energy distribution towards shorter wavelengths for the hotter stars (e.g. upper panel of Fig.~\ref{HB_models}), enabling observations to more distant clusters than currently possible. In addition to the He abundances, CUBES observations of HB stars will also provide important insights into the CNO abundances and distribution of s-process elements  (see Cases G6 and G7 above).

\subsubsection{G9: Key requirements}
\begin{itemize}
    \item{{\it Spectral resolution:} $R$\,$\gtrsim$\,10,000; Goal: $R$\,$\sim$\,20,000.}\smallskip
    \item {{\it Wavelength coverage:} Essential: 300-400\,nm; Desirable: 400-420\,nm.}\smallskip
    \item{{\it Simultaneous observations at longer \lam:} Observations with UVES are a useful complement.}
\end{itemize}

\subsection{G10: Early-type companions in binary Cepheids}

Most Cepheid variables (up to ~80\%) are expected to be members of binary systems, where their companions are main-sequence or slightly evolved early-type stars, which are typically much fainter in optical bands and hardly detectable. One of the methods to detect such companions is to look at the Ca~\2 H and K lines. Normally they should have similar depths, but when a hot companion is present its strong H-epsilon line is superimposed on the Ca~\2 H line, which appears significantly deeper than the Ca~\2 K line. Characterising the orbital properties of these systems beyond the Milky Way is challenging, with current spectroscopy limited to V\,$\sim$15.5\,mag (e.g. \cite{pilecki21}). CUBES will provide the greater sensitivity needed for efficient observations of larger samples in the Magellanic Clouds.

The motivation for this case is to better characterise Cepheid binary systems in the Clouds down to V\,$=$\,17-18\,mag. The improved efficiency of CUBES will enable us to better detect main-sequence companions beyond the reach of current facilities, either via the Ca~\2 H and K lines, or perhaps via the broadening function technique \cite{pilecki21}.
 The latter is less demanding in terms of the required S/N given the use of multiple lines, but tailored tests are needed for its applicability to the CUBES range.

\subsubsection{G10: Key requirements}
\begin{itemize}
    \item{{\it Spectral resolution:} $R$\,$\ge$\,20,000; Goal: $R$\,$\sim$\,40,000.}\smallskip
    \item{{\it Spectral coverage:} Essential: 380-400\,nm.}
\end{itemize}

\subsection{G11: Extragalactic massive stars}

With masses of more than ten times that of the Sun, massive OB-type stars burn brightly for just a few million years before exploding violently as supernovae. During their lives they manufacture the materials from which planets and life are made and, via their high-energy winds and explosive deaths, they shape the chemistry and evolution of their host galaxies \cite{a82,a13,ebm18}. Only by learning how these stars behave in the local Universe can we attempt to interpret distant, star-forming galaxies, whose light is dominated by vast numbers of these spectacular objects, and that are used to chart the history of the Universe from the Big Bang through to the present.

A focus of the past 20 years has been to explore the massive-star populations in galaxies beyond the Milky Way, to investigate their properties in the metal-poor environments of the Magellanic Clouds and dwarf galaxies in the Local Group and beyond. 

As expanded further in an article in this Special Issue \cite{ExA_Evans}, we have been able
to study the properties (including oxygen abundances) of B-type supergiants in metal-poor galaxies in targets at 1-2\,Mpc (e.g. \cite{evans07,castro55}) but we lack the sensitivity for comparable studies of O-type stars at such distances.

The ground-UV domain is relatively unexplored in the context of massive OB-type stars, with only a few published observations (e.g. \cite{lamers72,dufton,drissen95}). However, there are several He~\1 absorption lines shortwards of 400\,nm, as well as the higher members of the Balmer series up to the Balmer limit. The latter can provide constraints on the spectroscopic gravities, particularly for low-metallicity stars where wind effects are relatively minimal and comparisons with plane-parallel model atmospheres are sufficient for quantitative analysis. Moreover, above effective temperatures of T$_{\rm eff}$\,$\sim$\,27.5\,kK, absorption from He~\2 302.3\,nm becomes increasingly strong, offering a potential temperature diagnostic for O-type stars when compared with, e.g., the strength of the He~\1 318.8\,nm line. 

There are also a large number of O~\3 absorption lines in the ground UV (plus lines from O~\4 and O~\2 at the hotter and lower temperatures, respectively, of O-type stars). Given sufficient S/N ($\gtrsim$\,100), these O~\3 lines can be used to provide robust estimates of oxygen abundances (e.g. \cite{martins15}), that are required to test model predictions of mixing in stellar evolution models (complementing carbon and nitrogen abundances obtained from observations at longer wavelengths). 

When coupled with the rapidly increasing spectral energy distributions of O-type stars over the ground UV, observations with an optimised near-UV spectrograph can potentially bring targets into reach that are otherwise too faint at longer wavelengths. For instance, from calculations using the CUBES ETC, it should be possible to obtain high-quality (S/N\,$>$\,100) spectra of O-type stars in metal-poor systems at a distance of $\sim$1\,Mpc in a total integration of $\sim$3\,hrs (for more detailed performance estimates see \cite{ExA_Evans}). This will extend such abundance studies for O-type stars beyond the Galaxy and Magellanic Clouds for the first time, sampling a broad range of metallicities and galaxy types.

\subsubsection{G11: Key requirements}
\begin{itemize}
    \item{{\it Spectral resolution:} Essential: $R$\,$\ge$\,5,000; Desirable (for some applications): $R$\,$\sim$\,20,000.}\smallskip
    \item{{\it Spectral coverage:} Essential: 305-400\,nm; Goal: 300-415\,nm.}
\end{itemize}

\section{Extragalactic Cases}
\subsection{E1: Primordial deuterium abundance}

Cosmology is in a golden age. We have established a `Standard Model' of particle physics and cosmology, and we know the content of the Universe to within a few percent. However, we still have no accepted model of dark energy and dark matter. Somewhat remarkably, we do not know why the Universe contains baryons instead of antibaryons, nor why the Universe contains baryons at all (i.e. the baryon asymmetry). We are also missing crucial properties of neutrinos, such as their hierarchy, why they change flavour, and the number of neutrinos that existed during the earliest phases of the Universe. These limitations indicate there is missing physics in our understanding. Thus, new cosmological observations may allow us to discover something truly fundamental about the early properties of the Universe.

\begin{figure}[h]
\begin{center}
  \includegraphics[width=9cm]{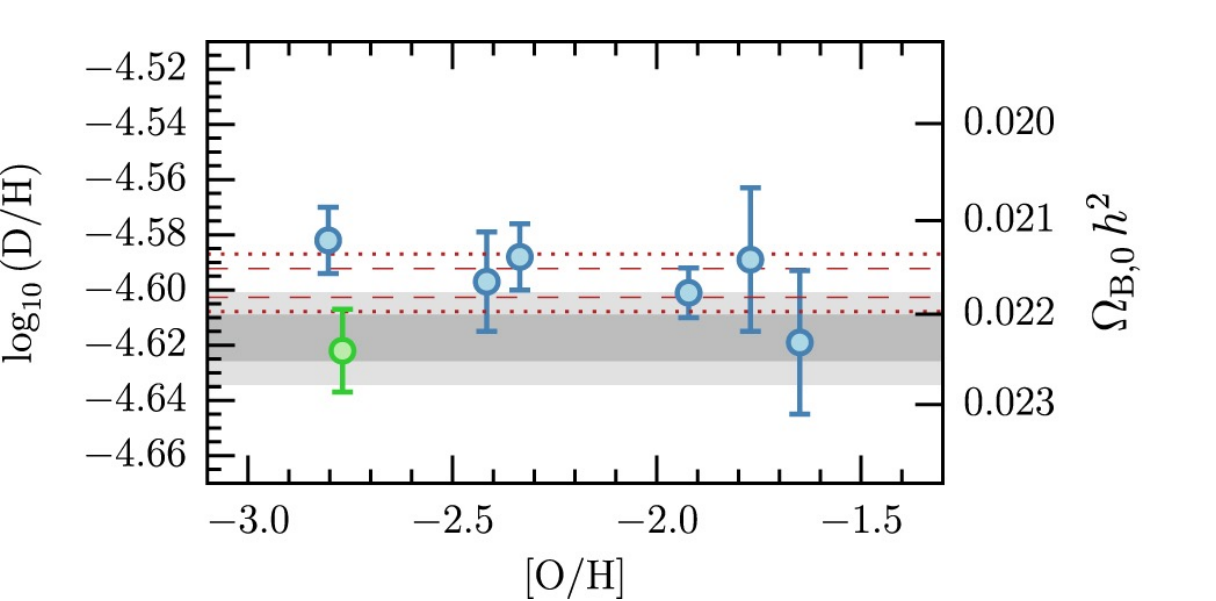}
  \caption{High-precision D/H measures as a function of oxygen abundance \cite{cooke18}). The weighted mean is shown by the red dashed and dotted lines, which represent the 68\% and 95\% confidence levels, respectively. Assuming the Standard Model of cosmology and particle physics, the right-hand axis shows the conversion from D/H to the universal baryon density. This conversion uses the theoretical determination of the d(p,$\gamma$)$^3$He cross-section from \cite{m16}. The dark and light shaded bands are the 68\% and 95\% confidence bounds on the baryon density derived from the CMB \cite{planck16}.}
\label{DH_cooke18}
\end{center}
\end{figure}

Some of the above questions can be studied and better understood by measuring the nuclides that were produced a few minutes after the Big Bang – the so-called ‘primordial elements’. The primordial deuterium abundance (D/H) is currently our most reliable probe of Big Bang Nucleosynthesis. It can be determined to $\sim$1\% precision using near-pristine gas clouds that imprint D~\1 and H~\1 absorption lines on the spectrum of a background quasar. The gas clouds where this measurement is possible are extremely rare, with just seven systems with precise measurements of D/H \cite{cooke18} (see Fig.~\ref{DH_cooke18}), most of which are at redshift $z$\,$\sim$2.8-3.1, placing the weak D~\1 Lyman series absorption lines at an observed wavelength of $<$400\,nm. The accessible redshift range is due to a combination of the efficiency of current spectrographs plus the atmospheric cut-off at UV wavelengths and, on the other hand, of the increasing thickness of the H~\1 Lyman-$\alpha$ forest (which can contaminate potential D/H systems) at high redshift.

To secure a measurement, the gas cloud needs to be near-pristine with unusually quiescent kinematics, and not blended by unrelated contamination from the forest of Lyman-$\alpha$ absorption lines due to the intergalactic medium along the line-of-sight. CUBES will deliver three key improvements over previous instruments that have been used for this measurement (specifically, VLT-UVES and Keck-HIRES):

\begin{itemize}
\item{Twice as many quasars are known at 2.5\,$<$\,$z$\,$<$\,2.8 than at 2.8\,$<$\,$z$\,$<$\,3.1 (the latter is the range used by current studies, the former is the range that will be enabled by CUBES). This increases the sample size by a factor of three.}\smallskip

\item{The impact of unrelated blending from the Lyman-$\alpha$ forest is much less significant at low redshift, leading to cleaner measurements.}\smallskip

\item{The increased sensitivity of CUBES at blue wavelengths will allow fainter quasars to be observed. Thanks to the steep shape of the quasar luminosity function, this will increase the sample size by a factor of at least 10 (simply because of the access to more quasars). Alternatively, the brightest quasars can be recorded in a fraction of the time compared to UVES or HIRES.}
\end{itemize}

CUBES should be able to improve the sample size by one-to-two orders of magnitude, while also providing cleaner measurements. UVES observations of the quasar spectrum at $>$400\,nm are also important to complement the CUBES data. This is because we need to record the H~\1 Lyman-$\alpha$ absorption line, in combination with the narrow metal absorption lines (with Doppler widths of $\sim$3-5\,\kms) belonging to the same redshift system, which reveal the structure of the velocity profiles. UVES has the most appropriate wavelength range and spectral resolution for this goal (i.e. it will resolve the key metal lines and has the ability to efficiently record the spectra of faint $u_{\rm AB}$\,$>$\,19 quasars). 

Calculations with the CUBES ETC predict that a quasar at $z=2.7$ of $u_{\rm AB}$\,$=$\,19, 19.5, 20\,mag could be recorded with the required S/N\,$=$\,20 at 320\,nm in $\sim$\,2, 5, and 13\,hr of observation, respectively (with 1\,hr exposures, airmass\,$=$\,1, two days from new moon and seeing of 0.8$''$). For this case the continuous wavelength coverage provided by the CUBES design is ideal, as there are 10-15 D~\1 absorption lines that could be detected in any given system, and these lines could fall anywhere in the 320-380\,nm wavelength range due to the redshift range of the targets.

\subsubsection{E1: Key requirements}

\begin{itemize}
    \item{{\it Wavelength range:} CUBES alone will not be able to simultaneously cover the high order D~\1 Lyman series lines and the H~\1 Lyman-$\alpha$ line. So the ideal wavelength range should be 320--380 nm to cover the D~\1 absorption lines between $2.5 < z < 3.1$, together with simultaneous UVES observations redwards of $\sim400$\,nm to cover the H~\1 Lyman-$\alpha$ line and resolve the metal absorption lines (the latter is needed to determine the component structure and requires the $R\sim40\,000$ provided by UVES).}\smallskip
    \item{{\it Spectral resolution:} For gas at $10^4$~K, the D~\1 lines have a width of $\sim$10~\kms. To resolve the line widths of deuterium absorption and assess line saturation, a spectral resolving power of $R\sim20\,000$ is required.}\smallskip
    \item{{\it Spectral sampling:} The line widths to be observed will be narrow (sometimes comparable or less than the instrument resolution). Therefore, this science case requires the data to at least be sampled at the Nyquist frequency (2--3 pixels FWHM) over the entire wavelength range.}
\end{itemize}

\subsection{E2: Missing baryonic mass in the high-$z$ CGM}

An active field of extragalactic research is the so-called ‘missing baryon’ problem, where the detected baryonic mass compared to the universal fraction suggests that we are missing more than 60\% in galaxies outside of clusters, rising to $\ge$90\% for gas-dominated galaxies and dwarfs (e.g. \cite{mcgaugh10}). Studies with {\em HST} of the CGM in galaxies at $z$\,$\sim$\,0.2 concluded that diffuse gas in their halos could account for at least half of the previously missing baryonic mass \cite{werk14}. A significant contribution to the baryonic mass is also expected from the hot diffuse gas (T\,$\sim$\,10$^{5.7}$ to 10$^{6.3}$\,K) traced by O~\7 and detectable in the soft X-rays \cite{nicastro18}. 

The dispersion observed in a sample of localized Fast Radio Bursts (FRBs, e.g. \cite{chatterjee17}) at $z$\,$\sim$\.0.1-0.5 has been used to measure the electron column density and accounts for every ionised baryon along the line of sight \cite{macquart20}. This independent measurement of the baryon content of the Universe with a first sample of five FRBs is consistent with the CMB and Big Bang Nucleosynthesis values, substantially solving the missing baryon problem at low redshift.  This result has established the total amount of baryons, but FRB observations cannot distinguish the share of the different gaseous components. In this respect, quasar absorption line studies will play a key role and, in particular, they could help understand whether a substantial fraction of the mass at high redshift is also contained in diffuse halo components. 

Several current and future instruments are being designed with the study of the high-redshift ($z$\,$>$\,2) CGM in mind. The Keck Cosmic Web Imager (KCWI) \cite{kcwi} was partly designed to address this topic (although its blueward coverage stops at 350\,nm), and one of the cases developed for the ELT-MOSAIC instrument (e.g. \cite{puech18}) is to use high-redshift Lyman-break galaxies as background sources to reconstruct the 3D density field of the IGM and to probe the CGM of galaxies at $z$\,$\sim$\,3 \cite{japelj19}. 

Observations in the ground-UV can play an important role in this field (complementing observations with {\em HST} and the ELT in the future), providing access to the baryons in halos at $z$\,$=$\,1.5 to 2, immediately after the era of peak star-formation in the Universe (e.g. \cite{md14}). The relatively low number density of lines in the Lyman-$\alpha$ forest at these redshifts implies that, even with observations at $R$\,$\sim$\,20,000, the impact of blending on the measurement of the CGM metal lines can be minimised (which is not the case at larger redshifts). Furthermore, the proposed wavelength range for CUBES will enable observations of the important O~\6 line (used to trace the warm-hot gas (see \cite{t11})) to redshifts $2$\,$<$\,$z$\,$<$\,2.8. In both cases, to derive the baryon contribution of the gas, complementary UVES (or ESPRESSO) spectra will be required to complete the coverage of the metal transitions and of the Lyman-$\alpha$ forest. 

A significant improvement in near-UV efficiency cf. UVES will open up background sources 1-2 magnitudes fainter than the quasars used at present, significantly increasing the number (and the space density) of available targets and/or the possibility to obtain very high signal-to-noise ratio spectra to detect the faint lines (e.g. \cite{vdo16}).   

To be able to characterise absorption systems associated with galaxies at impact parameters of up to a few hundreds of physical kpc (e.g. \cite{werk14}) we will need to carry out preparatory studies to identify and determine the properties of the galaxies in the fields of the quasars forming our sample. The recent QUBRICS survey \cite{calderone19,boutsia21} is being used to identify quasars in the southern hemisphere with $i_{\rm AB}$\,$<$\,18\,mag. and will provide the 30-40 quasars needed for this case. 

Detailed simulations of the proposed observations are presented elsewhere in this Special Issue \cite{ExA_DOdorico}.
These demonstrate that observations reaching S/N\,$\sim$\,15 per pixel would be possible for a sample of 10 targets at $z$\,$=$\,2, 20 at $z$\,$=$\,2.5 and 10 at $z$\,$=$\,3 (all with a conservative $u_{\rm AB}$\,$=$\,18\,mag.) in a total programme of 13\,hrs.

\subsubsection{E2: Key requirements}
\begin{itemize}
    \item{{\it Position angle:} Non-parallactic observations are desirable to observe quasar pairs or strongly-lensed quasar images (arcs) for tomographic studies.}\smallskip
    \item{{\it Observations at longer \lam:} Required, so simultaneous observations would give greater operational efficiency.}
\end{itemize}

\subsection{E3: Cold gas at high redshift}

Understanding the physical and chemical properties of the gas in and around galaxies is fundamental for a complete theory of star-formation along the history of the Universe. Molecular hydrogen (H$_2$) is a critical species to achieve this goal since it traces the cold (T\,$\sim$\,100\,K) component of the neutral gas, which is inclined to gravitational collapse through Jeans instabilities. Indeed, the cooling of the gas through atomic emission lines depends on the abundance of metals -- to which the formation of H$_2$ is also linked -- and both the gas heating and the dissociation of H$_2$ depend on the strength of the UV field. In addition, H$_2$ is vital for the formation of other molecules that participate in the gas cooling, eventually initiating the collapse of the cloud that gives birth to stars.

\begin{figure}[h]
\begin{center}
  \includegraphics[width=8cm]{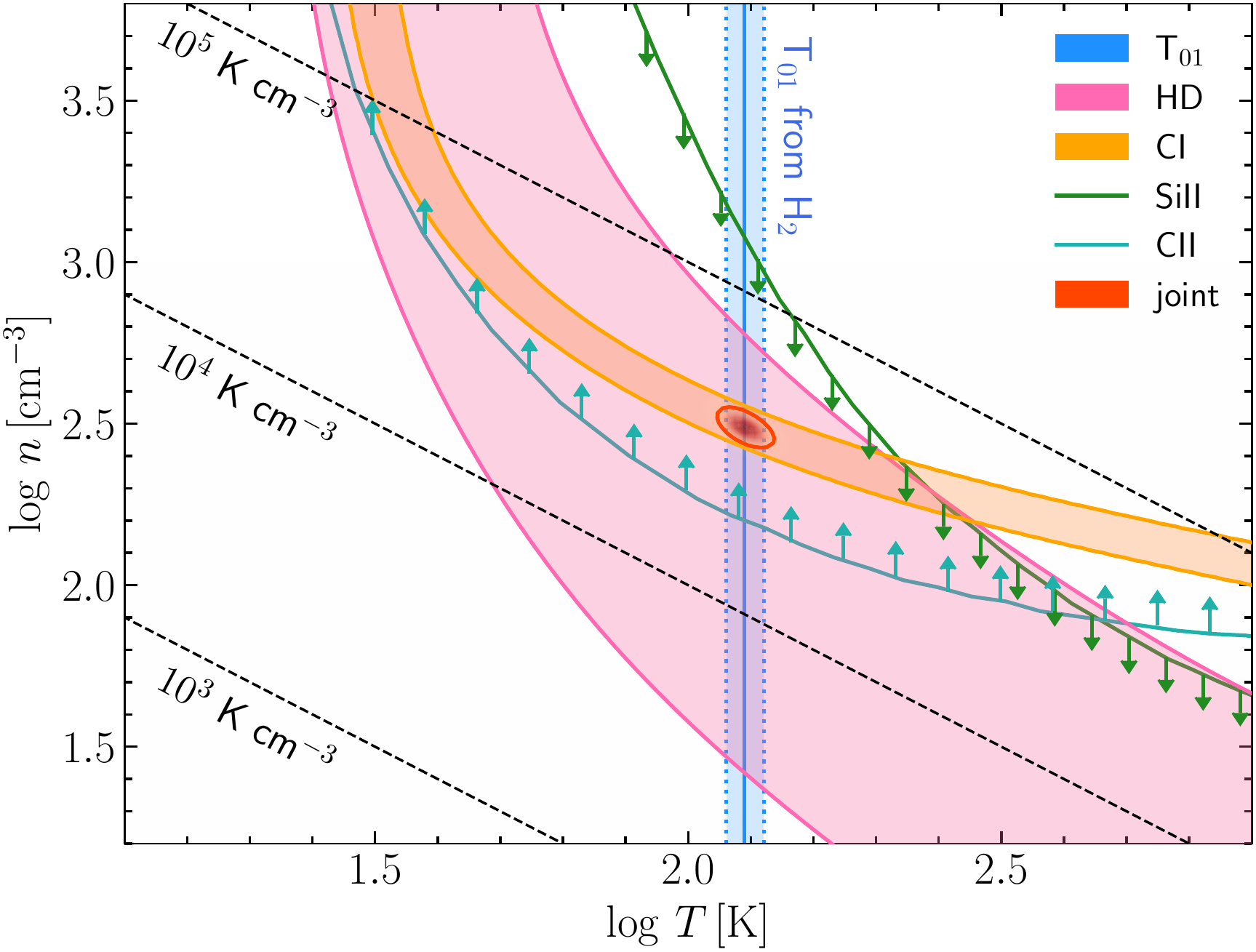}
  \caption{An example of tight constraints on the number density and temperature in cold H$_2$-bearing gas in galaxies at cosmological distances, that can be obtained from observations of molecular hydrogen (H$_2$ and HD) and fine-structure levels of associated metal species (adapted from \cite{balashev17}).}
\label{balashev}
\end{center}
\end{figure}

Molecular hydrogen can be detected directly through electronic absorption lines in the Lyman and Werner bands (90-110\,nm rest-frame). These have been observed since the 1970s towards bright nearby stars thanks to space-borne UV spectrographs and greatly contributed to the understanding of the local ISM. At $z$\,$>$\,2, H$_2$ lines are conveniently shifted into the optical domain and become accessible from the ground. UVES and X-Shooter have both played key roles in the investigation of diffuse molecular and translucent gas, probed by H$_2$ in the distant Universe towards quasars (e.g. \cite{nd08,balashev19}) and gamma-ray burst (GRB) afterglows (e.g. \cite{bolmer19}). More recently, it has also been shown that H$_2$ can serve as an independent and sensitive probe of AGN feedback when studied in the quasar environment itself \cite{nd19,nd21}. In short, H$_2$ together with other atomic (in particular C$^0$) and molecular (HD, CO) species, provide very sensitive and unique tools to investigate cold gas and the prevailing conditions such as temperature, density, UV flux and cosmic ray ionization rate (e.g. \cite{klimenko20,kosenko21}, and see Fig.~\ref{balashev}). 

\begin{figure}[h]
\begin{center}
  \includegraphics[width=9cm]{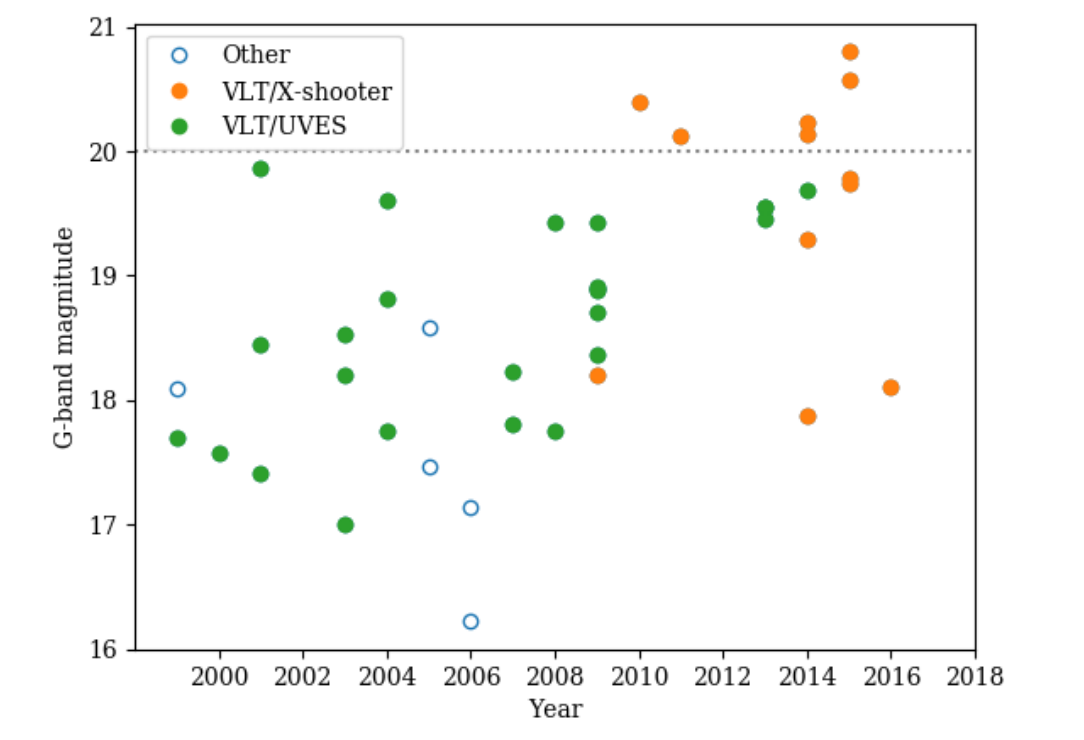}
  \caption{$G$-band magnitude of quasars with intervening H$_2$ systems observed over the past two decades. As progress has been made in selecting targets from large surveys (e.g. SDSS) and also those that are molecular rich, we have reached the magnitude limits of UVES (even with 10-15\,hr exposures and, even then, without much information on the physical conditions). X-Shooter has extended the limiting magnitude, but its resolution is not sufficient to estimate the physical conditions. CUBES will bring these fainter targets within our grasp, while also opening-up observations at lower redshifts and of absorption systems with more extreme properties (chemically enriched, high columns).}
\label{H2_systems}
\end{center}
\end{figure}

While good progress has been made in selecting targets for study from large spectroscopic surveys (e.g. SDSS), the field is now reaching a bottleneck with the vast majority of the (newly) identified systems being out of reach with current instruments (see Fig.~\ref{H2_systems}). This is due not only to the steep shape of the quasar luminosity function, but also to reddening of the quasar light by the absorbing (high column density and dusty) gas itself. 

Only a significant improvement in sensitivity in the ground-UV domain can provide the observations required to transform this field. The efficiencies envisaged for CUBES will:
\begin{itemize}
\item{Increase the sample size by 1-2 orders of magnitude. In addition to the systems that are already available from pre-selection in SDSS, other surveys in the coming years will also contribute to the discovery of further systems requiring follow-up.}
\item{Provide access to more extreme environments (e.g. more dusty/molecular rich).}
\item{Enable cleaner measurements. Indeed, in addition to the high blue throughput providing observations of fainter background sources (both quasars and GRB afterglows), it will also open-up observations of systems at lower redshift (reducing the huge gap in cosmic history that is still poorly explored).} 
\item{Cover more H$_2$ bands for a given system or redshift, enabling robust measurements.}
\end{itemize}

The observed population of H$_2$ (and HD) rotational levels, with proper modelling, enable the physical conditions prevailing in the associated diffuse molecular gas of high-$z$ galaxies to be derived. Observations of the H$_2$ (and HD) lines require a careful compromise between spectral resolution and system efficiency. For instance, X-Shooter (with $R$\,$\sim$\,5,000) has been used a few times to complement high-resolution ($R$\,$\sim$\,50,000) UVES observations (e.g. \cite{nd17}). However, this is extremely costly in observing time and the tenfold lower resolution of X-Shooter is far less suited to deblend the H$_2$ lines from the Lyman-$\alpha$ forest, impeding robust measurements of H$_2$ over various rotational levels which encode crucial information about collisional, radiative and turbulent processes. A spectral resolution of $R$\,$\sim$\,20,000 is the sweet spot between efficiency and sufficient information for quantitative analysis.

To illustrate this huge gain, we highlight example observations of a quasar at z = 2.3 with $V_{\rm AB}$\,$=$\,20\,mag. At an airmass of 1.3, with 1$''$ seeing and in a 1\,hr exposure, the ETC predicts a S/N per pixel at 350\,nm of: 1 (UVES, binning 2$\times$2), 5 (X-Shooter, binning 2$\times$2), and 6 (CUBES, no binning)\footnote{We note that the estimate provided in the paper on the H$_2$ science case \cite{ExA_Balashev} are unrealistically optimistic, due to the use of a preliminary version of the CUBES ETC.}.

Finally, we note that CUBES will be highly complementary to ELT observations with ANDES (formerly HIRES) to measure the evolution of the Cosmic Microwave Background temperature from the excitation of CO molecules. Indeed, recording of H$_2$ lines will be necessary to constrain the collisional contribution to the excitation of CO.

\subsubsection{E3: Key requirements}
\begin{itemize}
    \item{{\it Wavelength coverage:} Essential: 305-380\,nm; Desirable: 305-400\,nm.}\smallskip
    \item{{\it Simultaneous observations at longer \lam:} To maximise the impact of CUBES observations of H$_2 $ in the blue, observations at $>$\,400\,nm are also desirable to include the metal lines and associated fine-structure levels, ideally at higher spectral resolution. This will enable more precise modelling of the absorption profiles and robust derivation of the physical and chemical conditions. Simultaneous observations with UVES at $>$\,400nm are thus ideal for this case.}
\end{itemize}

\subsection{E4: Reionisation}

The search for sources responsible for cosmic reionization, likely to have occurred at z\,$>$\,6.5, is one of the most pressing questions in modern cosmology and galaxy evolution (e.g. \cite{f19}). Galaxies are commonly thought to be able to produce the bulk of the UV emissivity at high redshift (e.g. \cite{robertson15}) but the AGN population is also proposed as a relevant or dominant contributor (e.g. \cite{giallongo19}; but also see \cite{fontanot12,hs15} for different views), and more exotic possibilities such as decaying particles cannot be excluded (e.g. \cite{poulin15}). Due to the very high cosmic opacity (e.g. \cite{worseck14}), it is not possible to directly study the Lyman continuum (LyC) emission at the epoch of reionization or even in the aftermath (e.g., at $z$\,$\ge$\,5). As a consequence, analogues of the first sources have to be found and their properties studied at lower redshifts. 

Quasars are known to be efficient in producing UV photons with the fraction (f$_{\rm esc}$) of ionizing photons able to escape to the IGM close to 100\% \cite{cristiani16,grazian18}. In contrast, galaxies are much more numerous, but f$_{\rm esc}$ has large uncertainties. The census of LyC-galaxies is growing fast in the local Universe (e.g. \cite{izotov16a,izotov16b,izotov18}) and more slowly at high redshift, 1\,$<$\,$z$\,$<$\,4 (\cite{shapley16,vanzella16,bian17,vanzella18,saha20,rt19}), given the challenging observations required. Thus, the contribution of star-forming galaxies to the cosmic ionization budget must be inferred from other properties related to their ability to leak ionizing radiation into the IGM.

While the construction of a sample of LyC emitters serves as a reference for the identification of cosmic reionizers, the physical processes that made these galaxies transparent to LyC is interesting by itself. The way ionised channels are carved in the interstellar medium is matter of investigation, especially in the local Universe (e.g. \cite{bik18,micheva19}) where the detection of LyC radiation is also a tool to catch this phenomena in the act \cite{heckman11}. A recent work based on VLT/X-Shooter observations on a gravitationally-lensed object at $z$\,$=$\,2.37 shows that the LyC radiation can escape through ionized channels carved by massive stars hosted in young massive star clusters \cite{vanzella20}. The enhanced spatial resolution provided by lensing magnification allows us to observe such channels - aligned along the ionizing sources - with a better contrast, and consequently witness high f$_{\rm esc}$ values (as high as 90\%; \cite{rt19}).

It is not fully clear if the properties of local ($z$\,$<$\,0.4) LyC-galaxies really match those at high redshift ($z$\,$>$\,2), despite the fact they are often referred to as genuine analogues. It is therefore critical to investigate the LyC leakage at high redshift (e.g. $z$\,$>$\,2, 2\,Gyr after the Epoch of Reionization), to reduce the cosmic time interval from reionization and avoid any tantalising extrapolation from $z$\,$\sim$\,0. This needs to be done under the more favourable observational conditions: (1) using a spectroscopic approach closely probing the Lyman edge at 91.2\,nm
\cite{steidel18}; (2) reducing the cosmic opacity as much as possible while keeping $z$\,$>$\,2; (3) observing the enhanced signal and spatial resolution provided by strong lensing (e.g. \cite{vanzella20}). 

Combining these factors together for strongly-lensed, star-forming regions at $z$\,$>$\,2.35 argues for high-efficiency spectroscopy in the ground UV. In addition to ensuring high throughput, such an instrument needs to be sky limited at intermediate and high resolution (needed for disentangling in cases of contamination) and with a relatively long slit for background subtraction.

\subsubsection{Lyman continuum detection of a sample of faint AGN at z\,$\sim$\,3-4}
A component of the UV background case is to measure the escape fraction of a sample of $\sim$30 faint AGNs ($M_{\rm 1450}$\,$=$\,$–$23) at $z$\,$\sim$\,3-4. These AGN have been selected from an extensive spectroscopic campaign in the COSMOS field, where multiwavelength coverage from X-rays to radio ensures that the sample is complete and representative of the population of active supermassive black holes at $z$\,$\sim$\,3-4. Deep CUBES spectroscopy shortwards of 400\,nm (sampling shortwards of the restframe Lyman limit) will enable precise estimates of f$_{\rm esc}$ for these faint AGN and to derive the mean free path of H~\1 ionizing photons. These quantities are fundamental to derive the contribution of faint AGN to the ionizing background at $z$\,$\sim$\,4. These empirical results will then provide a solid foundation to explore the role of faint AGN in the epoch of reionisation.

\subsubsection{E4: Key requirements}
\begin{itemize}
    \item{{Spectral coverage:} Essential: 320-380\,nm; Goal: 300-400\,nm.}
    \item{{\it Spectral resolution:} Essential: $R$\,$>$\,1000; Goal: $R$\,$>$ 10,000. High resolution is not critical -- the ideal resolution is the highest value possible for which the detectors remain background limited (at all wavelengths) for 1\,hr observations of faint objects.} 
    \item{{\it Detector dark current:} The need to be background limited informs the requirements on the acceptable dark current of the detectors. The dark current should not generate more counts than a typical new moon sky. The variance in read-out noise should not exceed the dark current in a 1\,hr exposure (with 2\,$\times$ binning), giving a requirement on the dark current of $<$\,1e$^-$/hr and on the read-out noise of $<$2e$^-$/hr.} 
\end{itemize}

\section{Transients}
\subsection{T1: Gamma-ray bursts}
GRBs are extremely powerful explosions of cosmological origin lasting just a few seconds (see \cite{piran13} and references therein). Their $\gamma$-ray (or prompt) emission is followed by afterglow radiation at longer wavelengths (from X-rays to the radio), which can last several days (e.g. \cite{zhang07}). Such sources fade very quickly, so the capability to repoint narrow-field instruments to the $\gamma$-ray position and spot the afterglow has been critical to  understand these elusive phenomena. The {\em Swift} satellite (specifically designed to study GRBs \cite{gehrels04}) can detect the prompt event in $\gamma$-rays, repoint its X-ray and optical-UV instruments automatically, and disseminate arcsecond-precision coordinates of the afterglow in tens of seconds. This information is exploited by robotic telescopes that follow-up and gather information on the GRB just minutes after the explosion.

To follow-up such sources ESO developed a Rapid Response Mode (RRM), which allowed the VLT to react promptly to a GRB alert for the first time. The RRM procedure overrides execution of the current Observing Block and automatically repoints to the afterglow position (after checking for visibility constraints). This enabled UVES to obtain high-resolution spectra of GRB afterglows only a few minutes after the prompt event, when the sources were still bright. Indeed, the quickest spectrum was obtained for GRB060607 \cite{fox08}, just 6.5 minutes after the trigger, with better quality spectra (S/N\,$=$\,50) obtained 8 minutes after trigger for the naked-eye burst GRB080319B, observed when its magnitude was still $R$\,$=$\,12\,mag. \cite{delia09}. The capability to quickly point the VLT to such sources has uncovered a terrific wealth of information about the explosion sites of GRBs, their host galaxies and the gas along their line of sight (enabling detailed studies of abundances, metallicity and chemical enrichment up to cosmological scales, through the analysis of the hydrogen and metal absorption lines).

In the context of spectroscopy of the afterglow, the UVES+RRM combination made dissection of the close environment of GRBs possible via separation in velocity space of the absorbing components from the same metal feature. Even more importantly, time-resolved spectroscopy revealed a strong variation of fine-structure lines from excited levels. The overall decrease in time of the absorbing column density in these features was interpreted as the decay of the excited levels into the fundamental states as less and less flux was available for the gas to populate the higher energy states. This detection enabled determination of which gas component in velocity space was close to the GRB explosion site (as the burst flux directly influenced the gas physical state) and even determination of the GRB-absorber distance, by comparing the column density variation of the absorbing gas with time-dependent photo-excitation models \cite{vreeswijk07}. Remarkably, this distance is quite large, ranging from tens to several hundred parsecs. This shows how far the GRB flux can modify its surrounding medium and provides important constraints on radiative models of GRBs.

The primary diagnostics to determine the GRB-absorber distance are from Fe~\2. Its reddest lines are at $\sim$260\,nm (rest frame), so this kind of analysis is difficult for GRBs at $z$\,$<$\,0.8 because the throughput of UVES dramatically reduces below 400\,nm. CUBES will give a S/N that is at least five times better than UVES for the same exposure time over 300-400\,nm, so will significantly increase the redshift range for studies of GRB afterglow to investigate the distance between the GRB progenitor and the absorbing gas (UVES has enabled this for GRBs at $z$\,$\ge$\,1, but CUBES will bring events at 0.25\,$\lesssim$\,$z$\,$<$\,1 within reach). 

We note that such studies require a bright afterglow and a fast reaction, combined with clear weather and (in the past) the correct instrument mounted at Paranal. These demanding constraints mean that only 15 GRBs have been observed with UVES to date, with less than half yielding a spectrum with high enough S/N and with the correct set of lines falling in the instrumental range to warrant publication. The advent of CUBES will significantly increase these numbers, securing data for close-by GRBs, for which the flux level should be higher at constant luminosities. We add that, while the {\em Swift} satellite is presently in good health, there are other missions in development (e.g. SVOM) that will provide alerts once {\em Swift} is no longer operational.

For reference, $\sim$25\% of long GRBs with measured redshifts are at z\,$<$\,1, the fraction grows to $\sim$75\% for the short GRBs. In terms of absolute numbers, {\em Swift} typically detects 90-100 GRBs/yr, roughly 10 of which are short. So we expect $\sim$30 long/short GRBs at $z$\,$<$\,1 per year.  However, only (approx) 10\% will be promptly visible from Paranal during the night, i.e. events asking for possible RRM observations (although before the observation we also do not necessarily know the redshift if targetting an afterglow). In short, roughly one RRM per month is to be expected.

\subsubsection{Key requirements}
\begin{itemize}
    \item{{\it Simultaneous observations at longer \lam:} Simultaneous UVES observations are fundamental to provide:}
\begin{enumerate}
\item{Better handling of uncertainties such as skylines and intervening absorbers (which may only affect specific portions of the spectrum).} 
\item{Insights into how our understanding of the GRB environment is influenced by different diagnostic lines and spectral/velocity resolutions. In particular, it will be useful to determine how the different instruments separate the interstellar features, to study the saturation problem which arises as the instrumental resolving power goes down, and to evaluate the goodness of the fit with Voigt profiles (via data that could have a higher resolution but lower S/N, and vice versa).}
\item{Cross-matching the flux calibration of data from the two instruments, especially for observations at high airmass.}
\end{enumerate}
\item{{\it Acquisition images:} From past experience of GRB observations acquisition imaging has provided useful scientific information, particularly for instruments without imaging capabilities (e.g. X-Shooter). For transients, we often have an inaccurate idea of the target spectrum and/or are taking observations in sub-optimal conditions. Having one or more photometric points from the acquisition camera would be useful to support analysis of the CUBES data.}
\end{itemize}

\subsection{T2: Kilonovae}

GRBs with $\gamma$-ray durations of $<$2\,s (`short-GRBs') are likely produced by compact binary mergers involving either two neutron stars or a neutron star and a solar-mass black hole. The merger remnant ejects an ultra-relativistic jet, producing GRB emission. The GRB afterglow originates from synchrotron radiation produced by the interaction between the ultra-relativistic jet and the surrounding medium. The luminous, short-lived transient also ejects $\sim$0.01M$_\odot$ of neutron-rich matter that creates a radioactively-powered explosion called a kilonova or mini-supernova. The transient arises from the radioactive decay of freshly synthesised heavy r-process elements \cite{lp98}, that may be the dominant mechanism for their production throughout the Universe.

The short-GRB field was revolutionized by a kilonova event following the simultaneous detection of a gravitational-wave chirp (GW170817, \cite{abbott17,tanvir17}). At $z$\,$\sim$\,0, future detections of associated gravitational-wave events will allow us to constrain kilonova heavy-element yields. However, observations of short-GRBs up to $z$\,$\sim$\,1 or farther are required to constrain the evolving rate density of mergers to determine their contribution to the enrichment of heavy elements over cosmic history. Moreover, it will continue to be crucial to understand short-GRBs beyond the LIGO/Virgo horizon of $\sim$200\,Mpc (for neutron star-neutron star mergers) and $\sim$800 Mpc (for black hole-neutron star mergers). The kilonova spectra can tell us about the nuclear yields, neutrino flux, mass, velocity, and geometric orientations of the ejecta. Kilonova spectra are the least studied, with only one unambiguous case identified so far, and there is no broad consensus on the expected spectrum (e.g., \cite{bk13}). The theoretical models will remain unconstrained in the absence of spectroscopy, particularly of the kilonova UV emission. For AT~2017gfo, the electromagnetic counterpart to GW 170817, a bright UV/optical flash was detected during the first hours of the evolution of the event \cite{abbott17,evans17,arcavi18}. The flash evolved rapidly, and after a few hours the kilonova emission peaked in the infrared due to absorption by the heavy r-process elements.

The {\em Swift} satellite can detect the short-GRB in $\gamma$-rays and can localise the afterglow in a few seconds after the trigger. Short-GRB afterglows are faint and fade fast. Prompt reaction to a GRB alert with CUBES will maximise chances of obtaining a redshift from the transient and/or host and detection of elements present in its spectrum. The CUBES wavelength range is also crucial to study the early emission from kilonovae and to model the blackbody in the UV range. Multi-epoch observations will allow determination of the effective temperature of the source and the status of the photosphere at days and weeks after the event. CUBES will significantly increase the number of observations and detection of short-GRBs that essentially are at lower redshifts and cannot be captured by the sensitivity of UVES.

\subsubsection{Key requirements}
\begin{itemize}
    \item{{\it Simultaneous observations at longer \lam:} Simultaneous UVES observations are desirable to give clues on the environment of the kilonova and its host galaxy.}
\end{itemize}

\subsection{T3: Superluminous Supernovae}

Superluminous supernovae (SLSNe) are a class of extremely energetic SNe \cite{quimby07,gal-yam12,gal-yam19}, which are at least 10 and often 100 times more luminous than normal SNe, and evolve slowly over weeks/months \cite{decia18,lunnan18,angus19,chen22a,chen22b}. Because of their brightness, they can be observed out to high redshift (to $z\sim4$ so far, \cite{cooke12}). Their explosion mechanisms and progenitor stars are highly debated, and may be associated with very massive stars, potentially interacting with a surrounding medium, or with pair-instability SNe \cite{heger02,gal-yam09} or magnetars \cite{nicholl13,inserra17}. The host galaxies of SLSNe have low metallicities and extreme properties \cite{neill11,vreeswijk14,leloudas15,lunnan15,perley16,chen17,schulze18}. In addition, SLSNe have a potential use for cosmology \cite{inserra13,scovacricchi16} out to much higher $z$ than currently done with SNe Type Ia. Spectra of SLSNe are diverse, and fundamental to assess the properties of these explosions \cite{quimby18,gal-yam19b}. Current time-domain survey facilities such as the Zwicky Transient Facility (ZTF, \cite{graham19}) discover about 30 SLSNe per year, with $z<0.7$ \cite{chen22a,chen22b}. In the future, the Legacy Survey of Space and Time (LSST) at the Vera Rubin Telescope will discover about 1000 SLSNe per year \cite{scovacricchi16}, most of which will be fainter than 20 mag. in the $r$-band and out to higher $z$. 

Although SLSNe are much brighter than normal SNe, and have copious UV emission, they are also observed at typically larger distances, and thus are faint targets for near-UV spectroscopy. Spectral observations of SLSNe with 8-m telescopes are mostly limited to low-spectral-resolution studies, with a few notable exceptions where the SLSN host galaxies could be studied in absorption \cite{vreeswijk14,yan18}. X-Shooter spectra of one of the most luminous H-poor SLSNe, \cite{vreeswijk14} showed that the SLSN host galaxy has extreme properties in its absorption lines: they are very narrow, barely resolved at the X-Shooter resolution, among the narrowest in a larger sample of galaxies observed in absorption with UVES \cite{ledoux06}. In addition, the absorption lines in SLSN host galaxies (also measured at lower resolutions) are extremely week, much weaker than what is typically seen in other starforming galaxies such as GRB hosts. The reason for these extreme properties is not yet understood. Highly efficient observations in the near-UV regime with intermediate-high spectral resolution ($R > 10,000$) will allow us to shed light on this puzzling result. Most metal lines (e.g. from \feii{}, \znii{}, \niii{}, etc.) associated with the ISM of the SLSN host galaxies are in the rest-frame UV, and will be observable in the ground UV region for targets at $z>0.5$. The study of the kinematics and relative abundance of metals, regardless of the detection of the Ly$\alpha$ line, will enable the characterisation of the chemical and dynamical properties of the ISM of the exotic galaxies hosting SLSNe. 

In addition, ISM atoms that are close enough to the SLSN could, in principle, be excited by the SLSN radiation, and potentially show (variable) fine-structure lines of excited levels of \feii{} and other ions. This has been often observed in GRB afterglows (\cite{delia09,vreeswijk07,decia12}), but never in SLSNe. The detection of \feii{} fine-structure lines and their variability would enable the study of the environment surrounding SLSNe, and potentially reveal the nature of their progenitors. The strongest \feii{} fine-structure lines are in the UV (around 230, 240, and 260 nm), and are shifted into the ground UV for targets at $0.15 < z < 0.7$. The high near-UV efficiency of CUBES compared to X-Shooter will make it the ideal instrument to observe the \feii{} fine-structure lines and their variability. 

Finally, the emergence of \mgii{} ($\lambda\lambda$ 279.6, 280.3\,nm) emission was observed about 100 days after a SLSN from low-resolution spectroscopy \cite{lunnan18b}. This was interpreted as a line echo from a shell around the SLSN, showing how such observations can be very powerful to constrain the nature of the SLSN progenitors and the physics of the explosion. Thanks to its high efficiency in the near-UV, CUBES will be ideal to observe other potential \mgii{} echoes from SLSNe at $z < 0.4$, with about twice the efficiency of X-Shooter and at higher spectral resolution. In addition, the spectra of H-poor SLSNe show quite distinctive broad features at rest-frame 350 to 500\,nm, which seem to be due to a superposition of several individual \oii{} narrower components. This may indicate relatively narrow photospheres (in velocity space) that are nevertheless expanding rapidly, giving potentially new insights on the physics of the explosion \cite{gal-yam19,gal-yam19b}. While the velocity width of these narrower components could be of hundreds of \kms, they have yet to be resolved with higher-resolution spectroscopy. The spectral coverage of CUBES will enable a better characterisation of the near-UV spectra of the most nearby SLSNe ($z<0.1$), and further constrain the physics of these energetic and hotly debated explosions.

\subsubsection{T3: Key requirements}

\begin{itemize}
    \item{{\it Spectral coverage:} Essential: 300--405\,nm. Desirable: 300--510\,nm. A wider spectral coverage would enable the measurement of the ISM dust-corrected metallicity from simultaneous measurements of Ly$\alpha$ and metal lines from \feii{} and \znii{}, increase the $z$-range of the potential targets, and increase the variety of metals lines that can be detected.}\smallskip
    \item{{\it Spectral resolution:} Essential: $R>7,000$. Goal: $R > 10,000$. A higher resolution would resolve the narrow lines associated with the ISM of the SLSN host galaxies.}\smallskip
    \item{\it Operations:} Target of Opportunities observations of SLSNe can be non-disruptive (soft ToOs), given the slow evolution of their light curves. The Rapid Response Mode is not essential, but advisable in case the fine-structure lines turn out to vary on short timescales, or for rare discoveries of SLSNe very early after their explosion.  
\end{itemize}

\section{Defining cases for the top-level requirements}
To inform specification of the TLRs for the instrument, and to quantify their tolerances, specific cases were identified where they drive a particular requirement. These are now briefly summarised.

\subsection{Wavelength coverage: Blue end}
Several cases inform the extent of the wavelength coverage at the blue end:
\begin{itemize}
\item{Cometary Science (Case S1): Good coverage required on both sides of the OH line at 308\,nm; coverage down to 305\,nm is essential.}\smallskip
\item{Neutron-capture elements (Case G7): Specific absorption lines demand coverage shortwards of 305\,nm, namely: Bi~\1 at 302.5\,nm and Ge~\1 303.9\,nm. In metal-poor stars with `normal’ (solar-scaled) abundances, we note that Bi~\1 is very hard to detect, such that it should not drive the wavelength coverage requirements by itself. (Although there are some extremely metal-poor stars with an r-process excess, e.g. [Bi/Fe]\,$=$\,$-$1.0 for CS~31082-001 \cite{Bar11}).}\smallskip
\item{Reionization (Case E4): A general argument for this and other extragalactic cases is that by going to shorter wavelengths additional parameter space is opened-up in terms of accessing a broader range of redshifts.}\smallskip 
\item{Kilonovae (Case T2): To some extent we do not know what to expect for these (and some other) transients at the shortest wavelengths. Nevertheless, the opacity of lanthanides in the ejecta of kilonovae increases towards the blue. Maximising the instrument coverage to the atmospheric cut-off would therefore provide the best opportunities for ‘discovery space’.}
\end{itemize}

\subsection{Resolving power: High resolution}
Initial tests were presented of which elements had (potentially) useful diagnostics in stellar studies at $R$\,$=$\,$20,000$ compared to the $R$\,$=$\,$40,000$ from a 1” slit with UVES \cite{ernandes20}. In short, the conclusion was that for the large majority of the diagnostic lines the S/N was a more critical aspect than the resolution. Initial tests were also undertaken for $R$\,$=$\, $10,000$, but 
this resolving power is then too low to allow quantitative measurements of e.g. the Be doublet at 313\,nm. 

Given the technical design constraints, a significantly larger resolving power ($R$\,$>$\,30,000) would require cross-dispersion and blazed gratings, leading to a large reduction in efficiency. As such, and reinforcing analysis from the previous Phase A study (e.g. \cite{smiljanic14}), $R$\,$=$\,20,000 to 25,000 is an optimum balance between sufficient spectral resolution for quantitative analysis and maximum sensitivity. 

The baseline design \cite{ExA_Zanutta} uses 9K CCDs and neatly satisfies the requirement that $R$\,$>$\,20,000, combined with good spectral sampling ($>$2.3 pixels per resolution element) and with broad wavelength coverage of 300-405 nm. 

\subsection{Resolving power: Low resolution}
Some of the cases outlined above require the excellent throughput of CUBES but spectral resolution is less critical. For example, simulations were performed of the sky background at the shortest wavelengths in Case E4 to understand how far we can go to short wavelengths and remain background (rather than detector) limited. From this case (and others, e.g. S1), it was clear that a low-resolution (LR) option would provide a powerful complement to the primary high-resolution (HR) option. The technical team therefore investigated a LR option in the optical design, by exchanging the slicer unit with a second unit with physically larger slices (thus also sampling a larger on-sky area).

A couple of different possibilities were investigated using the end-to-end simulator, with 1$''$ and 0.75$''$ slits. The former of these was background limited with either 1\,$\times$\,1 or 1\,$\times$\,2 binning and was adopted as the LR option in the slicer design, giving  $R$\,$\sim$\,7,000 (comparable to that obtained with a 0.8$''$ slit with the UVB arm of X-Shooter).

\subsection{Position angle}
The cometary science case (Case S1) requires the ability to configure the slit to a position angle chosen by the user, without the complications of atmospheric dispersion in the data, i.e. an ADC is required such that observations can be obtained away from the parallactic angle. This enables observations of different parts of the cometary coma.

Atmospheric dispersion at the CUBES wavelengths quickly becomes significant when moving away from zenith. A removable ADC for observations near to zenith could give a  marginal 0.2-0.3\% gain in end-to-end efficiency (incl. telescope and atmosphere), but would entail additional mechanisms in the instrument, increase the operational overhead for daytime calibration, and complicate the data reduction (in the current design we have a fixed-format science product). As such, we did not pursue this further.

\subsection{Exposure times / duty cycle}
Case G3 was used to define the required duty-cycle of observations. The case is to monitor chromospheric activity in the Ca~\2 H\&K lines (potentially in parallel to ESPRESSO observations, obtained with a separate UT). UVES has been used for this type of monitoring programmes with a cadence of ~1 min (45s exposures with the ultrafast readout mode of the UVES arrays \cite{klocova17}). The goal is to achieve a similar $\sim$1 min cadence with CUBES (potentially with shorter 30s exposures given the better efficiency of CUBES).

In general for CUBES programmes we expect exposure times to be relatively long ($>10$ mins), but some bright targets might require observations of only 1 min or so. For instance, for a V=11 mag G2-type (solar) template at 313 nm (at airmass\,$=$\,1.2, 1$''$ seeing), we find S/N = 32 from the CUBES ETC in 60s, compared to $\sim$0.5 hr to achieve the same S/N with UVES.

In short, targets with V$<$11 mag can be done with reasonable UVES exposure times, but CUBES will still yield substantial efficiency gains in operational time even for relatively bright targets (and accounting for the overhead of the telescope preset etc).

\subsection{Flux-calibration requirements}
A few cases require flux calibration of the science spectra to a precision of 10\% or better (e.g. S1, G1). Approximate flux calibration is also beneficial to other cases (e.g. E2, T1, T2). 

The cometary cases (S1) are among the most demanding, with a requirement of 10\% precision, and a goal of 5\%. From recent experience with UVES observations of cometary targets, use of a spectrophometric standard observed during the night, combined with the instrument master-response curve, has been sufficient to achieve such results (assuming weather conditions remain comparable). This has been used to the shortest wavelengths possible with UVES (and X-Shooter) and we plan to adopt a similar strategy for CUBES.

\subsection{Fibre-link to UVES}
The primary motivation for a fibre-link to UVES is follow-up of transients (Cases T1, T2), although note that Case G1 to study accretion and outflows in YSOs also requires simultaneous observations. For these cases, the data must be taken at exactly the same time for astrophysical reasons, but many of the other cases outlined above would also benefit from simultaneous UVES observations (in the sense of obtaining important complementary or supporting data more efficiently than observing separately with the two instruments).

In general, at least half of the cases outlined above would benefit significantly from the proposed UVES fibre-link, bringing a powerful addition to CUBES, for a relatively small cost and with low risk. 
The fibre-link ultimately provides efficiencies by avoiding separate observations:
\begin{itemize}
\item{{\it Transient/YSO cases:} Simultaneous UVES observations are required for these cases to realise their scientific objectives. In absence of the fibre-link, separate observations would have to be undertaken with CUBES and UVES at the same time (implying that CUBES and UVES would have to be mounted on different UTs). Provision of the fibre-link, and assuming CUBES and UVES are mounted on the same UT, then means that only one UT is used by these programmes.}
\item{{\it Other cases:} Constraints on simultaneity are less critical, but there is a significant efficiency gain via not needing separate observations for many programmes.}
\end{itemize}

Of course, these gains only apply if the performance of UVES with the fibre-link is sufficiently good to deliver the observations required within the time needed for CUBES observations. Assuming the estimated 40\% throughput of the fibre-link at $>$500 nm
\cite{ExA_Zanutta}, calculations for the YSO case demonstrated the overall potential gain of CUBES$+$UVES observations compared to separate observations \cite{ExA_Alcala}.

\section{Comparison with other instruments}
Adopting the efficiency estimates for the end-to-end performance of CUBES (from \cite{ExA_Zanutta}), we now place its sensitivity in the context of existing instruments. In what follows we have assumed a telescope efficiency for one of the UTs of 76\%, which are then convolved by the {\sc skycalc} sky transmission model used in the ESO ETC  (which averages over the year and full nights).

\subsection{UVES}
\begin{figure}
\begin{center}
  \includegraphics[width=8.66cm]{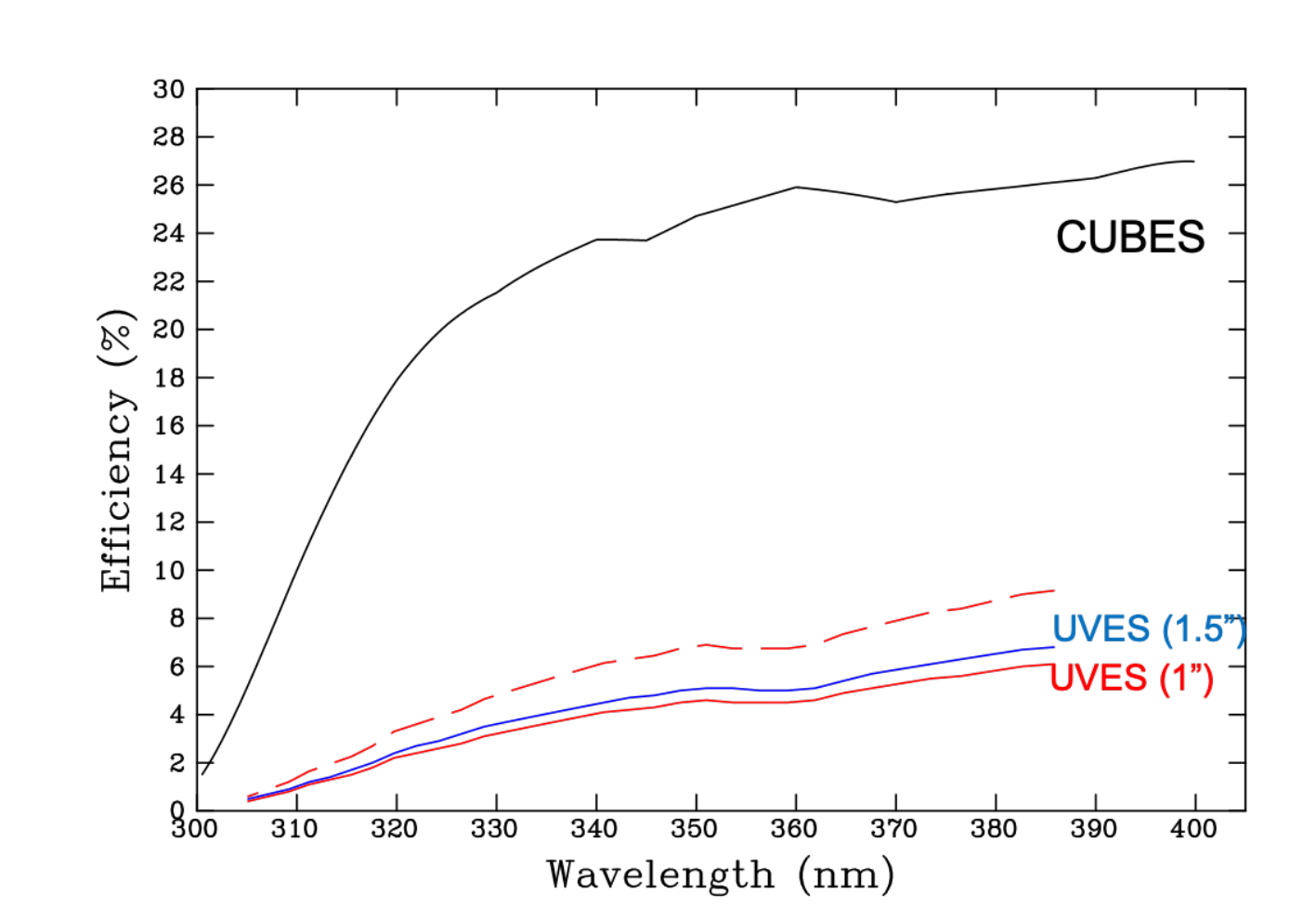}
  \caption{Comparison of predicted CUBES efficiency (including telescope and atmosphere) with those for the central wavelengths of the UVES echelle orders from the ESO ETC. The dashed red line shows the anticipated gain in performance ($\times$1.5) from a possible UVES upgrade.}
\label{cubes_cf_uves}
\medskip
  \includegraphics[width=8.66cm]{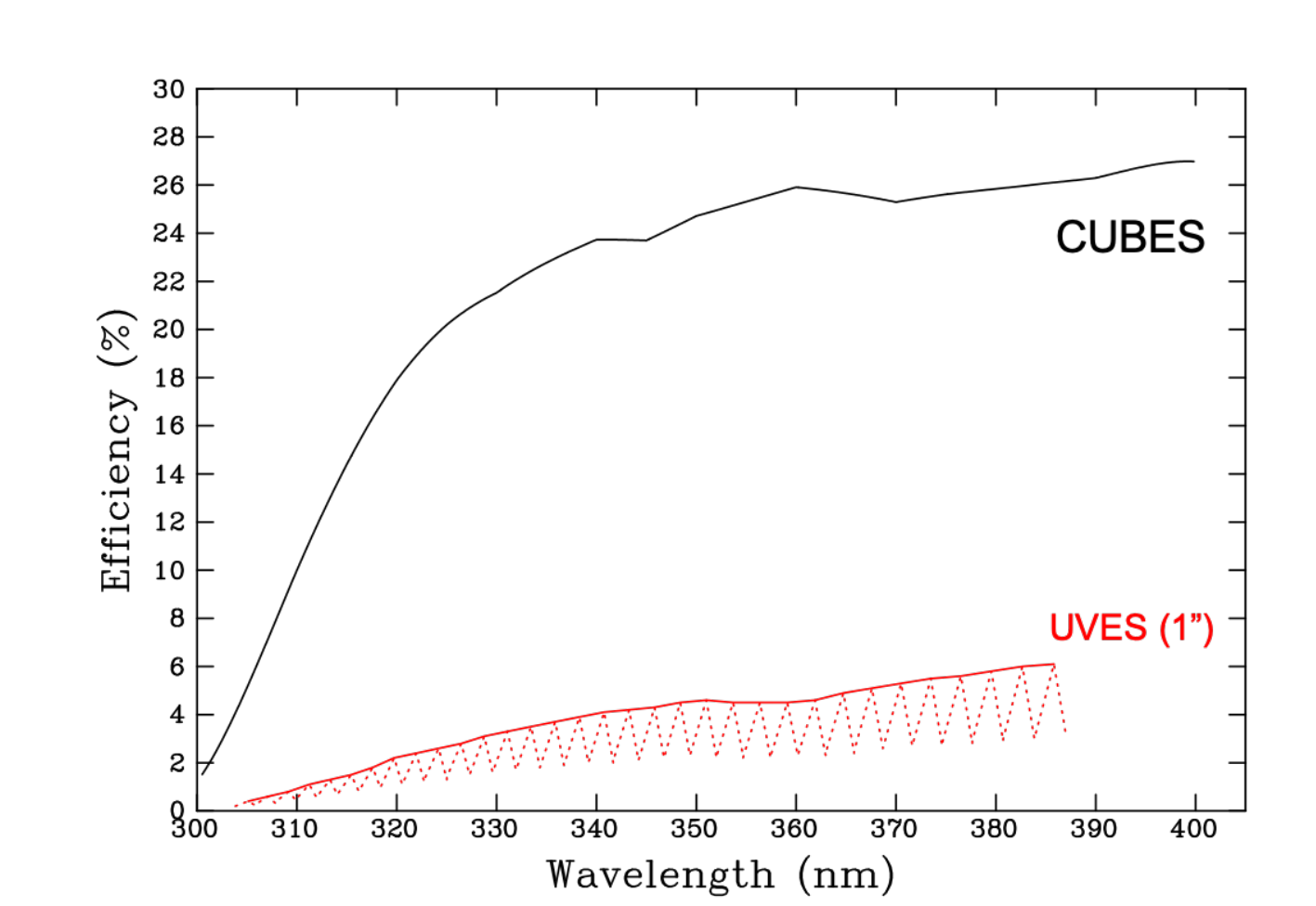}
  \caption{Comparison of predicted CUBES efficiency (including telescope and atmosphere) with those for the central wavelengths of the UVES echelle orders from the ESO ETC. The dashed red line shows variations of efficiency across each of the UVES orders.}
\label{cubes_cf_uves2}
\end{center}
\end{figure}

The end-to-end efficiencies of CUBES and UVES are compared in Fig.~\ref{cubes_cf_uves}, adopting values from the ESO ETC for the blue-arm 346 UVES template. In the figure we have assumed an airmass of 1, seeing of 0.8$''$ and two slit-widths for UVES (as indicated). The dashed red line shows the potential factor of 1.5 gain in performance that is anticipated if UVES were to be upgraded in the future.

Note that the UVES curves plotted in Fig.~\ref{cubes_cf_uves} are the peak efficiencies at the centre of the echelle orders – the mean (or typical) efficiencies away from the blaze peak are lower. To illustrate this point, Fig.~24 shows a similar comparison, but now with the addition of the UVES efficiencies at the min/max of the free spectral range of each echelle order. (In reality, the ‘saw-tooth’ in Fig.~\ref{cubes_cf_uves2} is a consequence of only having three points per order; the true response of each order is much smoother.) The comparison in Fig.~\ref{cubes_cf_uves2} also highlights another advantage of CUBES over higher-resolution (cross-dispersed) data such as those from UVES, as no blaze correction and order merging will be required for the final CUBES data product. 

As shown in the figures, CUBES represents a genuinely transformative difference in the end-to-end performance for high-resolution spectroscopy in the near UV at Paranal. For example, the transmission at 313 nm for CUBES (plus telescope and atmosphere) is ~12.5\%, as compared to 0.6-1.3\% across the corresponding UVES echelle order, i.e. factors of 10-20 times better with CUBES across the order.

\subsection{Other ESO instruments}
\begin{figure}
\begin{center}
  \includegraphics[width=8.66cm]{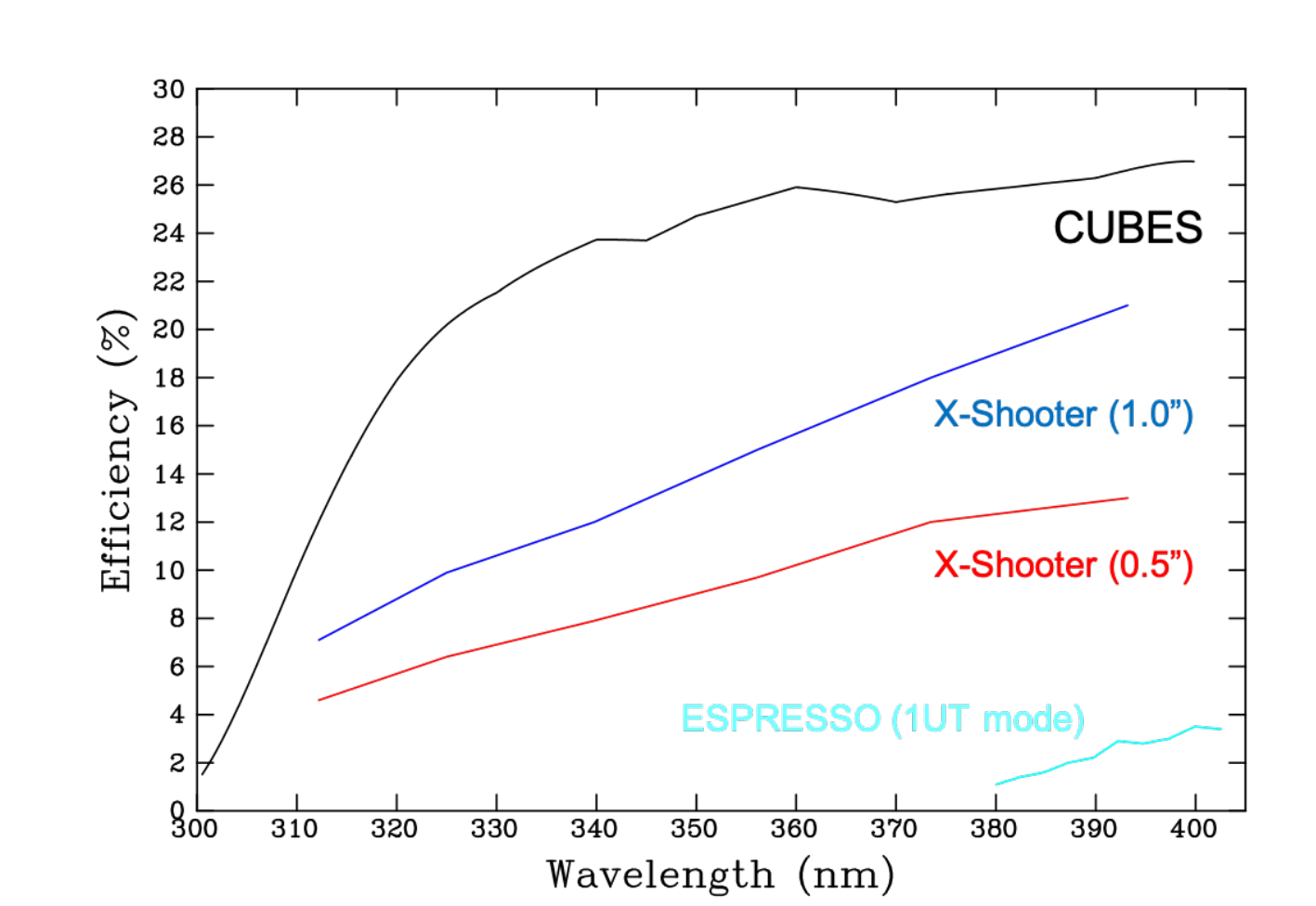}
  \caption{Comparison of predicted CUBES efficiency (including telescope and atmosphere) with those for the central wavelengths of the X-Shooter UVB echelle orders and the shortward orders of the 1~UT mode of ESPRESSO from the ESO ETC.}
\label{cubes_cf_xsh}
\end{center}
\end{figure}

In Fig.~\ref{cubes_cf_xsh} we present a similar comparison of the end-to-end efficiency of CUBES with results for the UVB arm of X-Shooter (again, where the curves are interpolated between the peak efficiencies of the echelle orders). We include results for a 1.0$''$ slit (giving $R$\,$\sim$\,5400) and a 0.5$''$ slit ($R$\,$\sim$\,9700). CUBES is factors of three (0.5$''$ X-Shooter slit) and two (1.0$''$ slit) better over the 320-340\,nm range, while also providing the greater resolving power required for quantitative spectral analysis (e.g. for the Be doublet). For completeness, we include the efficiency curve for the 1~UT mode of ESPRESSO in the overlapping spectral range with CUBES.

We also note possible synergies of CUBES LR observations with the LR-BLUE spectroscopic configuration of the MAVIS instrument now in development for the VLT \cite{mavis}. The LR-BLUE setting will deliver $R$\,$=$\,5,900 spectroscopy over 370-720\,nm (of an AO-corrected source).

Looking further ahead, the baseline spectral range of the high-resolution ANDES spectroscopy for the ELT is 400-1800\,nm, with a goal to extend its coverage shortwards to 350\,nm \cite{hires}. ANDES will provide R\,$\sim$\,100k spectroscopy of individual sources, but the overall system throughput of the ELT (with the current mirror coatings) drops-off steeply shortwards of 400\,nm (see the expected limiting magnitudes for ANDES from \cite{hires}). The broader UV coverage and competitive system efficiency of CUBES will mean that, in spite of the large difference in telescope apertures, it will remain unique well into ELT operations. Indeed, one can envisage strong synergies between CUBES and ANDES (e.g. observations of metal-poor stars).

\subsection{Other facilities}
It is harder to compare efficiencies of non-ESO instruments directly, given different sites (i.e. local atmospheric transmission) and the sparse data available for some instruments. We first consider two of the high-resolution spectrographs on Mauna Kea (Subaru-HDS, Keck-HIRES), then two of the other blue-sensitive Keck instruments (LRIS, KCWI):

\subsubsection{Subaru-HDS} 
The efficiencies for the blue set-up of the Subaru HDS instrument, including the effects of both telescope and atmosphere, range from $\sim$3 to 7\% across the 310-400\,nm range. However, the instrument provides $R$\,$\sim$\,100k so there is significant dispersion. We used the HDS ETC\footnote{https://subarutelescope.org/cgi-bin/hds$\_$etc.cgi}, considering a G0~V (Pickles) spectral template, for $V$\,$=$\,17 mag., seeing\,$=$\,0.8$''$, in a 1\,hr exposure using the ‘Ub’ standard set-up, which delivers $R$\,$=$\,90,000 over 298-458\,nm. At this high dispersion the S/N at 313\,nm (with 4\,$\times$\,4 pixel binning) is $\sim$1.5 so, even with significant further binning to the resolution of CUBES, the gains will be relatively limited. In contrast, for the same G0~V Pickles template with $V$\,$=$\,17\,mag. the CUBES ETC predicts S/N $\sim$15 in 1\,hr.

\subsubsection{Keck-HIRES} 
The HIRES instrument at Keck has been immensely productive over the years, and is well placed to take advantage of higher site at Mauna Kea and larger collecting area of Keck compared to the VLT. We used the HIRES ETC\footnote{http://etc.ucolick.org/web$\_$s2n/hires} to estimate its performance compared to CUBES, considering a $V$\,$=$\,18\,mag. flat input spectrum at airmass\,$=$\,1 and  seeing\,$=$\,0.8$''$ at 313\,nm in a 1\,hr exposure:
\begin{itemize}
    \item{HIRES: slit-width\,$=$\,1.15$''$ ($R$\,$=$\,37k): S/N\,$=$\,2.3/pix.}
    \item{HIRES: slit-width\,$=$\,1.70$''$ ($R$\,$=$\,24k): S/N\,$=$\,2.7/pix.}
    \item{CUBES HR: S/N\,$=$\,31/pix.}
\end{itemize}
Similar calculations for 302.5\,nm (near the Bi~\1 line) yielded S/N\,$=$\,11/pix for CUBES, compared with S/N\,$=$ 0.2 and 0.3/pix for 1.15$''$ and 1.7$''$ slits (respectively) with HIRES. A similar gain is also obtained at longer wavelengths with e.g. S/N\,$=$\,40/pix for CUBES compared to S/N\,$\sim$\,4/pix with HIRES at 340\,nm.

\medskip
Compared with existing high-resolution instruments on Mauna Kea, our above estimates demonstrate the significant performance gain that CUBES will provide.

\subsubsection{Keck-LRIS:} 

The publicly-available LRIS ETC does not include the two configurations that are most relevant to be compared to the CUBES LR option, namely the 1200 l/mm grating (blazed at 340\,nm) and the 600 l/mm grating (blazed at 400\,nm). With a 0.7$''$ slit the 1200 grating provides R\,$\sim$\,3,600, and the total system throughput (excluding atmosphere) is reported as $\sim$25\% at 350\,nm \cite{steidel04}. The end-to-end efficiency of CUBES at 350\,nm is $\sim$25\% (see Fig.~\ref{cubes_cf_uves}), but this also includes the effect of the atmosphere. Taking this into account, and considering that the Keck telescopes have a larger collecting area than the UTs, the performance of LRIS and CUBES are expected to be roughly comparable, but with CUBES LR delivering greater resolving power.

\subsubsection{Keck-KCWI:} 
KCWI was designed as a blue-optimised, integral-field spectrograph, with a range of resolving powers (from $R$\,$=$\,1250 up to 20,000) over the 350-560\,nm range \cite{kcwi}. As with LRIS, it is hard to directly compare the on-sky performance with CUBES. From comparison of the published KCWI efficiency curves \cite{kcwi}, while taking into account the larger collecting area of Keck (cf. the VLT), CUBES will be very competitive with KCWI at 350-400\,nm, with the important addition that only CUBES covers the 300-350\,nm range.

\section{Summary}

We have presented the scientific cases that were developed to motivate
the design and construction of the new CUBES instrument. This broad
range of cases illustrates the strong demand for a new, optimised
near-UV spectroscopic capability at ESO's VLT, that will provide a
unique facility for the community for many years to come.  

The Phase~A design of CUBES \cite{ExA_Zanutta} provides unprecedented
spectroscopic sensitivity at ground UV (300-405\,nm) wavelengths.
This will enable observation of much fainter sources than currently
possible, opening-up new opportunities and parameter space across
Solar System, Galactic and extragalactic targets, including rapid
follow-up of several classes of explosive transients.
The cases introduced here included performance estimates calculated
with simulation tools developed during the Phase~A study
\cite{ExA_Genoni}; for more detailed examples of the anticipated
observational samples, we refer to the expanded papers on specific
cases as indicated in Table~\ref{sci_summary}.

To bolster its observational capabilities, the CUBES design
includes exchangable imager slicers \cite{ExA_Calcines} to provide two spectral resolving powers: $R$\,$\sim$\,24,000 for high-quality quantitative spectroscopy, and $R$\,$\sim$\,7,000 to maximise the efficiency for observations of
faint targets.  The design also includes a fibre-link to UVES to give
the powerful capability of simultaneous, high-resolution
($R$\,$\sim$ 40,000) observations at longer wavelengths. This will
provide critical information for studies of transients and YSOs, while
delivering significant operational savings for many cases where
longer-wavelength observations are also required.

\begin{acknowledgements}
A.~R.~da Silva, R.~E.~Giribaldi and R.~Smiljanic acknowledge support by the National Science Centre, Poland, under project 2018/31/B/ST9/01469.
\end{acknowledgements}

%
\section*{Conflict of interest}
The authors declare that they have no conflict of interest.

\section*{Data availability statement}
Data sharing is not applicable to this article as no datasets were generated or analysed during the current study.


\newpage
\begin{thebibliography}{}

\bibitem{bar14}
Barbuy, B., Bawden Macanhan, V., Bristow, P. et al.: CUBES: Cassegrain U-Band Brasil-ESO Spectrograph,  Ap\&SS, 354, 191 (2014)

\bibitem{ExA_Zanutta}
Zanutta, A., Cristiani, S., Atkinson, D. et al.:
CUBES Phase A design overview, ExA, arXiv:2203.15352, in press as part of the Special Issue

\bibitem{ExA_Opitom} Opitom, C., Snodgrass, C., La Forgia, F. et al.: Cometary Science with CUBES, ExA, arXiv:2203.15579, in press as part of the Special Issue

\bibitem{ExA_Alcala} Alcal\'{a}, J. M., Cupani, G., Evans, C. et al.: Accretion and outflows in young stars with CUBES, ExA, arXiv:2203.15581, in press as part of the Special Issue

\bibitem{ExA_Smiljanic} Smiljanic, R., da Silva, A. R. \& Giribaldi, R. E.:
Detecting weak beryllium lines with CUBES, ExA, arXiv:2203.16158, in press as part of the Special Issue

\bibitem{ExA_Giribaldi} Giribaldi, R. E. \& Smiljanic, R.: Beryllium abundances in turn-off stars of globular clusters with the CUBES spectrograph, ExA, arXiv:2203.15604, in press as part of the Special Issue

\bibitem{ExA_Bonifacio} Bonifacio, P.: Metal-poor stars: The role of cubes,
ExA, in press as part of the Special Issue

\bibitem{ExA_Hansen} Hansen, C. J.: Heavy elements -- they came out of the blue, ExA, in press as part of the Special Issue

\bibitem{ExA_Ernandes} Ernandes, H., Barbuy, B., Castilho, B. et al.:
Simulated observations of heavy elements with CUBES, ExA, arXiv:2203.15693, 
in press as part of the Special Issue

\bibitem{ExA_Evans} Evans, C., Marcolino, W., Bouret, J.-C. \& Garcia, M.:
A near-UV reconnaissance of metal-poor massive stars, ExA, in press
as part of the Special Issue

\bibitem{ExA_DOdorico} D'Odorico, V.: Portraying the missing baryonic mass at the cosmic noon: the contribution of CUBES, ExA, in press as part of the Special Issue

\bibitem{ExA_Balashev} Balashev, S. \& Noterdaeme, P.:
"Molecular hydrogen in absorption at high redshifts -- Science cases for CUBES", ExA, in press as part of the Special Issue

\bibitem{snodgrass17a}
Snodgrass, C., Agarwal, J., Combi, M. et al.: The Main Belt Comets and ice in the Solar System, A\&ARv, 25, 5 (2017)

\bibitem{kramer17}
Kramer, T., L\"{a}uter, M., Rubin, M. \& Altwegg, K.: Seasonal changes of the volatile density in the coma and on the surface of comet 67P/Churyumov-Gerasimenko, MNRAS, 469, S20

\bibitem{snodgrass17b}
Snodgrass, C., Yang, B. \& Fitzsimmons, A.: X-shooter search for outgassing from main belt comet P/2012 T1 (Pan-STARRS), A\&A, 605, A56 (2017)

\bibitem{obrien18}
O'Brien, D. P., Izidoro, A., Jacobson, S. A., Raymond, S. N. \& Rubie, D. C.:
The Delivery of Water During Terrestrial Planet Formation, SSRv, 214, 47 (2018)

\bibitem{snodgrass19}
Snodgrass, C. \& Jones, G. H.: The European Space Agency's Comet Interceptor lies in wait, Nat. Comms., 10, 5418 (2019)




\bibitem{lane81}
Lane, A. L., Nelson, R. M. \& Matson, D. L.:
Evidence for sulphur implantation in Europa's UV absorption band, Nature, 292, 38 (1981)




\bibitem{jm70}
Johnson, T. V. \& McCord, T. B.: Galilean Satellites. The Spectral Reflectivity 0. 30-1.10 Micron, Icar, 13, 37 (1970)

\bibitem{wamsteker72} 
Wamsteker, W.: Narrow-band photometry of the Galilean satellites. Comm. Lunar Planet. Lab. 9 (No. 167), 171 (1972)

\bibitem{mcfadden80}
McFadden, L. A., Bell, J. F. \& McCord, T. B.:
Visible spectral reflectance measurements (0.33-1.1\,$\mu$m) of the Galilean satellites at many orbital phase angles, Icar, 44, 410 (1980)

\bibitem{nelson87}
Nelson, R. M., Lane, A. L., Matson, D. L. et al.L
Spectral geometric albedos of the Galilean satellites from 0.24 to 0.34 micrometers: Observations with the international ultraviolet explorer, Icar, 72, 358 (1987)

\bibitem{noll95}
Noll, K. S., Weaver, H. A. \& Gonnella, A. M.:
The albedo spectrum of Europa from 2200\,\AA\ to 3300\,\AA, JGR, 100, 19057 (1995)

\bibitem{spencer95}
Spencer, J. R., Calvin, W. M. \& Person, M. J.:
CCD Spectra of the Galilean Satellites: Molecular Oxygen on Ganymede, JGR, 100, 19049 (1995)





\bibitem{hendrix08}
Hendrix, A. R. \& Johnson, R. E.:
Callisto: New Insights from Galileo Disk-resolved UV Measurements, ApJ, 687, 706 (2008)







\bibitem{hendrix16}
Hendrix, A. R., Vilas, F. \& Li, J.-Y.: The UV signature of carbon in the solar system, M\&PS, 51, 105 (2016)

\bibitem{mcgrath04}
McGrath, M. A., Jia, X., Retherford, K. et al.: Aurora on Ganymede, JGR, 118, 2043 (2004)







\bibitem{bk17}
Bida, T. \& Killen, R. M.: Observations of the minor species Al and Fe in Mercury's exosphere, Icar, 289, 227 (2017)






\bibitem{giannini15}
Giannini, T., Antoniucci, S., Nisini, B., Bacciotti, F. \& Podio, L.:
Solving the Excitation and Chemical Abundances in Shocks: The Case of HH 1, ApJ, 814, 52 (2015)

\bibitem{alcala17}
Alcal\'{a}, J. M., Manara, C. F., Natta, A. et al.:
X-shooter spectroscopy of young stellar objects in Lupus. Accretion properties of class II and transitional objects, A\&A, 600, A20 (2017)

\bibitem{manara17}
Manara, C. F., Testi, L., Herczeg, G. J. et al.: X-shooter study of accretion in Chamaeleon I. II. A steeper increase of accretion with stellar mass for very low-mass stars? A\&A, 604, A127
(2017)

\bibitem{nisini18}
Nisini, B., Antoniucci, S., Alcal\'{a}, J. M. et al.: Connection between jets, winds and accretion in T Tauri stars. The X-shooter view, A\&A, 609, A87 (2018)

\bibitem{alencar12}
Alencar, S. H. P., Bouvier, J., Walter, F. M. et al.:
Accretion dynamics in the classical T Tauri star V2129 Ophiuchi, A\&A, 541, A116 (2012)

\bibitem{alcala19}
Alcal\'{a}, J. M., Manara, C. F., France, K., et al.:
HST spectra reveal accretion in MY Lupi, A\&A, 629, A108 (2019)

\bibitem{fang18}
Fang, M., Pascucci, I., Edwards, S. et al.: A New Look at T Tauri Star Forbidden Lines: MHD-driven Winds from the Inner Disk, ApJ, 868, 28 (2018)

\bibitem{liu14}
Liu, C.-F., Shang, H., Walter, F. M. \& Herczeg, G. J.: Velocity-resolved [Ne~{\scriptsize III}] from X-Ray Irradiated Sz~102 Microjets, ApJ, 786, 99 (2014)

\bibitem{rs10}
Rogers, L. A. \& Seager, S.: A Framework for Quantifying the Degeneracies of Exoplanet Interior Compositions, ApJ, 712, 974 (2010)

\bibitem{z07}
Zuckerman, B., Koester, D., Melis, C., Hansen, B. M., Jura, M.:
The Chemical Composition of an Extrasolar Minor Planet, ApJ, 671, 872 (2007)

\bibitem{jura03}
Jura, M.: A Tidally Disrupted Asteroid around the White Dwarf G29-38, ApJ, 594, L91 (2003)

\bibitem{klein10}
Klein, B., Jura, M., Koester, D., Zuckerman, B. \& Melis, C.:
Chemical Abundances in the Externally Polluted White Dwarf GD 40: Evidence of a Rocky Extrasolar Minor Planet, ApJ, 709, 950 (2010)

\bibitem{gf19}
Gentile Fusillo, N. P, Tremblay, P.-E., G\"{a}nsicke, B. T. et al.:
A Gaia Data Release 2 catalogue of white dwarfs and a comparison with SDSS, MNRAS, 482, 4570



\bibitem{tsd15}
Testa, P., Saar, S. H. \& Drake, J. J.
Stellar activity and coronal heating: an overview of recent results, RSPTA, 373, 40259 (2015)


\bibitem{klocova17}
Klocov\'{a}, T., Czesla, S., Khalafinejad, S., Wolter, U. \& Schmitt, J. H. M. M.:
Time-resolved UVES observations of a stellar flare on the planet host HD 189733 during primary transit, 
A\&A, 607, A66 (2017)

\bibitem{kowalski15}
Kowalski, A. F., Cauzzi, G. \& Fletcher, L.:
Optical Spectral Observations of a Flickering White-light Kernel in a C1 Solar Flare, ApJ, 798, 107 (2015)



\bibitem{k13}
Kowalski, A. F., Hawley, S. L., Wisniewski, J. P. et al.:
Time-resolved Properties and Global Trends in dMe Flares from Simultaneous Photometry and Spectra, ApJS, 207, 15 (2013)

\bibitem{yu17}
Yu, L., Donati, J.-F., H\'{e}brard, E. M. et al.: A hot Jupiter around the very active weak-line T Tauri star TAP 26, MNRAS, 467, 1342 (2017)

\bibitem{donati97}
Donati, J.-F., Semel, M., Carter, B. D. et al. Spectropolarimetric observations of active stars, MNRAS, 291, 658 (1997)

\bibitem{donati16}
Donati, J. F., Moutou,C., Malo, L. et al.:
A hot Jupiter orbiting a 2-million-year-old solar-mass T Tauri star, 
Nature, 534, 662 (2016)



\bibitem{spite19}
Spite, M., Bonifacio, P., Spite, F. et al.:
Be and O in the ultra metal-poor dwarf 2MASS J18082002-5104378: the Be-O correlation, A\&A, 624, A44 (2019)

\bibitem{smiljanic14}
Smiljanic, R.: Stellar abundances of beryllium and CUBES, Ap\&SS, 354, 55 (2014)

\bibitem{reeves70}
Reeves, H., Fowler, W. A. \& Hoyle, F.: Galactic Cosmic Ray Origin of Li, Be and B in Stars, Nature, 226, 727 (1970)

\bibitem{beers00}
Beers, T. C., Chiba, M., Yoshii, Y. et al.: Kinematics of Metal-poor Stars in the Galaxy. II. Proper Motions for a Large Nonkinematically Selected Sample, AJ, 119, 2866 (2000)

\bibitem{sy01}
Suzuki, T. K. \& Yoshii, Y.: A New Model for the Evolution of Light Elements in an Inhomogeneous Galactic Halo, ApJ, 549, 303 (2001)

\bibitem{pasquini05}
Pasquini, L., Bonifacio, P., Molaro, P., et al.: Li in NGC 6752 and the formation of globular clusters, A\&A, 441, 549 (2005)

\bibitem{pasquini07}
Pasquini, L., Bonifacio, P., Randich, S. et al.: Beryllium abundance in turn-off stars of NGC 6752, A\&A, 464, 601 (2007)

\bibitem{smiljanic09}
Smiljanic, R., Pasquini, L., Bonifacio, P. et al.: Beryllium abundances and star formation in the halo and in the thick disk, A\&A, 499, 103 (2009)

\bibitem{molaro20}
Molaro, P., Cescutti, G., Fu, X.: Lithium and beryllium in the Gaia-Enceladus galaxy, MNRAS, 496, 2902 (2020)

\bibitem{smiljanic21}
Smiljanic, R., Zych, M. G. \& Pasquini, L.:
Inhomogeneity in the early Galactic chemical enrichment exposed by beryllium abundances in extremely metal-poor stars, A\&A, 646, A70 (2021)

\bibitem{smiljanic10}
Smiljanic, R., Pasquini, L., Charbonnel, C., Lagarde, N.: Beryllium abundances along the evolutionary sequence of the open cluster IC 4651 - A new test for hydrodynamical stellar models, A\&A, 510, A50 (2010)

\bibitem{smiljanic11}
Smiljanic, R., Randich, S., Pasquini, L.: Mixing at young ages: beryllium abundances in cool main-sequence stars of the open clusters IC 2391 and IC 2602, A\&A, 535, A75 (2011)

\bibitem{boesgaard20}
Boesgaard, A.~M., Lum, M.~G., Deliyannis, C.~P.:  Correlated Depletion and Dilution of Lithium and Beryllium Revealed by Subgiants in M67, ApJ, 888, 28 (2020)

\bibitem{boesgaard22}
Boesgaard, A.~M., Lum, M.~G., Chontos, A., Deliyannis, C.~P.: Lithium and Beryllium in NGC 752 - an Open Cluster Twice the Age of the Hyades, ApJ, 927, 118 (2022)  

\bibitem{boesgaard99}
Boesgaard, A. M., Deliyannis, C. P., King, J. R. et al.: 
Beryllium Abundances in Halo Stars from Keck/HIRES Observations, AJ, 117, 1549 (1999)

\bibitem{primas00a}
Primas, F., Molaro, P., Bonifacio, P. \& Hill, V.:
First UVES observations of beryllium in very metal-poor stars, A\&A, 362, 666 (2000)

\bibitem{primas00b}
Primas, F., Asplud, M., Nissen, P. E. \& Hill, V.:
The beryllium abundance in the very metal-poor halo star G 64-12 from VLT/UVES observations, A\&A, 364, L42 (2000)

\bibitem{boesgaard11}
Boesgaard, A. M., Rich, J. A., Levesque, E. M. \& Bowler, B. P.:
Beryllium and Alpha-element Abundances in a Large Sample of Metal-poor Stars, ApJ, 743, 140 (2011)

\bibitem{pristine}
Starkenburg, E., Martin, N., Youakim, K. et al.:
The Pristine survey - I. Mining the Galaxy for the most metal-poor stars, MNRAS, 471, 2587 (2017)

\bibitem{skymapper}
Wolf, C., Onken, C. A., Luvaul, L. C. et al.:
SkyMapper Southern Survey: First Data Release (DR1), PASA, 35, 10 (2018)

\bibitem{gratton12}
Gratton, R. G., Carretta, E. \& Bragaglia, A.:
Multiple populations in globular clusters. Lessons learned from the Milky Way globular clusters, A\&ARv, 20, 50 (2012)

\bibitem{ymc08}
Yong, D., Mel\'{e}ndez, J., Cunha, K. et al.: Chemical Abundances in Giant Stars of the Tidally Disrupted Globular Cluster NGC 6712 from High-Resolution Infrared Spectroscopy, ApJ, 689, 1020 (2008)

\bibitem{gcs19}
Guer\c{c}o, R., Cunha, K, Smith, V. et al.: Fluorine Abundances in the Globular Cluster M4, ApJ, 876, 43 (2019)

\bibitem{tajitsu15}
Tajitsu, A., Sadakane, K., Naito, H., Arai, A. \& Aoki, W.:
Explosive lithium production in the classical nova V339 Del (Nova Delphini 2013), Nature, 518, 381 (2015)

\bibitem{ExA_Izzo}
Izzo, L. , Molaro, P., Bonifcaio, P. et al.: Classical Novae with CUBES, ExA, in press as part of the CUBES Special Issue

\bibitem{bc05}
Beers, T. C. \& Christlieb, N.: The Discovery and Analysis of Very Metal-Poor Stars in the Galaxy, 
ARA\&A, 43, 531 (2005)

\bibitem{plez12}
Plez, B.: Turbospectrum: Code for spectral synthesis, ASCL:1205.004, (2012)

\bibitem{cayrel01}
Cayrel, R., Hill, V., Beers, T. C. et al.: Measurement of stellar age from uranium decay, Nature, 409, 691 (2001)

\bibitem{hill02}
Hill, V., Plez, B., Cayrel, R. et al.: First stars. I. The extreme r-element rich, iron-poor halo giant CS~31082-001. Implications for the r-process site(s) and radioactive cosmochronology, A\&A, 387, 560 (2002)

\bibitem{chiappini13}
Chiappini, C.: First stars and reionization: Spinstars, AN, 334, 595 (2013)

\bibitem{choplin16}
Choplin, A., Maeder, A., Meynet, G. \& Chiappini, C.:
Constraints on CEMP-no progenitors from nuclear astrophysics, A\&A, 593, A36 (2016)

\bibitem{un02}
Umeda, H. \& Nomoto, K.: Nucleosynthesis of Zinc and Iron Peak Elements in Population III Type II Supernovae: Comparison with Abundances of Very Metal Poor Halo Stars, ApJ, 565, 385 (2002)

\bibitem{un05}
Umeda, H. \& Nomoto, K.: Variations in the Abundance Pattern of Extremely Metal-Poor Stars and Nucleosynthesis in Population III Supernovae, ApJ, 619, 427 (2005)

\bibitem{t14}
Tominaga, N., Iwamoto, N. \& Nomoto, K.: Abundance Profiling of Extremely Metal-poor Stars and Supernova Properties in the Early Universe, ApJ, 785, 98 (2014)

\bibitem{abate15}
Abate, C., Pols, O., Izzard, R. G. \& Karakas, A. I.:
Modelling the observed properties of carbon-enhanced metal-poor stars using binary population synthesis, 
A\&A, 581, A62

\bibitem{placco15}
Placco, V., Frebel, A., Lee, Y. S. et al.:
Metal-poor Stars Observed with the Magellan Telescope. III. New Extremely and Ultra Metal-poor Stars from SDSS/SEGUE and Insights on the Formation of Ultra Metal-poor Stars, ApJ, 809, 136 (2015)

\bibitem{hansen15}
Hansen, T. T., Andersen, J., Nordstr\"{o}m, B. et al.:
The role of binaries in the enrichment of the early Galactic halo. I. r-process-enhanced metal-poor stars, 
A\&A, 583, A49 (2015)

\bibitem{hansen19}
Hansen, C. J., Hansen, T. T., Koch, A. et al.:
Abundances and kinematics of carbon-enhanced metal-poor stars in the Galactic halo. A new classification scheme based on Sr and Ba, A\&A, 623, A128 (2019)

\bibitem{lodders03}
Lodders, K.: Solar System Abundances and Condensation Temperatures of the Elements, ApJ, 591, 1220 (2003)

\bibitem{busso99}
Busso, M., Gallino, R. \& Wasserburg, G. J.: Nucleosynthesis in Asymptotic Giant Branch Stars: Relevance for Galactic Enrichment and Solar System Formation, ARA\&A, 37, 239 (1999)

\bibitem{p10}
Pignatari, M., Gallino, R., Heil, M. et al.: The Weak s-Process in Massive Stars and its Dependence on the Neutron Capture Cross Sections, ApJ, 710, 1557 (2010)

\bibitem{pian17}
Pian, E., D'Avanzo, P., Benetti, S. et al.:
Spectroscopic identification of r-process nucleosynthesis in a double neutron-star merger, Nature, 551, 67 (2017)

\bibitem{smartt17}
Smartt, S. J., Chen, T.-W., Jerkstrand, A. et al.: A kilonova as the electromagnetic counterpart to a gravitational-wave source, Nature, 551, 75 (2017)

\bibitem{watson19}
Watson, D., Hansen, C. J., Selsing, J. et al.:
Identification of strontium in the merger of two neutron stars, Nature, 574, 497 (2019)

\bibitem{bovard17}
Bovard, L., Martin, D., Guercilena, F., Arcones, A. et al.:
r-process nucleosynthesis from matter ejected in binary neutron star mergers
PhRvD, 96, 124005 (2017)

\bibitem{matteucci14}
Matteucci, F., Romano, D., Arcones, A., Korobkin, O. \& Rosswog, S.:
Europium production: neutron star mergers versus core-collapse supernovae
MNRAS, 438, 2177 (2014)

\bibitem{cescutti15}
Cescutti, G., Romano, D., Matteucci, F., Chiappini, C. \& Hirschi, R.:
The role of neutron star mergers in the chemical evolution of the Galactic halo, 
A\&A, 577, A139 (2015)

\bibitem{mc90}
Mathews, G. J. \& Cowan, J. J.: New insights into the astrophysical r-process, Nature, 345, 491 (1990)

\bibitem{iw99}
Ishimaru, Y. \& Wanajo, S.: Enrichment of the R-Process Element Europium in the Galactic Halo, 
ApJ, 511, L33 (1999)

\bibitem{arcones07}
Arcones, A., Janka, J.-Th. \& Scheck, L.:
Nucleosynthesis-relevant conditions in neutrino-driven supernova outflows. I. Spherically symmetric hydrodynamic simulations, A\&A, 467, 1227 (2007)

\bibitem{winteler12}
Winteler, C., K\"{a}ppeli, R., Perego, A. et al.:
Magnetorotationally Driven Supernovae as the Origin of Early Galaxy r-process Elements? ApJ, 750, L22 (2012)

\bibitem{ntt15}
Nishimura, N., Takiwaki, T. \& Thielemann, F.-K.:
The r-process Nucleosynthesis in the Various Jet-like Explosions of Magnetorotational Core-collapse Supernovae, ApJ, 810, 109 (2015)

\bibitem{r18}
Roederer, I. U., Sakari, C. M., Placco, V. M. et al.:
The R-Process Alliance: A Comprehensive Abundance Analysis of HD~222925, a Metal-poor Star with an Extreme R-process Enhancement of [Eu/H]\,$=$\,$-$0.14, ApJ, 865, 129 (2018)

\bibitem{ernandes20}
Ernandes, H., Evans, C. J., Barbuy, B. et al.:
Stellar astrophysics in the near-UV with VLT-CUBES, Proc. SPIE, 11447, E60 (2020)

\bibitem{swt20}
Sandford, N. R., Weisz, D. R. \& Ting, Y.-S.:
Forecasting Chemical Abundance Precision for Extragalactic Stellar Archaeology, ApJS, 249, 24 (2020)

\bibitem{riess19}
Riess, A., Casertano, S., Yuan, W., Macri, L. \& Scolnic, D.:
Large Magellanic Cloud Cepheid Standards Provide a 1\% Foundation for the Determination of the Hubble Constant and Stronger Evidence for Physics beyond $\Lambda$CDM, ApJ, 876, 85 (2019)

\bibitem{verde19}
Verde, L., Treu, T. \& Riess, A. G.: Tensions between the early and late Universe, NatAs, 3, 891 (2019)

\bibitem{marconi05}
Marconi, M., Musella, I. \& Fiorentino, G.:
Cepheid Pulsation Models at Varying Metallicity and $\Delta$Y/$\Delta$Z, 
ApJ, 632, 590 (2005)

\bibitem{marconi10}
Marconi, M., Musella, I., Fiorentino, G. et al.:
Pulsation Models for Ultra-low (Z = 0.0004) Metallicity Classical Cepheids, ApJ, 713, 615 (2010)

\bibitem{ripepi12}
Ripepi, V., Moretti, M. I., Marconi, M. et al.:
The VMC survey - V. First results for classical Cepheids, MNRAS, 424, 1807 (2012)

\bibitem{fiorentino13}
Fiorentino, G., Musella, I. \& Marconi, M.:
Cepheid theoretical models and observations in HST/WFC3 filters: the effect on the Hubble constant H$_0$, MNRAS, 434 2866 (2013)

\bibitem{ripepi17}
Ripepi, V., Cioni, M.-R., Moretti, M. I. et al.:
The VMC survey - XXV. The 3D structure of the Small Magellanic Cloud from Classical Cepheids, 
MNRAS, 472, 808 (2017)

\bibitem{ripepi19}
Ripepi, V., Molinaro, R., Musella, I. et al.:
Reclassification of Cepheids in the Gaia Data Release 2. Period-luminosity and period-Wesenheit relations in the Gaia passbands, A\&A, 625, A14 (2019)

\bibitem{romaniello08}
Romaniello, M., Primas, F., Mottini, M. et al.:
The influence of chemical composition on the properties of Cepheid stars. II. The iron content, 
A\&A, 488, 731 (2008)

\bibitem{bono10}
Bono, G., Caputo, F., Marconi, M. \& Musella, I.:
Insights into the Cepheid Distance Scale, ApJ, 715, 277 (2010)

\bibitem{bono01}
Bono, G., Caputo, F., Castellani, V., Marconi, M. \& Storm, J.:
Theoretical insights into the RR Lyrae K-band period-luminosity relation, MNRAS, 326, 1183

\bibitem{bono03}
Bono, G., Caputo, F., Castellani, V. et al.:
A pulsational approach to near-infrared and visual magnitudes of RR Lyr stars, MNRAS, 344, 1097 (2003)

\bibitem{sollima06}
Sollima, A., Cacciari, C. \& Valenti, E.:
The RR Lyrae period-K-luminosity relation for globular clusters: an observational approach, MNRAS, 372, 1675 (2006)

\bibitem{marconi15}
Marconi, M., Coppola, G., Bono, G. et al.:
On a New Theoretical Framework for RR Lyrae Stars. I. The Metallicity Dependence, ApJ, 808, 50 (2015)

\bibitem{piotto07}
Piotto, G., Bedin, L. R., Anderson, J. et al.: A Triple Main Sequence in the Globular Cluster NGC 2808, ApJ, 661, L53 (2007)

\bibitem{bragaglia10}
Bragaglia, A., Carretta, E., Gratton, R. G. et al.:
X-shooter Observations of Main-sequence Stars in the Globular Cluster NGC 2808: First Chemical Tagging of a He-normal and a He-rich Dwarf, ApJ, 720, L41 (2010)

\bibitem{dercole08}
D'Ercole, A., Vesperini, E., D'Antona, F., McMilan, S. L. W. \& Recchi, S.:
Formation and dynamical evolution of multiple stellar generations in globular clusters, MNRAS, 391, 825 (2008)

\bibitem{decressin07}
Decressin, T., Charbonnel, C. \& Meynet, G.:
Origin of the abundance patterns in Galactic globular clusters: constraints on dynamical and chemical properties of globular clusters, A\&A, 475, 859 (2007)

\bibitem{dh14}
Denissenkov, P. A. \& Hartwick, F. D. A.:
Supermassive stars as a source of abundance anomalies of proton-capture elements in globular clusters, MNRAS, 437, L21 (2014)

\bibitem{sdm09}
de Mink, S. E., Pols, O. R., Langer, N. \& Izzard, R.:
Massive binaries as the source of abundance anomalies in globular clusters, A\&A, 507, L1 (2009)

\bibitem{renzini15}
Renzini, A., D'Antona, F., Cassisi, S. et al.:
The Hubble Space Telescope UV Legacy Survey of Galactic Globular Clusters - V. Constraints on formation scenarios, MNRAS, 454, 4197 (2015)

\bibitem{sarajedini07}
Sarajedini, A., Bedin, L. R., Chaboyer, B. et al.:
The ACS Survey of Galactic Globular Clusters. I. Overview and Clusters without Previous Hubble Space Telescope Photometry, AJ, 133, 1658 (2007)

\bibitem{bedin00}
Bedin, L. R., Piotto, G., Zoccali, M. et al.: The anomalous Galactic globular cluster NGC 2808. Mosaic wide-field multi-band photometry, A\&A, 363, 159 (2000)

\bibitem{dc15}
Di Criscienzo, M., Tailo, M., Milone, A. P. et al.:
An HST/WFC3 view of stellar populations on the horizontal branch of NGC 2419, MNRAS, 446, 1469 (2015)

\bibitem{piotto02}
Piotto, G., King, I. R., Djorgovski, S. G. et al.:
HST color-magnitude diagrams of 74 Galactic globular clusters in the HST F439W and F555W bands, A\&A, 391, 945 (2002)

\bibitem{dantona02}
D'Antona, F., Caloi, V., Montalb\'{a}n, J., Ventura, P. \& Gratton, R.:
Helium variation due to self-pollution among Globular Cluster stars. Consequences on the horizontal branch morphology, A\&A, 395, 69 (2002)

\bibitem{cda07}
Caloi, V. \& D'Antona,F.: NGC 6441: another indication of very high helium content in globular cluster stars, A\&A, 463, 959 (2007)


\bibitem{pilecki21}
Pilecki, B., Pietrzy\'{n}ski, G., Anderson, R. I. et al.: Cepheids with Giant Companions. I. Revealing a Numerous Population of Double-lined Binary Cepheids, 
ApJ, 910, 118 (2021)


\bibitem{a82}
Abbott, D. C.: The return of mass and energy to the interstellar medium by winds from early-type stars, ApJ, 273, 723 (1982)

\bibitem{a13}
Agertz, O., Kravtsov, A. V., Leitner, S. N. \& Gnedin, N. Y.: Toward a Complete Accounting of Energy and Momentum from Stellar Feedback in Galaxy Formation Simulations, ApJ, 770, 25 (2013)

\bibitem{ebm18}
Emerick, A., Bryan, G. L. \& Mac Low, M.-M.: Stellar Radiation Is Critical for Regulating Star Formation and Driving Outflows in Low-mass Dwarf Galaxies, ApJ, 865, L22 (2018)


\bibitem{evans07} Evans, C. J., Bresolin, F., Urbaneja, M. A. et al.:
The ARAUCARIA Project: VLT-FORS Spectroscopy of Blue Supergiants in NGC 3109 -- Classifications, First Abundances, and Kinematics.
ApJ, 659, 1198 (2007)

\bibitem{castro55} Castro, N., Urbanejea, M. A., Herrero, A. et al.:
The ARAUCARIA project: Grid-based quantitative spectroscopic study of massive blue stars in NGC 55.
A\&A, 542, A79 (2012)

\bibitem{lamers72} Lamers, H. J.:
The spectrum of the supergiant $\epsilon$ Orionis (BO Ia). I. Identifications, equivalent-widths, line profiles.
A\&AS, 7, 113 (1972)

\bibitem{dufton} Dufton, P. L. \& McKeith, C. D.:
Copernicus observations of neutral helium lines in early-type stars.
A\&A, 81, 8 (1980)

\bibitem{drissen95} Drissen, L., Moffat, A. F. J., Walborn, N. R. \& Shara, M. M.:
The Dense Galactic Starburst NGC 3603. I. HST/FOS Spectroscopy of Individual Stars in the Core and the source of Ionization and Kinetic Energy.
AJ, 110, 2235 (1995)

\bibitem{martins15} Martins, F., Herv\'{e}, A., Bouret, J.-C. et al.:
The MiMeS survey of magnetism in massive stars: CNO surface abundances of Galactic O stars.
A\&A, 575, A34 (2015)

\bibitem{cooke18}
Cooke, R., Pettini, M. \& Steidel, C. C.: One Percent Determination of the Primordial Deuterium Abundance, ApJ, 855, 102 (2018)

\bibitem{m16}
Marcucci, L. E., Mangano, G., Kievsky, A. \& Viviani, M.:
Implication of the Proton-Deuteron Radiative Capture for Big Bang Nucleosynthesis, 
PhRvL, 116, 102501 (2016)

\bibitem{planck16}
Planck Collaboration, Ade, P. A. R., Aghanim, N. et al.: Planck 2015 results. XIII. Cosmological parameters, A\&A, 594, A13 (2016)

\bibitem{mcgaugh10}
McGaugh, S., S., Schombert, J. M., de Blok, W. J. G. \& Zagursky, M. J.:
The Baryon Content of Cosmic Structures, ApJ, 708, L14 (2010)

\bibitem{werk14}
Werk, J. K., Prochaska, J. X., Tumlinson, J. et al.:
The COS-Halos Survey: Physical Conditions and Baryonic Mass in the Low-redshift Circumgalactic Medium,
ApJ, 792, 8 (2014)

\bibitem{nicastro18}
Nicastro, F., Kaastra, J., Krongold, Y. et al.:
Observations of the missing baryons in the warm-hot intergalactic medium, Nature, 558, 406 (2018)

\bibitem{chatterjee17}
Chatterjee, S., Law, C. J., Wharton, R. S. et al.:
A direct localization of a fast radio burst and its host, Nature, 541, 58 (2017)
 
 \bibitem{macquart20}
Macquart, J.-P., Prochaska, J. X., McQuinn, M. et al.: A census of baryons in the Universe from localized fast radio bursts, Nature, 581, 391 (2020)

\bibitem{kcwi}
Morrissey, P., Matuszewki, M., Martin, C. D. et al.:
The Keck Cosmic Web Imager Integral Field Spectrograph, ApJ, 864, 93 (2018)

\bibitem{puech18}
Puech, M., Evans, C. J., Disseau, K. et al.:
Simulating surveys for ELT-MOSAIC: status of the MOSAIC science case after phase A, Proc. SPIE, 10702, 8R (2018)

\bibitem{japelj19}
Japelj, J., Laigle, C., Puech, M. et al.: Simulating MOS science on the ELT: Ly$\alpha$ forest tomography, A\&A, 632, A94 (2019)

\bibitem{md14}
Madau, P. \& Dickinson, M.: Cosmic Star-Formation History, ARA\&A, 52, 415 (2014)

\bibitem{t11}
Tumlinson, J., Thom, C., Werk, J. K. et al.: 
The Large, Oxygen-Rich Halos of Star-Forming Galaxies Are a Major Reservoir of Galactic Metals, 
Sci, 334, 948 (2011)

\bibitem{vdo16}
D'Odorico, V., Cristiani, S., Pomante, E. et al.:
Metals in the z\,$\sim$\,3 intergalactic medium: results from an ultra-high signal-to-noise ratio UVES quasar spectrum, MNRAS, 463, 2690 (2016)

\bibitem{calderone19}
Calderone, G., Boutsia, K., Cristiani, S. et al.: Finding the Brightest Cosmic Beacons in the Southern Hemisphere, ApJ, 887, 268 (2019)

\bibitem{boutsia21}
Boutsia, K., Grazian, A., Fontanot, F. et al.: The Luminosity Function of Bright QSOs at z\,$\sim4$\,4 and Implications for the Cosmic Ionizing Background, ApJ, 912, 111 (2021)


\bibitem{nd08}
Noterdaeme, P., Petitjean, P., Ledoux, C., Srianand, R. \& Ivanchik, A.:
HD molecules at high redshift. A low astration factor of deuterium in a solar-metallicity DLA system at z = 2.418, A\&A, 491, 397 (2008)
(2008)

\bibitem{balashev19}
Balashev, S., Klimenko, V. V., Noterdaeme, P. et al.: X-shooter observations of strong H$_2$-bearing DLAs at high redshift, MNRAS, 490, 2668 (2019)

\bibitem{bolmer19}
Bolmer, J, Ledoux, C., Wiseman, P. et al.:
Evidence for diffuse molecular gas and dust in the hearts of gamma-ray burst host galaxies. Unveiling the nature of high-redshift damped Lyman-$\alpha$ systems, A\&A, 623, A43 (2019)

\bibitem{nd19}
Noterdaeme, P., Balashev, S., Krogager, J.-K. et al.:
Proximate molecular quasar absorbers. Excess of damped H$_2$ systems at $z_{\rm abs}$\,$\approx$\,zQSO in SDSS DR14, A\&A, 627, A32 (2019)

\bibitem{nd21}
Noterdaeme, P., Balashev, S., Krogager, J.-K. et al.:
Down-the-barrel observations of a multi-phase quasar outflow at high redshift. VLT/X-shooter spectroscopy of the proximate molecular absorber at $z$\,$=$\,2.631 towards SDSS J001514$+$184212, A\&A, 646, A108 (2021)

\bibitem{klimenko20}
Klimenko, V.V., Balashev, S.A.:
Physical conditions in the diffuse interstellar medium of local and high-redshift galaxies: measurements based on the excitation of H$_2$ rotational and CI fine-structure levels, MNRAS, 498, 1531, (2020)

\bibitem{kosenko21}
Kosenko, D.N., Balashev, S.A., Noterdaeme, P. et al.:
HD molecules at high redshift: cosmic ray ionization rate in the diffuse interstellar medium, MNRAS, 505, 3810, (2021)


\bibitem{nd17}
Noterdaeme, P., Krogager, J.-K., Balashev, S. et al.:
Discovery of a Perseus-like cloud in the early Universe. H~{\scriptsize I}-to-H$_2$ transition, carbon monoxide and small dust grains at $z_{\rm abs}$\,$\approx$\,2.53 towards the quasar J0000$+$0048, A\&A, 597, A82 (2017)

\bibitem{balashev17}
Balashev, S. A., Noterdeame, P., Rahmani, H. et al.:
CO-dark molecular gas at high redshift: very large H$_2$ content and high pressure in a low-metallicity damped Lyman alpha system, MNRAS, 470, 2890 (2017)

\bibitem{f19}
Finkelstein, S. L., D'Aloisio, A., Paardekooper, J.-P. et al.:
Conditions for Reionizing the Universe with a Low Galaxy Ionizing Photon Escape Fraction,
ApJ, 879, 36 (2019)

\bibitem{robertson15}
Robertson, B. E., Ellis, R. S., Furlanetto, S. R. \& Dunlop, J. S.: Cosmic Reionization and Early Star-forming Galaxies: A Joint Analysis of New Constraints from Planck and the Hubble Space Telescope, ApJ, 802, L19

\bibitem{giallongo19}
Giallongo, E., Grazian, A., Fiore, F. et al.:
Space Densities and Emissivities of Active Galactic Nuclei at $z$\,$>$\,4, ApJ, 884, 19 (2019)

\bibitem{fontanot12}
Fontanot, F., Cristiani, S. \& Vanzella, E.:
On the relative contribution of high-redshift galaxies and active galactic nuclei to reionization, MNRAS, 425, 1413 (2012)

\bibitem{hs15}
Haardt, F. \& Salvaterra, R.: High-redshift active galactic nuclei and H I reionisation: limits from the unresolved X-ray background, A\&A, 575, L16 (2015)

\bibitem{poulin15}
Poulin, V., Serpico, P. D. \& Lesgourgues, J.: Dark Matter annihilations in halos and high-redshift sources of reionization of the universe, JCAP, 12, 41 (2015)

\bibitem{worseck14}
Worseck, G., Prochaska, J. X., O'Meara, J. M. et al.:
The Giant Gemini GMOS survey of zem > 4.4 quasars - I. Measuring the mean free path across cosmic time, MNRAS, 445, 1745 (2014)

\bibitem{cristiani16}
Cristiani, S., Serrano, L. M., Fontanot, F., Vanzella, E. \& Monaco, P.:
The spectral slope and escape fraction of bright quasars at $z$\,$\sim$\,3.8: the contribution to the cosmic UV background, MNRAS, 462, 2478 (2016)

\bibitem{grazian18}
Grazian, A., Giallongo, E., Boutsia, K. et al.:
The contribution of faint AGNs to the ionizing background at $z$\,$\sim$\,4, A\&A, 613, A44
(2018)

\bibitem{izotov16a}
Izotov, Y. I., Orlitov\'{a}, I., Schaerer, D. et al.: Eight per cent leakage of Lyman continuum photons from a compact, star-forming dwarf galaxy, Nature, 529, 178 (2016)

\bibitem{izotov16b}
Izotov, Y. I., Schaerer, D., Thuan, T. X. et al.:
Detection of high Lyman continuum leakage from four low-redshift compact star-forming galaxies, MNRAS, 461, 3683 (2016)

\bibitem{izotov18}
Izotov, Y. I., Schaerer, D., Worseck, G. et al.:
J1154$+$2443: a low-redshift compact star-forming galaxy with a 46 per cent leakage of Lyman continuum photons, MNRAS, 474, 4514 (2018)

\bibitem{shapley16}
Shapley, A. E., Steidel, C. C., Strom, A. L. et al.:
Q1549-C25: A Clean Source of Lyman-Continuum Emission at $z$\,$=$\,3.15, ApJ, 826, L24 (2016)

\bibitem{vanzella16}
Vanzella, E., de Barros, S., Vasei, K. et al.:
Hubble Imaging of the Ionizing Radiation from a Star-forming Galaxy at $z$\,$=$\,3.2 with f$_{\rm esc}$\,$>$\,50\%, ApJ, 825, 41 (2016)

\bibitem{bian17}
Bian, F., Fan, X., McGreer, I., Cai, Z. \& Jiang, L.:
High Lyman Continuum Escape Fraction in a Lensed Young Compact Dwarf Galaxy at $z$\,$=$\,2.5, ApJ, 837, L12 (2017)

\bibitem{vanzella18}
Vanzella, E., Nonino, M., Cupani, G. et al.:
Direct Lyman continuum and Ly $\alpha$ escape observed at redshift 4, MNRAS, 476, L15 (2018)

\bibitem{saha20}
Saha, K., Tandon, S. N., Simmonds, C. et al.:
AstroSat detection of Lyman continuum emission from a $z$\,$=$\,1.42 galaxy, NatAs, 4, 1185 (2020)

\bibitem{rt19}
Rivera-Thorsen, T., Emil, D., H\r{a}kon, C. et al.: Gravitational lensing reveals ionizing ultraviolet photons escaping from a distant galaxy, Sci, 366, 738  (2019)

\bibitem{bik18}
Bik, A., \"{O}stlin, G., Menacho, V. et al.:
Super star cluster feedback driving ionization, shocks and outflows in the halo of the nearby starburst ESO 338-IG04, A\&A, 619, A131 (2018)

\bibitem{micheva19}
Micheva,G., Christian Herenz, E., Roth, M. M., \"{O}stlin, G. \& Girichidis, P.:
IFU investigation of possible Lyman continuum escape from Mrk 71/NGC 2366, A\&A, 623, A145 (2019)

\bibitem{heckman11}
Heckman, T. M., Borthakur, S., Overzier, R. et al.:
Extreme Feedback and the Epoch of Reionization: Clues in the Local Universe, ApJ, 730, 5 (2011)

\bibitem{vanzella20}
Vanzella, E., Caminha, G. B., Calura, F. et al.: Ionizing the intergalactic medium by star clusters: the first empirical evidence, MNRAS, 491, 1093 (2020)

\bibitem{steidel18}
Steidel, C. C., Bogosavljevi\'{c}, M., Shapley, A. E. et al.:
The Keck Lyman Continuum Spectroscopic Survey (KLCS): The Emergent Ionizing Spectrum of Galaxies at $z$\,$\sim$\,3, ApJ, 869, 123 (2018)

\bibitem{piran13}
Piran, T., Bromberg, O., Nakar, E. \& Sari, R.:
The long, the short and the weak: the origin of gamma-ray bursts, RSPTA, 371, 0273 (2013)

\bibitem{zhang07}
Zhang, B.: Gamma-ray burst afterglows, AdSpR, 40, 1186 (2007)

\bibitem{gehrels04}
Gehrels, N., Chincarini, G., Giommi, P. et al.:
The Swift Gamma-Ray Burst Mission, ApJ, 611, 1005 (2004)

\bibitem{fox08}
Fox, A. J., Ledoux, C., Vreeswijk, P. M., Smette, A. \& Jaunsen, A. O.:
High-ion absorption in seven GRB host galaxies at z = 2-4. Evidence for both circumburst plasma and outflowing interstellar gas, A\&A, 2008, 491, 189 (2008)

\bibitem{delia09}
D'Elia, V., Fiore, F., Perna, R. et al.:
The Prompt, High-Resolution Spectroscopic View of the `Naked-Eye' GRB080319B, ApJ, 694, 332 (2009)

\bibitem{vreeswijk07}
Vreeswijk P.~M., Ledoux C., Smette A., et al.: Rapid-response mode VLT/UVES spectroscopy of GRB{\,}060418. Conclusive evidence for UV pumping from the time evolution of Fe II and Ni II excited- and metastable-level populations, A\&A, 468, 83 (2007)



\bibitem{lp98}
Li, L.-X. \& Paczy\,{n}ski, B.: Transient Events from Neutron Star Mergers, ApJ, 507, L59 (1998)

\bibitem{abbott17}
Abbott, B. P., Abbott, R., Abbott, T. D. et al.:
GW170814: A Three-Detector Observation of Gravitational Waves from a Binary Black Hole Coalescence,
PhRvL, 119, 1101 (2017)

\bibitem{tanvir17}
Tanvir, N., Levan, A. J., Gonz\'{a}lez-Fern\'{a}ndez, C. et al.:
The Emergence of a Lanthanide-rich Kilonova Following the Merger of Two Neutron Stars, ApJ, 848, L27 (2017)

\bibitem{bk13}
Barnes, J. \& Kasen, D.:
Effect of a High Opacity on the Light Curves of Radioactively Powered Transients from Compact Object Mergers,
ApJ, 775, 18 (2013)

\bibitem{evans17}
Evans, P. A., Cenko, S. B., Kennea, J. A. et al.: Swift and NuSTAR observations of GW170817: Detection of a blue kilonova, Sci, 358, 1565 (2017)

\bibitem{arcavi18}
Arcavi, I.: The First Hours of the GW170817 Kilonova and the Importance of Early Optical and Ultraviolet Observations for Constraining Emission Models, ApJ, 855, L23 (2018)




\bibitem{quimby07}
Quimby R.~M., Aldering G., Wheeler J.~C., et al.: SN 2005ap: A Most Brilliant Explosion, ApJL, 668, L99 (2007)

\bibitem{gal-yam12}
Gal-Yam, A.: Luminous Supernovae, Sci, 337, 927 (2012)

\bibitem{gal-yam19}
Gal-Yam, A.: The Most Luminous Supernovae, ARA\&A, 57, 305 (2019a)

\bibitem{decia18}
De Cia A., Gal-Yam A., Rubin A., et al.: Light Curves of Hydrogen-poor Superluminous Supernovae from the Palomar Transient Factory, ApJ, 860, 100 (2018)

\bibitem{lunnan18}
Lunnan R., Chornock R., Berger E., et al.: Hydrogen-poor Superluminous Supernovae from the Pan-STARRS1 Medium Deep Survey, ApJ, 852, 81 (2018)

\bibitem{angus19}
Angus C.~R., Smith M., Sullivan M., et al.: Superluminous supernovae from the Dark Energy Survey, MNRAS, 487, 2215 (2019)

\bibitem{chen22a}
Chen Z.~H., Yan L., Kangas T., et al.: The Hydrogen-Poor Superluminous Supernovae from the Zwicky Transient Facility Phase-I Survey: I. Data, arXiv:2202.02059 (2022a)

\bibitem{chen22b}
Chen Z.~H., Yan L., Kangas T., et al.: The Hydrogen-Poor Superluminous Supernovae from the Zwicky Transient Facility Phase-I Survey: II. Light Curve Modeling and Analysis, arXiv:2202.02060 (2022b)

\bibitem{cooke12}
Cooke J., Sullivan M., Gal-Yam A., et al.: Superluminous supernovae at redshifts of 2.05 and 3.90, Nature, 491, 228 (2012)

\bibitem{heger02}
Heger \& Woosley: The Nucleosynthetic Signature of Population III, ApJ, 567, 532 (2002)

\bibitem{gal-yam09}
Gal-Yam A., Mazzali P., Ofek E.~O., et al.: Supernova 2007bi as a pair-instability explosion, Nature, 462, 624 (2009)

\bibitem{nicholl13}
Nicholl M., Smartt S.~J., Jerkstrand A., et al.: Slowly fading super-luminous supernovae that are not pair-instability explosions, Nature, 502, 346 (2013)

\bibitem{inserra17}
Inserra C., Nicholl M., Chen T.-W., et al.: Complexity in the light curves and spectra of slow-evolving superluminous supernovae, MNRAS, 468, 4642 (2017)

\bibitem{neill11}
Neill J.~D., Sullivan M., Gal-Yam A., et al.: The Extreme Hosts of Extreme Supernovae, ApJ, 727, 15 (2011)

\bibitem{vreeswijk14}
Vreeswijk P.~M., Savaglio S., Gal-Yam A., et al.: The Hydrogen-poor Superluminous Supernova iPTF 13ajg and its Host Galaxy in Absorption and Emission, ApJ, 797, 24 (2014)

\bibitem{leloudas15}
Leloudas G., Schulze S., Kr{\"u}hler T., et al.: Spectroscopy of superluminous supernova host galaxies. A preference of hydrogen-poor events for extreme emission line galaxies, MNRAS, 449, 917 (2015)

\bibitem{lunnan15}
Lunnan R., Chornock R., Berger E., Rest A., Fong W., Scolnic D., Jones D.~O., et al.: Zooming In on the Progenitors of Superluminous Supernovae With the HST, ApJ, 804, 90 (2015)

\bibitem{perley16}
Perley D.~A., Quimby R.~M., Yan L., et al.: Host-galaxy Properties of 32 Low-redshift Superluminous Supernovae from the Palomar Transient Factory, ApJ, 830, 13 (2016)

\bibitem{chen17}
Chen T.-W., Smartt S.~J., Yates R.~M., et al.: Superluminous supernova progenitors have a half-solar metallicity threshold, MNRAS, 470, 3566 (2017)

\bibitem{schulze18}
Schulze S., Kr{\"u}hler T., Leloudas G., et al.: Cosmic evolution and metal aversion in superluminous supernova host galaxies, MNRAS, 473, 1258 (2018)

\bibitem{inserra13}
Inserra C., Smartt S.~J., Jerkstrand A., et al.: Super-luminous Type Ic Supernovae: Catching a Magnetar by the Tail, 2013, ApJ, 770, 128 (2013)

\bibitem{scovacricchi16}
Scovacricchi D., Nichol R.~C., Bacon D., et al..: Cosmology with superluminous supernovae, MNRAS, 456, 1700 (2016)

\bibitem{quimby18}
Quimby R.~M., De Cia A., Gal-Yam A., et al.: Spectra of Hydrogen-poor Superluminous Supernovae from the Palomar Transient Factory, ApJ, 855, 2 (2018)

\bibitem{gal-yam19b} 
Gal-Yam A.: A Simple Analysis of Type I Superluminous Supernova Peak Spectra: Composition, Expansion Velocities, and Dynamics, ApJ, 882, 102 (2019b)

\bibitem{graham19}
Graham M.~J., Kulkarni S.~R., Bellm E.~C., et al.: The Zwicky Transient Facility: Science Objectives, PASP, 131, 078001 (2019)

\bibitem{yan18}
Yan L., Perley D.~A., De Cia A., et al.: Far-UV HST Spectroscopy of an Unusual Hydrogen-poor Superluminous Supernova: SN2017egm, ApJ, 858, 91 (2018)

\bibitem{ledoux06}
Ledoux C., Petitjean P., Fynbo J.~P.~U., et al.: Velocity-metallicity correlation for high-z DLA galaxies: evidence of a mass-metallicity relation?, A\&A, 457, 71 (2006)



\bibitem{decia12}
De Cia A., Ledoux C., Fox A.~J., et al.: Rapid-response mode VLT/UVES spectroscopy of super iron-rich gas exposed to GRB 080310. Evidence of ionization in action and episodic star formation in the host, A\&A, 545, A64 (2012)

\bibitem{lunnan18b}
 Lunnan R., Fransson C., Vreeswijk P.~M., et al.: A UV resonance line echo from a shell around a hydrogen-poor superluminous supernova, NatAs, 2, 887 (2018)




\bibitem{mavis}
Rigaut, F., McDermid, R., Cresci, G. et al.:
MAVIS on the VLT: A Powerful, Synergistic ELT Complement in the Visible, Msngr, 185, 7 (2021)

\bibitem{hires}
Marconi, A., Abreu, M., Adibekyan, V. et al.:
HIRES, the High-resolution Spectrograph for the ELT, Msngr, 192, 27 (2021)



\bibitem{Bar11} Barbuy, B., Spite, M., Hill, V. et al. A\&A, 435, A60 (2011)


\bibitem{steidel04}
Steidel, C. C., Shapley, A. E., Pettini, M. et al. "A Survey of Star-forming Galaxies in the 1.4\,$\lesssim$\,$z$\,$\lesssim$\,2.5 Redshift Desert: Overview", ApJ, 604, 534 (2004)


\bibitem{ExA_Genoni}
Genoni, M., Landoni, M., Cupani, G. et al.: The CUBES Instrument model and simulation tools, ExA, arXiv:2203.15477, in press as part of the CUBES Special Issue.

\bibitem{ExA_Calcines}
Calcines, A., Wells, M., O'Brien, K. et al., "Design of the VLT-CUBES image slicers", ExA, in press as part of the CUBES Special Issue.

\end{thebibliography}


\end{document}